\documentclass[aps,pra,amsmath,amssymb,reprint,floatfix]{revtex4-1}
\usepackage{graphicx}
\usepackage{bm}
\usepackage{array}
\usepackage{dcolumn}

\newcounter{gg}

\newcommand{\zhat}{\bf \hat{z}}

\newcommand{\wt}[1]{\widetilde{#1}}
\newcommand{\rt}{\tilde{\rho}}
\newcommand{\zt}{\tilde{z}}
\newcommand{\Vt}{\widetilde{V}}
\newcommand{\At}{\widetilde{A}}
\newcommand{\gt}{\widetilde{w}}

\begin{document}

\newcommand{\w}{\columnwidth}

\newcommand{\QuantumJumpsFigure}{
\begin{figure}[htbp!]
\centering
\begin{tabular}{m{0.4\w}m{0.6\w}}
    \includegraphics*[width=0.38\w,]{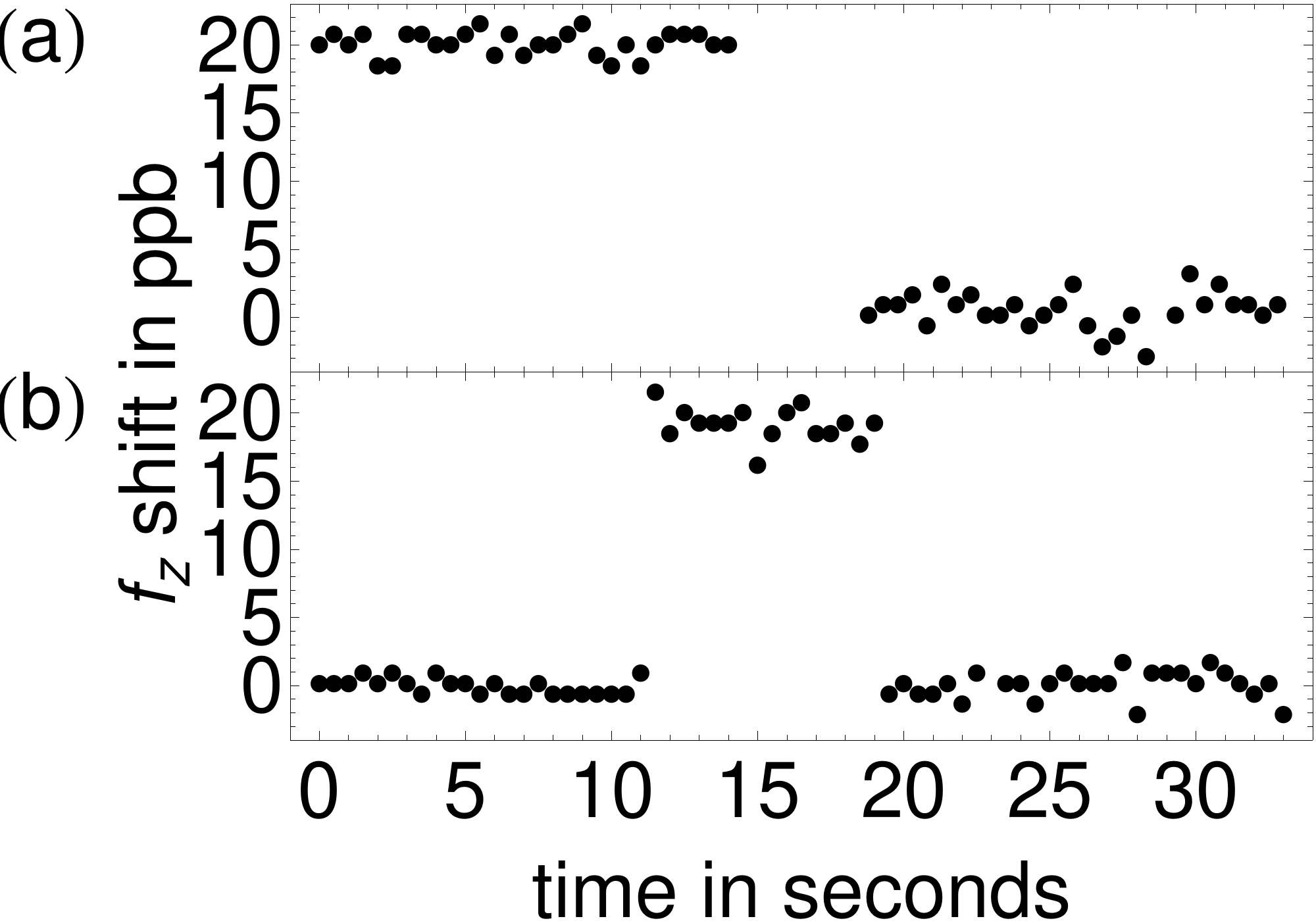}&
    \includegraphics*[width=0.58\w]{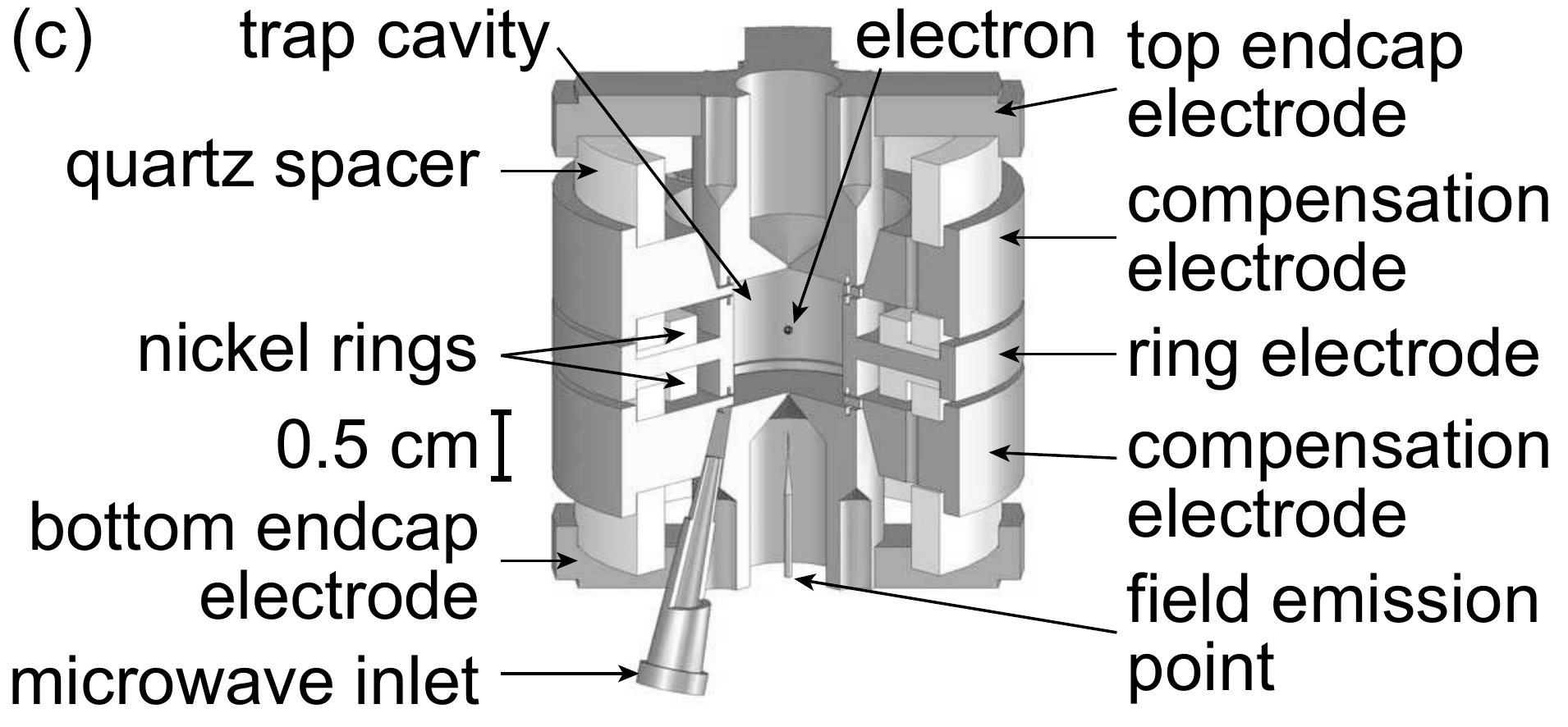}\\
\end{tabular}
\caption{(a) QND observation of a spin flip of one trapped electron.  (b) QND observation of a one-quantum cyclotron transition for one electron.  (c) Cylindrical Penning trap within which the electron is suspended.}
\label{fig:QuantumJumps}
\end{figure}
}

\newcommand{\PlanarTrapThreeDimensionsFigure}{
\begin{figure}[htbp!]
\centering
\includegraphics*[width=\columnwidth]{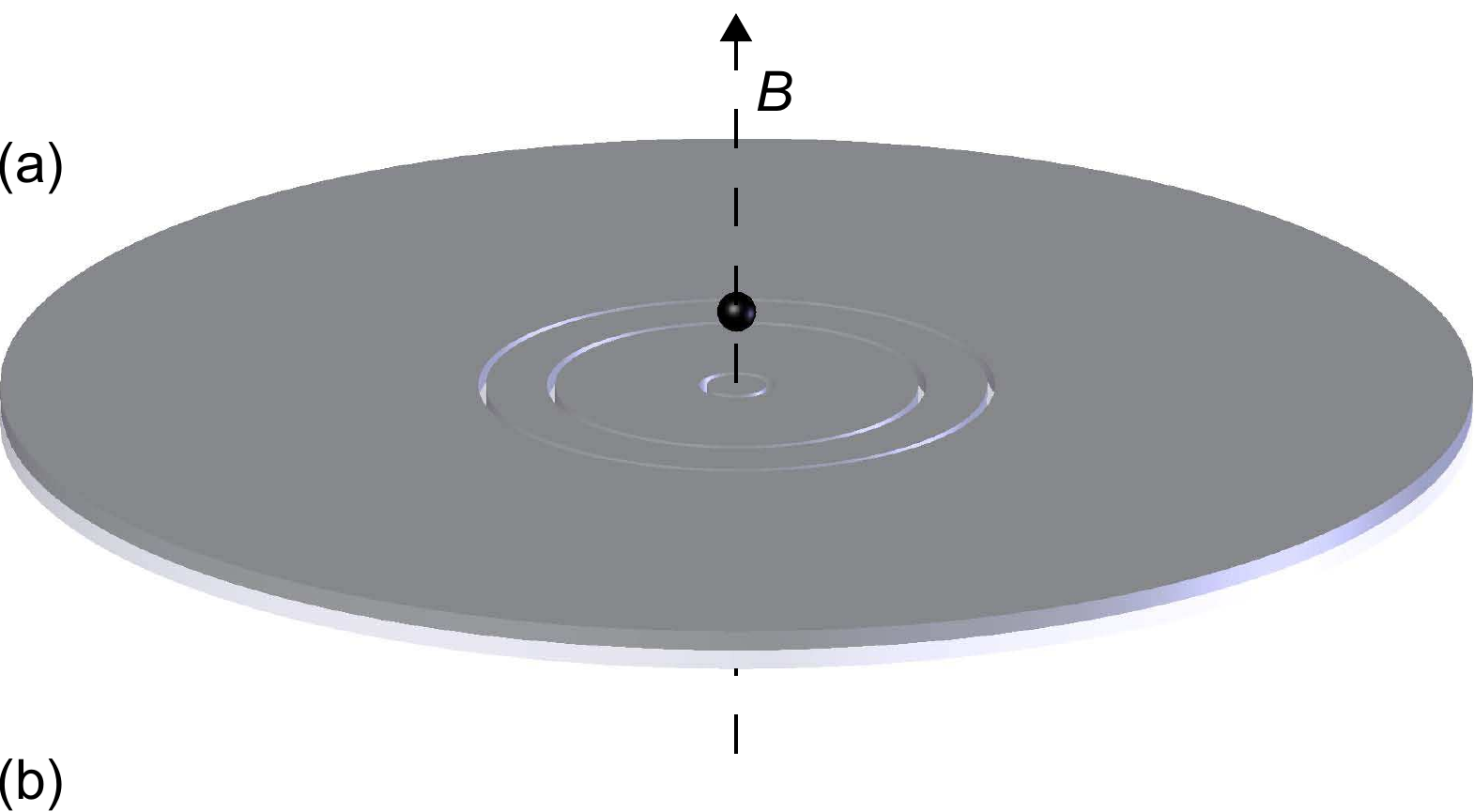}
\includegraphics*[width=\columnwidth]{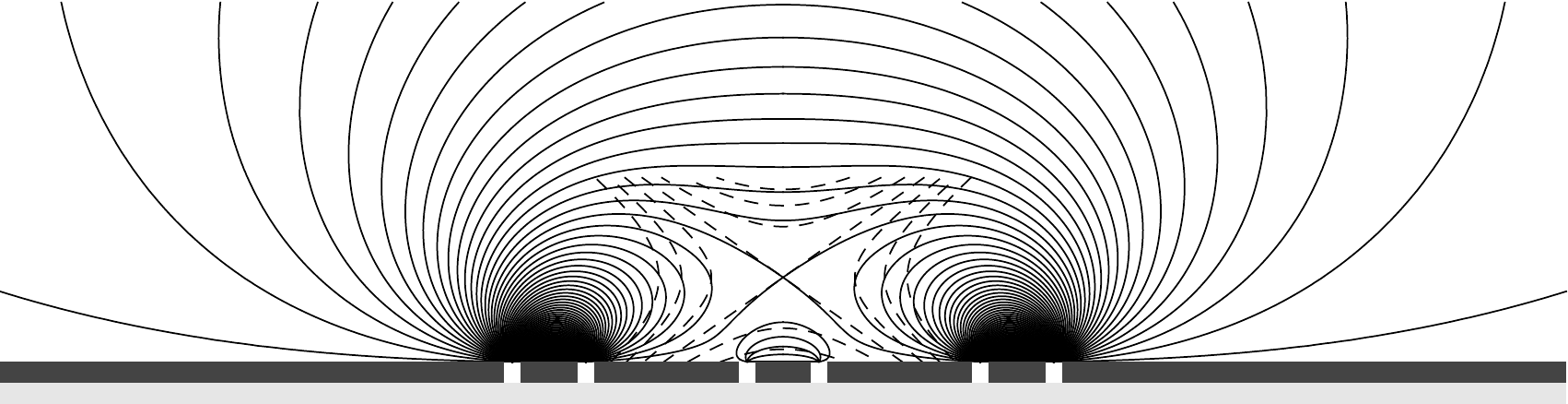}
\caption{(a) Three-gap planar trap with a trapped particle suspended above an electrode plane that extends to infinity.   (b) Side view of trap electrodes and equipotentials spaced by $V_0$, with the infinitesimal gaps between the electrodes widened to make them visible. The equipotentials extend into the gaps between electrodes. The dashed equipotentials of an ideal quadrupole are superimposed near the trap center.}
\label{fig:PlanarTrapThreeDimensions}
\end{figure}
}

\newcommand{\PlanarTrapFigure}{
\begin{figure}[htbp!]
\centering
\includegraphics*[width=\w]{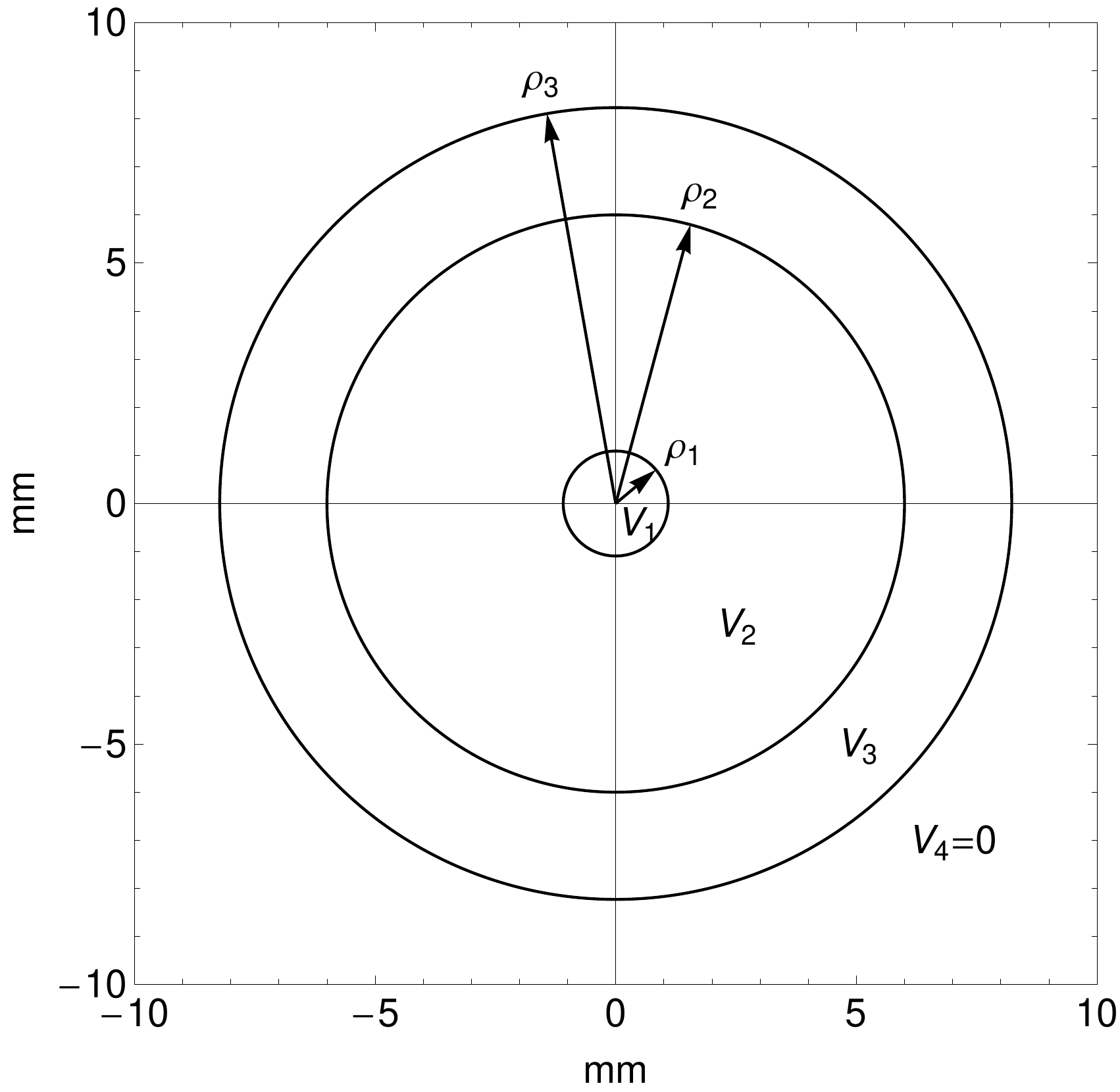}
\caption{Rings of a planar Penning trap.  The relative geometry of the electrodes is that of the
sample trap used to illustrate the general features of planar
traps.  Infinitesimal gaps between electrodes are assumed until Sec.~\ref{sec:Gaps}.}
\label{fig:PlanarTrap}
\end{figure}
}

\newcommand{\SamplePlanarTrapTable}{
\begin{table}
\newcommand\T{\rule{0pt}{2.6ex}}
\newcommand\B{\rule[-1.2ex]{0pt}{0pt}}
\begin{tabular}{|c|rr|rr|}
\cline{2-5}
\multicolumn{1}{c}{} & \multicolumn{4}{|c|}{\T $\{\wt{\rho}_i \} = \{1, 5.5, 7.5426\}$ \B}\\
\cline{2-5}
\multicolumn{1}{c}{~} \T & \multicolumn{2}{|c}{$a_2=a_4=0$} & \multicolumn{2}{|c|}{$C_3=C_4=0$} \\
\multicolumn{1}{c}{} & \multicolumn{1}{|c}{I} & \multicolumn{1}{c}{II} &\multicolumn{1}{|c}{III} & \multicolumn{1}{c|}{IV} \\
\multicolumn{1}{c}{} & \multicolumn{1}{|c}{Eq.~\ref{eq:Constraintsa2a4}} & \multicolumn{1}{c}{Eqs.~\ref{eq:Constraintsa2a4}, \ref{eq:OptimizationHarmonic}~} &\multicolumn{1}{|c}{~Eqs.~\ref{eq:OptimizationC3C4}, \ref{eq:OptimizationHarmonic}} & \multicolumn{1}{c|}{Eq.~\ref{eq:OptimizationC3C4}} \\
\hline
$~\wt{V}_1~$ &\T $-$12.2615 & $-$26.4192 & $-$26.4192 & $-$31.0353\\
$\wt{V}_2$ & $-$16.4972 & $-$27.0861 & $-$27.0861 & $-$31.6642\\
$\wt{V}_3$ &\B $-$79.7942 & $-$111.1415 & $-$111.1415 & $-$120.1261\\
\hline
$\wt{z}_0$ & \T 2.3469 & 1.4351 & 1.4351 & 1.0250\\
\hline
$C_3$ & \T $-$0.1516 & 0.0000 & 0.0000 & 0.0000\\
$C_4$ & 0.0287 & 0.0000 & 0.0000 & 0.0000\\
$C_5$ & $-$0.0156 & $-$0.0112 & $-$0.0112 & 0.0213\\
$C_6$ &\B 0.0064 & 0.0000 & 0.0000 & $-$0.0366\\
\hline
$a_2$ &\T 0.0000 & 0.0000 & 0.0000 & 0.0000\\
$a_3$ & 0.0000 & 0.0000 & 0.0000 & 0.0000\\
$a_4$ & 0.0000 & 0.0000 & 0.0000 & $-$0.0343\\
$a_5$ & 0.0000 & 0.0000 & 0.0000 & 0.0000\\
$a_6$ &\B $-$0.0003 & $-$0.0039 & $-$0.0039 & $-$0.0095\\
\hline
$C_{11}$ &\T $-$0.1205 & $-$0.3737 & $-$0.3737 & $-$0.6810 \\
$C_{12}$ & $-$0.1625 & 0.0443 & 0.0443 & 0.3356 \\
$C_{13}$ & 0.0521 & 0.0780 & 0.0789 & 0.0875 \\
$C_{1d}^{(opt)}$ & $-$0.3280 & $-$0.5364 & $-$0.5364 & $-$0.7510 \\
$\rt_d^{(opt)}$ & \B 3.3191 & 2.0295 & 2.0295 & 1.4496 \\
\hline
$\gamma_1$ &\T $-194.15$ & 3.45 & 3.45 & 3.50 \\
$\gamma_2$ & 8.14 & 2.64 & 2.64 & 4.15 \\
$\gamma_3$ & \B $-1.56$ & $-3.12$ & $-3.12$ & 1.61\\
\hline
\multicolumn{1}{c}{~} \T & \multicolumn{2}{|c}{Fig.~\ref{fig:ThreeGapaTwoAndaFourVanishing} points} & \multicolumn{2}{|c|}{Fig.~\ref{fig:ThreeGapCThreeCFourVanishing} points} \\
\multicolumn{1}{c}{} & \multicolumn{1}{|c}{right} & \multicolumn{1}{c}{left} &\multicolumn{1}{|c}{right} & \multicolumn{1}{c|}{left} \\
\cline{2-5}
\end{tabular}
\caption{Scaled parameters for the sample planar trap geometry.}
\label{table:SampleTrap}
\end{table}
}

\newcommand{\SamplePlanarTrapTableAbsolute}{
\begin{table}
\newcommand\T{\rule{0pt}{2.6ex}}
\newcommand\B{\rule[-1.2ex]{0pt}{0pt}}
\begin{tabular}{|c|rr|rr|c|}
\cline{2-5}
\multicolumn{1}{c}{~} & \multicolumn{4}{|c|}{\T $\{\rho_i \} = \{1.0909, 6, 8.2283\}$ mm} & \multicolumn{1}{c}{ \B}\\
\cline{2-5}
\multicolumn{1}{c}{~} \T & \multicolumn{2}{|c}{$a_2=a_4=0$} & \multicolumn{2}{|c|}{$C_3=C_4=0$} & \multicolumn{1}{c}{} \\
\multicolumn{1}{c}{~} & \multicolumn{1}{|c}{I} & \multicolumn{1}{c}{II} &\multicolumn{1}{|c}{III} & \multicolumn{1}{c|}{IV} & \multicolumn{1}{c}{~}\\
\multicolumn{1}{c}{} & \multicolumn{1}{|c}{Eq.~\ref{eq:Constraintsa2a4}} & \multicolumn{1}{c}{Eqs.~\ref{eq:Constraintsa2a4}, \ref{eq:OptimizationHarmonic}~} &\multicolumn{1}{|c}{~Eqs.~\ref{eq:OptimizationC3C4}, \ref{eq:OptimizationHarmonic}} & \multicolumn{1}{c|}{Eq.~\ref{eq:OptimizationC3C4}} \\
\hline
$\rho_1$ \T & 1.0909 & 1.0909 & 1.0909 & 1.0909 & mm\\
$z_0$ & 2.5603 & 1.5655 & 1.5655 & 1.1182 & mm\\
$\rho_d^{(opt)}$ \B & 3.6208 & 2.2140 & 2.2140 & 1.5814 & mm\\
\hline
$f_z$ \T & 64 & 64 & 64 & 64 & MHz\\
$V_0$ \B & $-$1.0941 & $-$1.0941 & $-$1.0941 & $-$1.0941 & V\\
\hline
$V_1$ & \T 13.4158 & 28.9064 & 28.9064 & 33.9572 & V\\
$V_2$ & 18.0504 & 29.6361 & 29.6361 & 34.6452 & V\\
$V_3$ & \B 87.3065 & 121.6051 & 121.6051 & 131.4355 & V\\
\hline
$\Delta f_z$ \T & 0.0 & 0.0 & 0.0 & 0.0 & Hz \B\\
\hline
$1:~\gamma_z$ & \T 2$\pi\,$1.37  & 2$\pi\,$13.16 & 2$\pi\,$13.16 & 2$\pi\,$43.70 & s$^{-1}$\\
$2:~\gamma_z$ & 2$\pi\,$2.49 & 2$\pi\,$0.19 & 2$\pi\,$0.19 & 2$\pi\,$10.61 & s$^{-1}$\\
$3:~\gamma_z$ & 2$\pi\,$0.26 & 2$\pi\,$0.57 & 2$\pi\,$0.57 & 2$\pi\,$0.72 & s$^{-1}$\\
d:~$\gamma_z^{(opt)}$ & \B 2$\pi\,$10.14 & 2$\pi\,$27.11 & 2$\pi\,$27.11 & 2$\pi\,$53.14 & s$^{-1}$\\
\hline
\multicolumn{1}{c}{~} \T & \multicolumn{2}{|c}{Fig.~\ref{fig:ThreeGapaTwoAndaFourVanishing} points} & \multicolumn{2}{|c|}{Fig.~\ref{fig:ThreeGapCThreeCFourVanishing} points} \\
\multicolumn{1}{c}{} & \multicolumn{1}{|c}{right} & \multicolumn{1}{c}{left} &\multicolumn{1}{|c}{right} & \multicolumn{1}{c|}{left} \\
\cline{2-5}
\end{tabular}
\caption{One set of absolute values for the sample planar trap geometry. The broadening $\Delta f_z$ is for a 5 K thermal
distribution of axial energies.  The damping widths $\gamma_z/(2\pi)$ are for the numbered electrode connected to $R =
100~\rm{k}\Omega$. Thermal frequency spreads $\Delta f_z$ below 1 Hz will be very difficult to realize in practice owing to
imperfections in real traps.} \label{table:SampleTrapAbsolute}
\end{table}
}

\newcommand{\AxialPotentialFigure}{
\begin{figure}[htbp!]
\centering
\includegraphics*[width=\w]{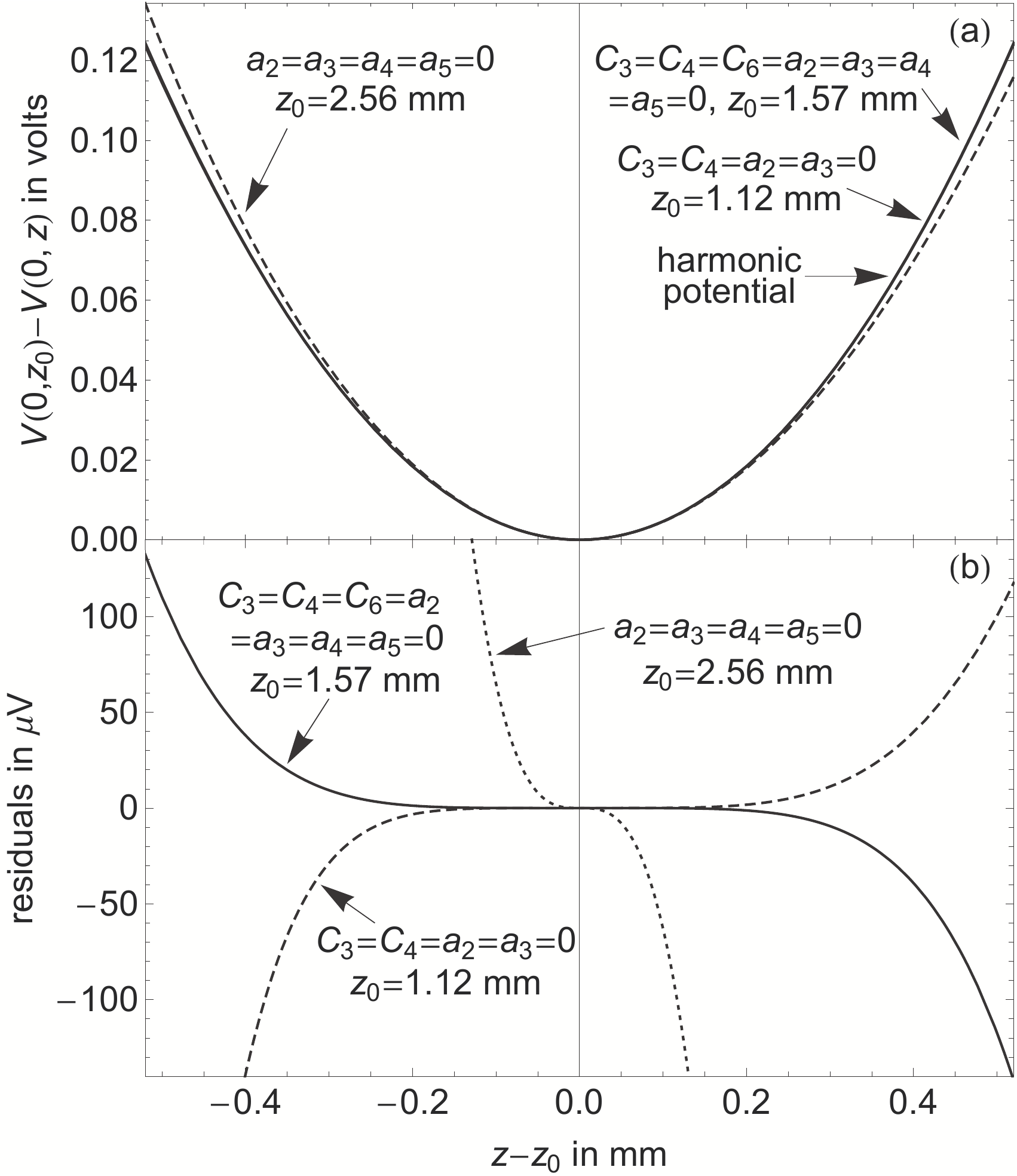}
\caption{(a) Trap potential on axis.  (b) Difference between the trap potential and a perfect harmonic potential on axis.  The labels identify optimized  configurations of the sample trap (Tables~\ref{table:SampleTrap}-\ref{table:SampleTrapAbsolute}), using $C_k$ from Eq.~\ref{eq:ExpandV} and $a_k$ from Eq.~\ref{eq:AmplitudeDependentFrequency}.}
\label{fig:AxialPotential}
\end{figure}
}

\newcommand{\TwoGapFigure}{
\begin{figure}[htbp!]
\includegraphics*[width=\w]{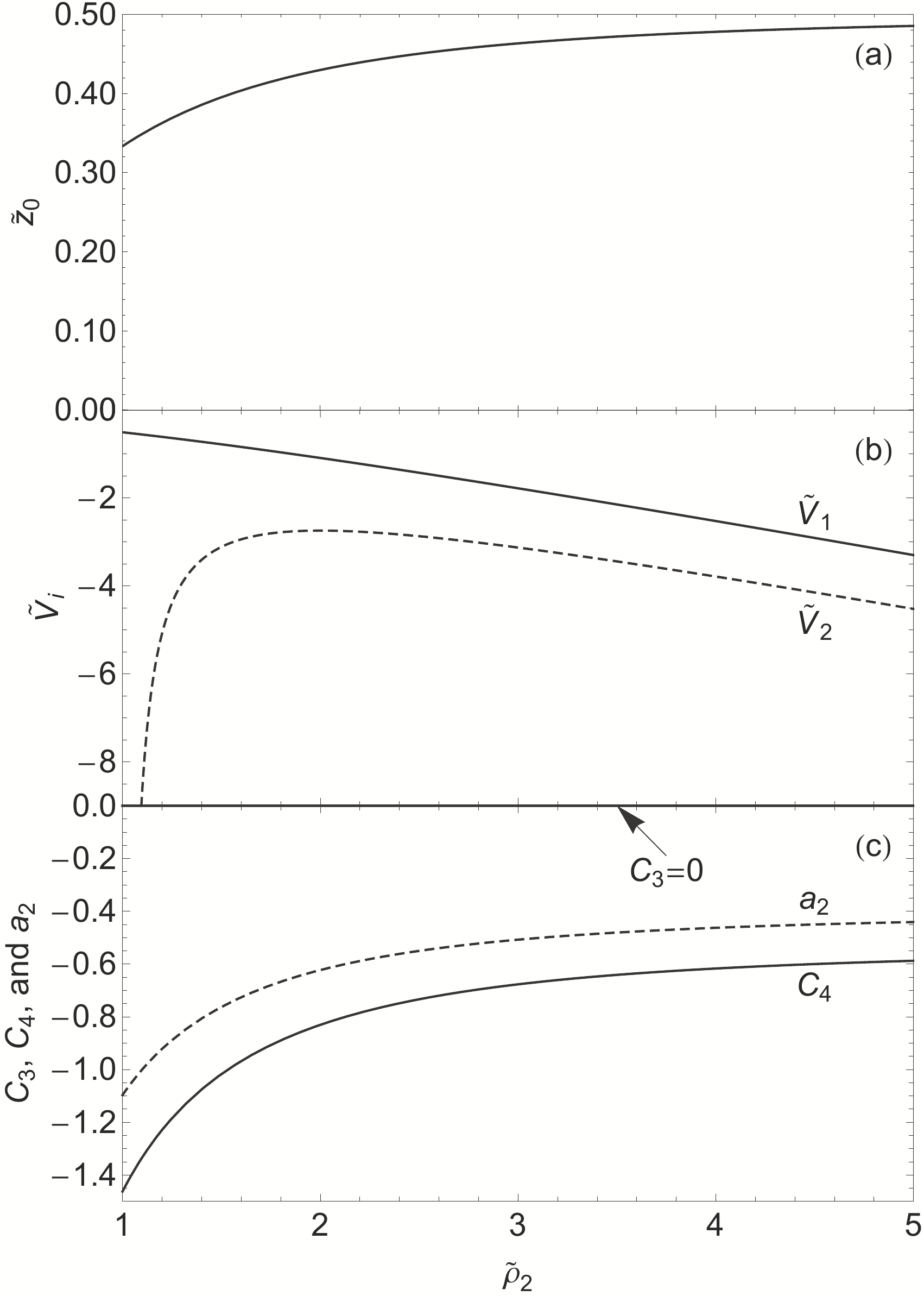}
\caption{Parameters of two-gap trap that make $C_3=0$ as a function of $\rt_2$. }
\label{fig:TwoGap}
\end{figure}
}

\newcommand{\OptimizedTrapFrequencyShiftFigure}{
\begin{figure}[htbp!]
\centering
\includegraphics*[width=\w]{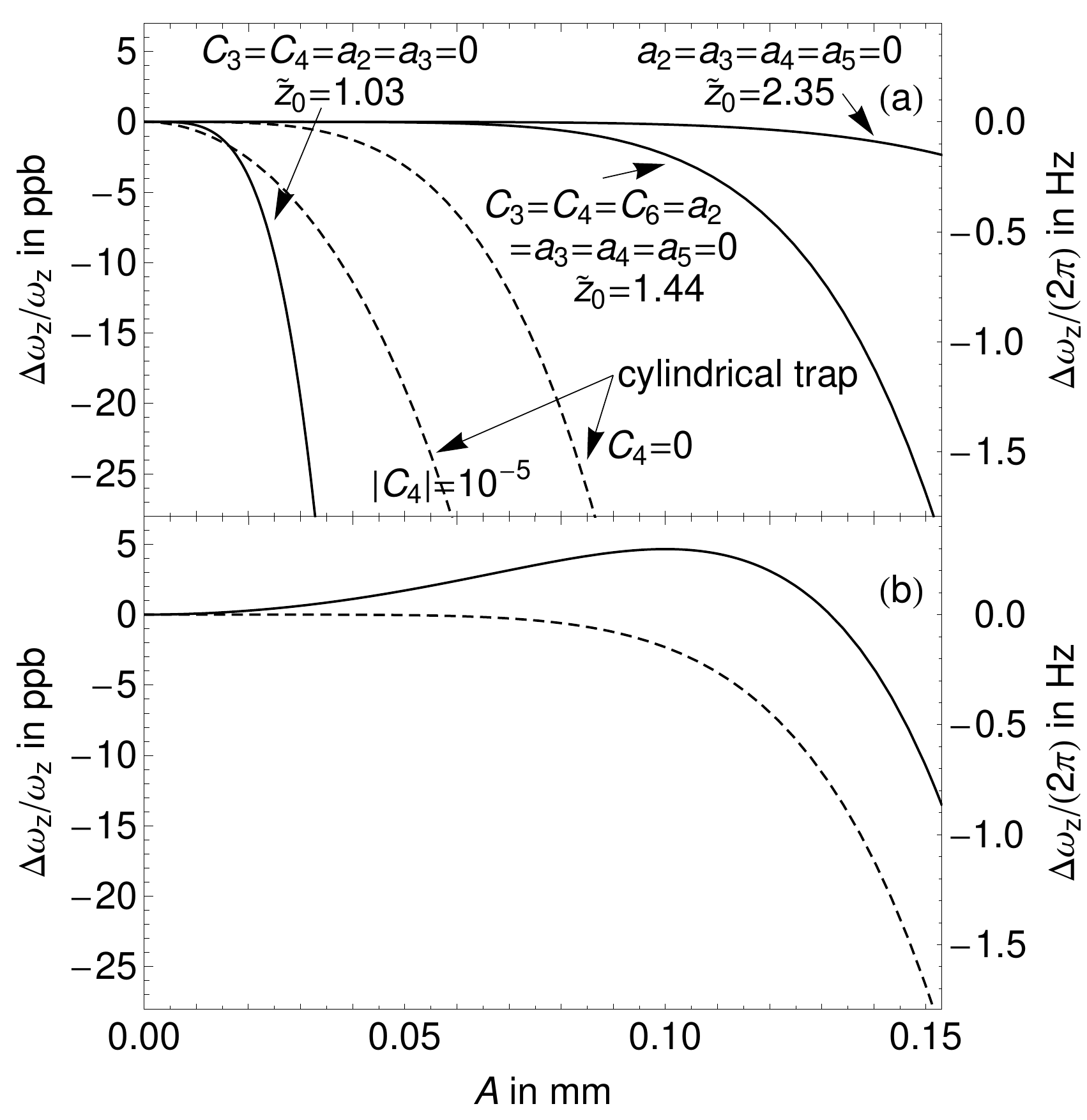}
\caption{(a) Amplitude dependence of $\omega_z(A) \approx 2\pi \,64$ MHz for optimized configurations of the sample trap (Table~\ref{table:SampleTrap}) is comparable or smaller than for the cylindrical trap.  (b) Slight adjustments in the applied potentials minimizes the dependence of $\omega_z$ upon fluctuations about a large oscillation amplitude (solid) rather than for small oscillation amplitudes (dashed).}
\label{fig:OptimizedTrapFrequencyShift}
\end{figure}
}

\newcommand{\PossibleThreeRingGeometriesFigure}{
\begin{figure}[htbp!]
\centering
\includegraphics*[width=\w]{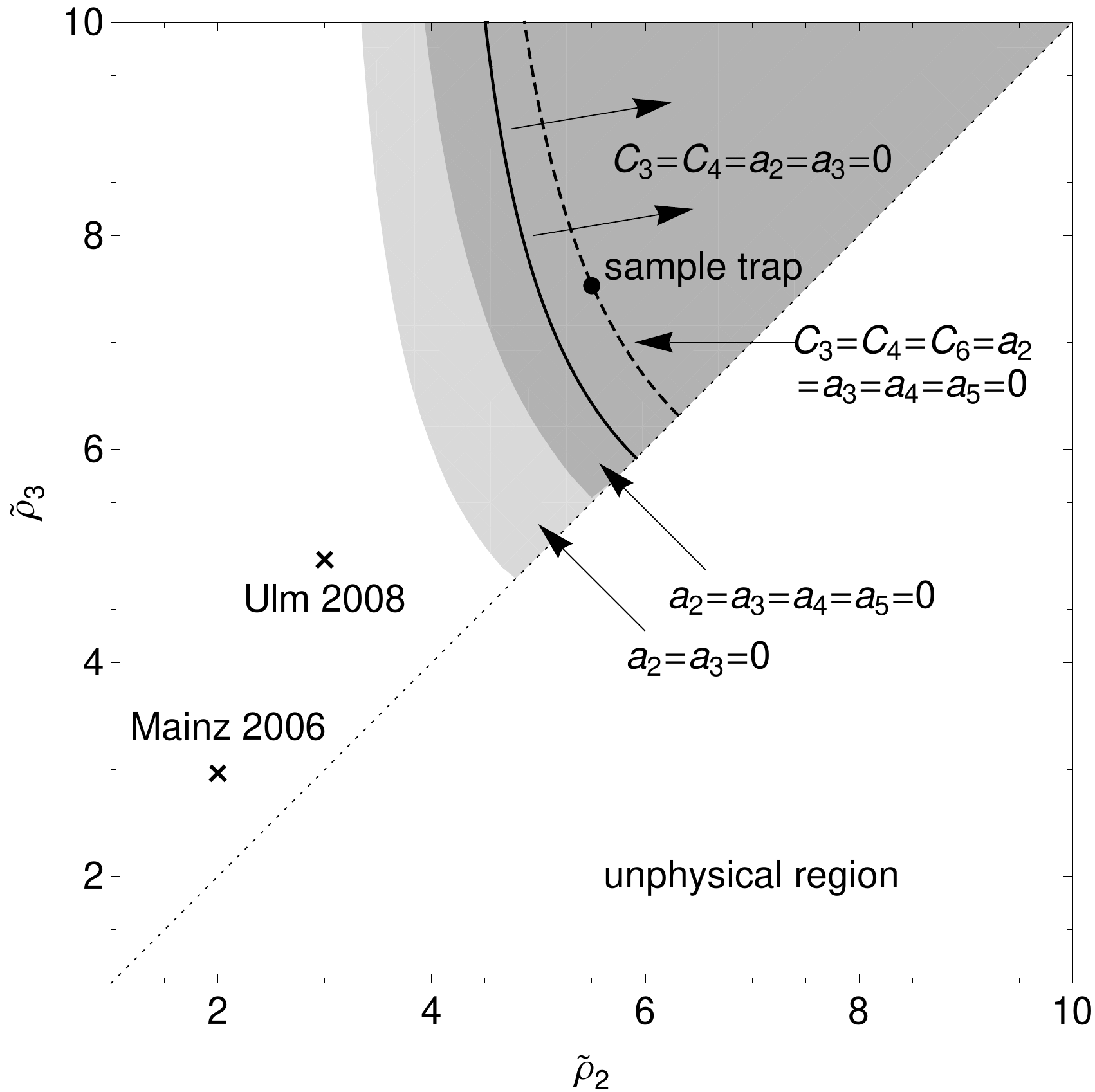}
\caption{The shaded regions for which the indicated $a_k$ can be made to vanish for a three-gap planar Penning trap, along with the region and the curve for
which the indicated $C_k$ can alternatively be made to vanish.  To be avoided is the shaded area near the diagonal boundary
${\rt}_2 = {\rt}_3$ where there is a rather strong and sensitive cancelation between the effect of the potentials $V_2$
and $V_3$. No optimized traps are possible in the unshaded region, with the the earlier traps (crosses) at Mainz
\cite{ElectronsInPlanarPenningTrapMainz} and Ulm \cite{ElectronsInPlanarPenningTrapUlm} as examples.}
\label{fig:PossibleThreeRingGeometries}
\end{figure}
}

\newcommand{\ThreeGapaTwoAndaFourVanishingFigure}{
\begin{figure}[htbp!]
\centering
\includegraphics*[width=\w]{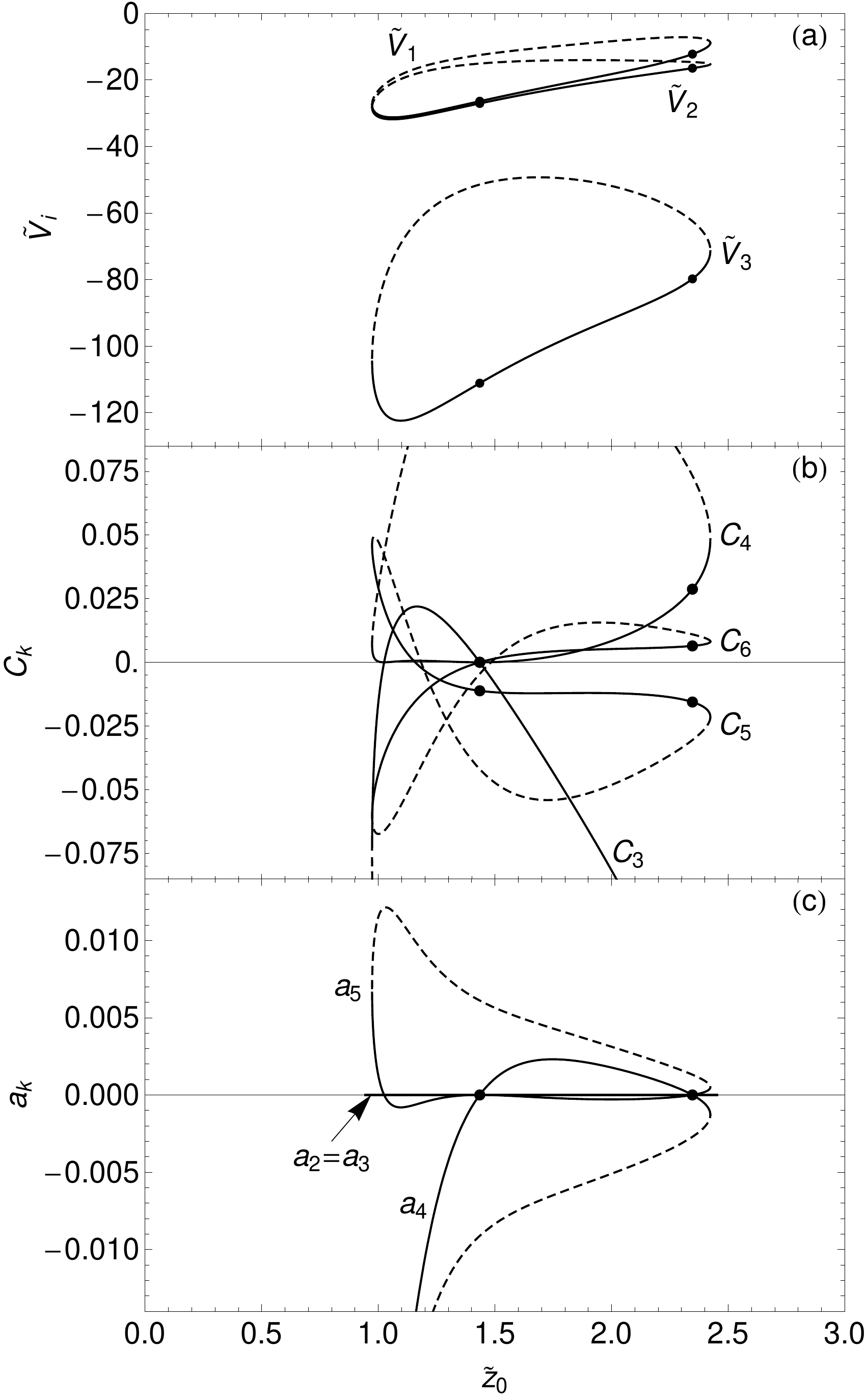}
\caption{(a) Scaled potentials applied to the sample trap electrodes to make a trap with $a_2=0$ as a function of the position
$\zt_0$ of the axial potential minimum.  (b) The corresponding $C_k$.  (c) The corresponding $a_k$.} \label{fig:ThreeGapaTwoAndaFourVanishing}
\end{figure}
}

\newcommand{\ThreeGapCThreeCFourVanishingFigure}{
\begin{figure}[htbp!]
\centering
\includegraphics*[width=\w]{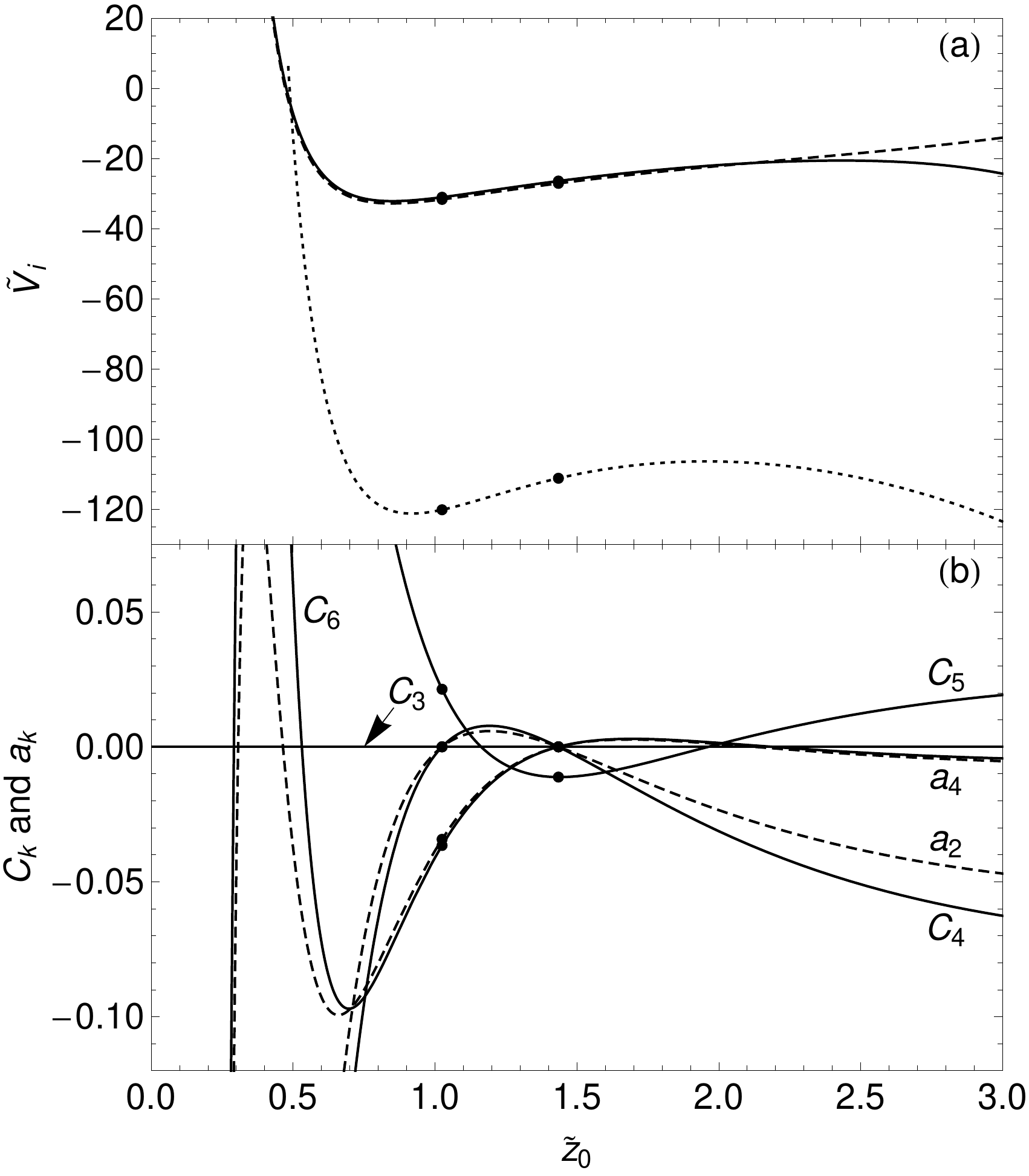}
\caption{(a) Scaled potentials applied to our sample trap electrodes to produce a trap with $C_3=0$ as a function of $\zt_0$.  (b) The resulting $C_k$ and $a_k$.} \label{fig:ThreeGapCThreeCFourVanishing}
\end{figure}
}

\newcommand{\OptimizedHarmonicFigure}{
\begin{figure}[htbp!]
\centering
\includegraphics*[width=\w]{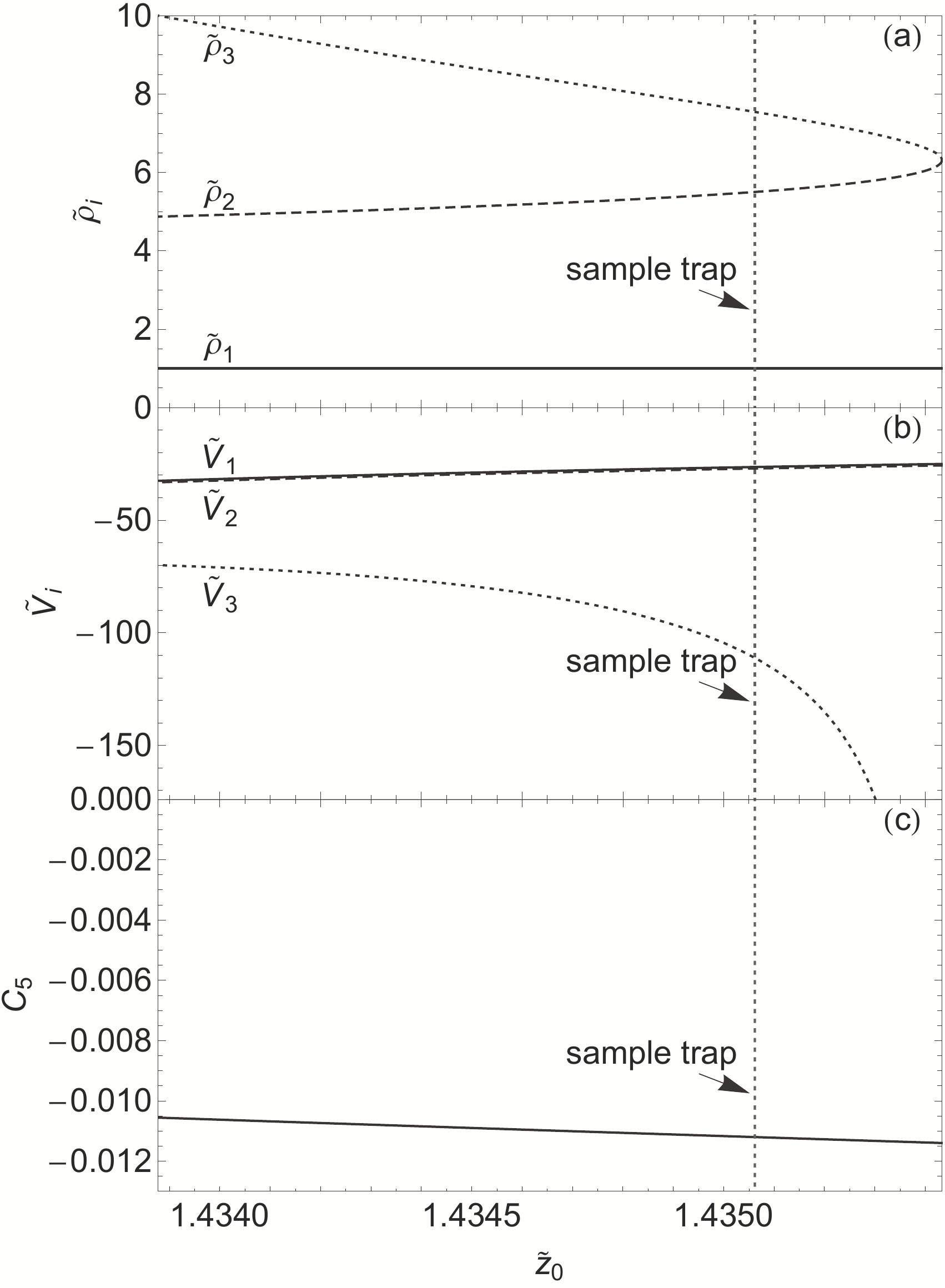}
\caption{(a) An optimized harmonic trap (Eq.~\ref{eq:OptimizationHarmonic}), as illustrated using our sample trap geometry, is possible with only a small range of $\zt_0$ values and relative geometries.  (b) The required scaled potentials.  (c) Values of the non-zero $C_5$.} \label{fig:OptimizedHarmonic}
\end{figure}
}

\newcommand{\FrequencyWidthComparisonFigure}{
\begin{figure}[htbp!]
\centering
\includegraphics*[width=\w]{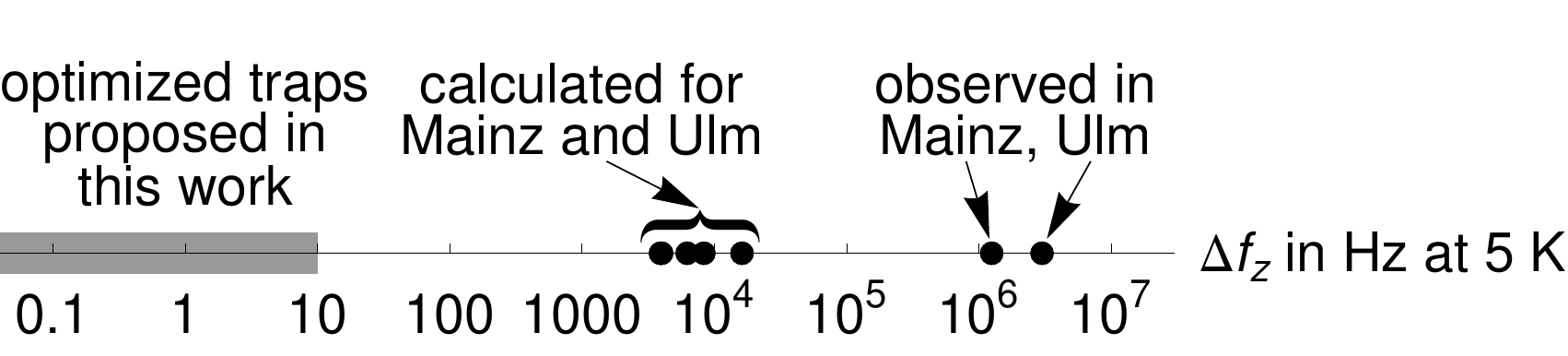}
\caption{Comparison of the frequency widths calculated for optimized planar traps with a 5 K axial temperature.  The imperfections of a real planar trap will likely make it difficult to realize widths below 1 Hz for an axial frequency of 64 MHz.} \label{fig:FrequencyWidthComparison}
\end{figure}
}

\newcommand{\SamplePlanarTrapWithGapsBiasedForNoGapsTable}{
\begin{table}
\newcommand\T{\rule{0pt}{2.6ex}}
\newcommand\B{\rule[-1.2ex]{0pt}{0pt}}
\begin{tabular}{|c|rr|rr|r|}
\cline{2-5}
\multicolumn{1}{c}{} & \multicolumn{4}{|c|}{\T $\{\wt{\rho}_i \} = \{1, 5.5, 7.5426\}$ \B}\\
\cline{2-5}
\multicolumn{1}{c}{~} \T & \multicolumn{2}{|c}{$a_2=a_4=0$} & \multicolumn{2}{|c|}{$C_3=C_4=0$} \\
\multicolumn{1}{c}{} & \multicolumn{1}{|c}{I} & \multicolumn{1}{c}{II} &\multicolumn{1}{|c}{III} & \multicolumn{1}{c|}{IV} \\
\multicolumn{1}{c}{} & \multicolumn{1}{|c}{Eq.~\ref{eq:Constraintsa2a4}} & \multicolumn{1}{c}{Eqs.~\ref{eq:Constraintsa2a4}, \ref{eq:OptimizationHarmonic}~} &\multicolumn{1}{|c}{~Eqs.~\ref{eq:OptimizationC3C4}, \ref{eq:OptimizationHarmonic}} & \multicolumn{1}{c|}{Eq.~\ref{eq:OptimizationC3C4}} \\
\cline{1-5}
$\T \wt{z}_0$ & 2.3469 & 1.4351 & 1.4351 & 1.0250 \\
\cline{1-5}
$C_3$ & \T $-0.1516$ & 0.0000 & 0.0000 & 0.0000 \\
$C_4$ & 0.0287 & 0.0000 & 0.0000 & 0.0001 \\
$C_5$ & $-0.0156$ & $-0.0112$ & $-0.0112$ & 0.0213 \\
$C_6$ &\B 0.0064 & 0.0000 & 0.0000 & $-0.0365$ \\
\cline{1-5}
$a_2$ &\T 0.0000 & 0.0000 & 0.0000 & 0.0000 \\
$a_3$ & 0.0000 & 0.0000 & 0.0000 & 0.0000 \\
$a_4$ & 0.0000 & 0.0000 & 0.0000 & $-0.0342$ \\
$a_5$ & 0.0000 & 0.0000 & 0.0000 & 0.0000 \\
$a_6$ &\B $-0.0003$ & $-0.0039$ & $-0.0039$ & $-0.0095$ \\
\hline
$\T \Delta f_z$ @ 5 K & 0.3 & 0.3 & 0.3 & 2.7 & Hz\\
\hline
\end{tabular}
\caption{Gaps between the electrodes of the sample trap when the potentials for an ideal planar trap (from Tables~\ref{table:SampleTrap}-\ref{table:SampleTrapAbsolute}) are applied
produce different $C_k$ and $a_k$.  The gap width is $w\!=\! 0.002~\rm{in} \!=\! 50~\mu\mathrm{m}$. }
\label{table:SampleTrapWithGapsBiasedForNoGaps}
\end{table}
}

\newcommand{\SamplePlanarTrapWithGapsTable}{
\begin{table}
\newcommand\T{\rule{0pt}{2.6ex}}
\newcommand\B{\rule[-1.2ex]{0pt}{0pt}}
\begin{tabular}{|c|rr|rr|c|}
\cline{2-5}
\multicolumn{1}{c}{} & \multicolumn{4}{|c|}{\T $\{\wt{\rho}_i \} = \{1, 5.5, 7.5426\}$ \B}\\
\cline{2-5}
\multicolumn{1}{c}{~} \T & \multicolumn{2}{|c}{$a_2=a_4=0$} & \multicolumn{2}{|c|}{$C_3=C_4=0$} \\
\multicolumn{1}{c}{} & \multicolumn{1}{|c}{I} & \multicolumn{1}{c}{II} &\multicolumn{1}{|c}{III} & \multicolumn{1}{c|}{IV} \\
\multicolumn{1}{c}{} & \multicolumn{1}{|c}{Eq.~\ref{eq:Constraintsa2a4}} & \multicolumn{1}{c}{Eqs.~\ref{eq:Constraintsa2a4}, \ref{eq:OptimizationHarmonic}~} &\multicolumn{1}{|c}{~Eqs.~\ref{eq:OptimizationC3C4}, \ref{eq:OptimizationHarmonic}} & \multicolumn{1}{c|}{Eq.~\ref{eq:OptimizationC3C4}} \\
\cline{1-5}
$~\delta\wt{V}_1~$ & \T $-0.0008$ & 0.0015 & 0.0015 & $-0.0076$ \\
$\delta\wt{V}_2$ & $-0.0006$ & 0.0014 & 0.0014 & $-0.0076$ \\
$\delta\wt{V}_3$ &\B 0.0034 & 0.0048 & 0.0048 & $-0.0157$ \\
\hline
~$\delta V_1$ \T & 0.0009 & $-0.0016$ & $-0.0016$ & 0.0083 & ~V\\
$\delta V_2$ & 0.0006 & $-0.0015$ & $-0.0015$ & 0.0083 & ~V \\
~$\delta V_3$ \B & $-0.0037$ & $-0.0053$ & $-0.0053$ & 0.0172 & ~V\\
\hline
$\T \Delta f_z$ @ 5 K & 0.0 & 0.0 & 0.0 & 0.0 & Hz \\
\hline
\end{tabular}
\caption{Gaps of width $w\!=\!0.002~\rm{in}\!=\!50~\mu\mathrm{m}$ between the electrodes of our sample trap shift the scaled and absolute potentials that must be applied to obtain the optimized trap configurations. The shifts are with respect to the values calculated for a vanishing gap width in Tables~\ref{table:SampleTrap}-\ref{table:SampleTrapAbsolute}.}
\label{table:SampleTrapWithGaps}
\end{table}
}

\newcommand{\SamplePlanarTrapWithEnclosureBiasedForNoEnclosureTable}{
\begin{table}
\newcommand\T{\rule{0pt}{2.6ex}}
\newcommand\B{\rule[-1.2ex]{0pt}{0pt}}
\begin{tabular}{|c|rr|rr|}
\cline{2-5}
\multicolumn{1}{c}{} & \multicolumn{4}{|c|}{\T $\{\wt{\rho}_i \} = \{1, 5.5, 7.5426\}$ \B}\\
\cline{2-5}
\multicolumn{1}{c}{~} \T & \multicolumn{2}{|c}{$a_2=a_4=0$} & \multicolumn{2}{|c|}{$C_3=C_4=0$} \\
\multicolumn{1}{c}{} & \multicolumn{1}{|c}{I} & \multicolumn{1}{c}{II} &\multicolumn{1}{|c}{III} & \multicolumn{1}{c|}{IV} \\
\multicolumn{1}{c}{} & \multicolumn{1}{|c}{Eq.~\ref{eq:Constraintsa2a4}} & \multicolumn{1}{c}{Eqs.~\ref{eq:Constraintsa2a4}, \ref{eq:OptimizationHarmonic}~} &\multicolumn{1}{|c}{~Eqs.~\ref{eq:OptimizationC3C4}, \ref{eq:OptimizationHarmonic}} & \multicolumn{1}{c|}{Eq.~\ref{eq:OptimizationC3C4}} \\
\hline
$\T \wt{z}_0$ & 2.2406 & 1.2750 & 1.2750 & 0.8481 \\
\hline
$C_3$ & \T $-0.1586$ & $-0.0034$ & $-0.0034$ & 0.0110 \\
$C_4$ & 0.0365 & 0.0082 & 0.0082 & $-0.0414$ \\
$C_5$ & $-0.0193$ & $-0.0081$ & $-0.0081$ & 0.0785 \\
$C_6$ &\B 0.0075 & $-0.0075$ & $-0.0075$ & $-0.0712$ \\
\hline
$a_2$ &\T 0.0038 & 0.0062 & 0.0062 & $-0.0312$ \\
$a_3$ & $-0.0006$ & 0.0000 & 0.0000 & $-0.0003$ \\
$a_4$ & $-0.0003$ & $-0.0072$ & $-0.0072$ & $-0.0701$ \\
$a_5$ & 0.0001 & 0.0000 & 0.0000 & $-0.0040$ \\
$a_6$ &\B $-0.0004$ & $-0.0069$ & $-0.0069$ & $-0.0064$ \\
\hline
$\T \Delta f_z$ @ 5 K & 190 Hz & 310 Hz & 310 Hz & 1600 Hz \B \\
\hline
\end{tabular}
\caption{A conducting enclosure changes the $C_k$ and $a_k$ when the optimal potentials for an ideal planar trap (from Tables~\ref{table:SampleTrap}-\ref{table:SampleTrapAbsolute}) are applied.  The enclosure for the sample trap is shown to scale in Fig.~\ref{fig:FinitePlanarTrapThreeDimensions} with $\rho_c =  19.05$ mm and $z_c = 45.72$ mm.}
\label{table:SamplePlanarTrapWithEnclosureBiasedForNoEnclosure}
\end{table}
}

\newcommand{\SamplePlanarTrapWithEnclosureTable}{
\begin{table}
\newcommand\T{\rule{0pt}{2.6ex}}
\newcommand\B{\rule[-1.2ex]{0pt}{0pt}}
\begin{tabular}{|c|rr|rr|c|}
\cline{2-5}
\multicolumn{1}{c}{} & \multicolumn{4}{|c|}{\T $\{\wt{\rho}_i \} = \{1, 5.5, 7.5426\}$ \B}\\
\cline{2-5}
\multicolumn{1}{c}{~} \T & \multicolumn{2}{|c}{$a_2=a_4=0$} & \multicolumn{2}{|c|}{$C_3=C_4=0$} \\
\multicolumn{1}{c}{} & \multicolumn{1}{|c}{I} & \multicolumn{1}{c}{II} &\multicolumn{1}{|c}{III} & \multicolumn{1}{c|}{IV} \\
\multicolumn{1}{c}{} & \multicolumn{1}{|c}{Eq.~\ref{eq:Constraintsa2a4}} & \multicolumn{1}{c}{Eqs.~\ref{eq:Constraintsa2a4}, \ref{eq:OptimizationHarmonic}~} &\multicolumn{1}{|c}{~Eqs.~\ref{eq:OptimizationC3C4}, \ref{eq:OptimizationHarmonic}} & \multicolumn{1}{c|}{Eq.~\ref{eq:OptimizationC3C4}} \\
\cline{1-5}
$~\delta\wt{V}_1~$ & 1.2769 & 1.4585 & 1.6122 & 1.6247 \\
$\delta\wt{V}_2$ & 1.1724 & 1.4626 & 1.6088 & 1.6239 \\
$\delta\wt{V}_3$ &\B 1.9777 & 2.2179 & 2.6248 & 2.5540 \\
\hline
~$\delta V_1$ \T & $-1.3971$ & $-1.5958$ & $-1.7640$ & $-1.7777$ & ~V\\
$\delta V_2$ & $-1.2828$ & $-1.6003$ & $-1.7603$ & $-1.7768$ & ~V \\
~$\delta V_3$ \B & $-2.1639$ & $-2.4267$ & $-2.8719$ & $-2.7944$ & ~V\\
\hline
$\wt{z}_0$ & 2.3625 & 1.4333 & 1.4436 & 1.0225 \\
\cline{1-5}
$C_3$ & \T $-0.1520$ & 0.0013 & 0.0000 & 0.0000 \\
$C_4$ & 0.0289 & 0.0000 & 0.0000 & 0.0000 \\
$C_5$ & $-0.0155$ & $-0.0111$ & $-0.0113$ & 0.0220 \\
$C_6$ &\B 0.0064 & 0.0000 & 0.0002 & $-0.0371$ \\
\cline{1-5}
$a_2$ &\T 0.0000 & 0.0000 & 0.0000 & 0.0000 \\
$a_3$ & 0.0000 & 0.0000 & 0.0000 & 0.0000 \\
$a_4$ & 0.0000 & 0.0000 & 0.0002 & $-0.0347$ \\
$a_5$ & 0.0000 & 0.0000 & 0.0000 & 0.0000 \\
$a_6$ &\B $-0.0003$ & $-0.0039$ & $-0.0038$ & $-0.0095$ \\
\hline
$\Delta \T f_z$ @ 5 K & 0.0 & 0.0 & 0.0 & 0.0 & Hz\B\\
\hline
\end{tabular}
\caption{For a conducting enclosure around the sample trap, optimized trap configurations can be obtained by shifting the applied potentials by $\delta V_i$ and $\delta \Vt_i$ from the values for an ideal planar trap in Tables~\ref{table:SampleTrap}-\ref{table:SampleTrapAbsolute}.  The enclosure for the sample trap is shown to scale in Fig.~\ref{fig:FinitePlanarTrapThreeDimensions} with $\rho_c =  19.05$ mm and $z_c = 45.72$ mm.  Note that configurations II and III are now distinct.}
\label{table:SamplePlanarTrapWithEnclosure}
\end{table}
}

\newcommand{\OptimizedHarmonicTrapWithChangeRadiiTable}{
\begin{table}
\newcommand\T{\rule{0pt}{2.6ex}}
\newcommand\B{\rule[-1.2ex]{0pt}{0pt}}
\begin{tabular}{|c|r|r|r|r|r|}
\hline
$\delta \rho_2$ & $-25$ & 25 & 0 & 0 & $\mu$m\\
$\delta \rho_3$ & 0 & 0 & $-25$ & 25 & $\mu$m \\
\hline
$\wt{z}_0$ & 1.4809 & 1.3876 & 1.3966 & 1.4726 \\
\cline{1-5}
$C_3$ & \T 0.0017 & $-0.0027$ & $-0.0017$ & 0.0012 \\
$C_4$ & $-0.0036$ & 0.0038 & 0.0024 & $-0.0024$ \\
$C_5$ & $-0.0108$ & $-0.0112$ & $-0.0111$ & $-0.0110$ \\
$C_6$ &\B 0.0011 & $-0.0016$ & $-0.0012$ & 0.0009 \\
\cline{1-5}
$a_2$ &\T $-0.0027$ & 0.0028 & 0.0018 & $-0.0018$ \\
$a_3$ & 0.0000 & 0.0000 & 0.0000 & 0.0000 \\
$a_4$ & 0.0011 & $-0.0016$ & $-0.0012$ & 0.0009 \\
$a_5$ & 0.0000 & 0.0000 & 0.0000 & 0.0000 \\
$a_6$ &\B $-0.0031$ & $-0.0048$ & $-0.0046$ & $-0.0033$ \\
\hline
$\T \Delta f_z$ @ 5 K & 130 & 140 & 93 & 89 & Hz\B\\
\hline
\end{tabular}
\caption{Changes of $0.001~\rm{in}= 25~\mu \rm{m}$ for the electrode radii of the sample trap deteriorate its performance when the potentials for an ideal optimized harmonic configuration (from Tables~\ref{table:SampleTrap}-\ref{table:SampleTrapAbsolute}) are applied.}
\label{table:OptimizedHarmonicTrapWithChangeRadii}
\end{table}
}

\newcommand{\FinitePlanarTrapThreeDimensionsFigure}{
\begin{figure}[htbp!]
\centering
\includegraphics[width=\w]{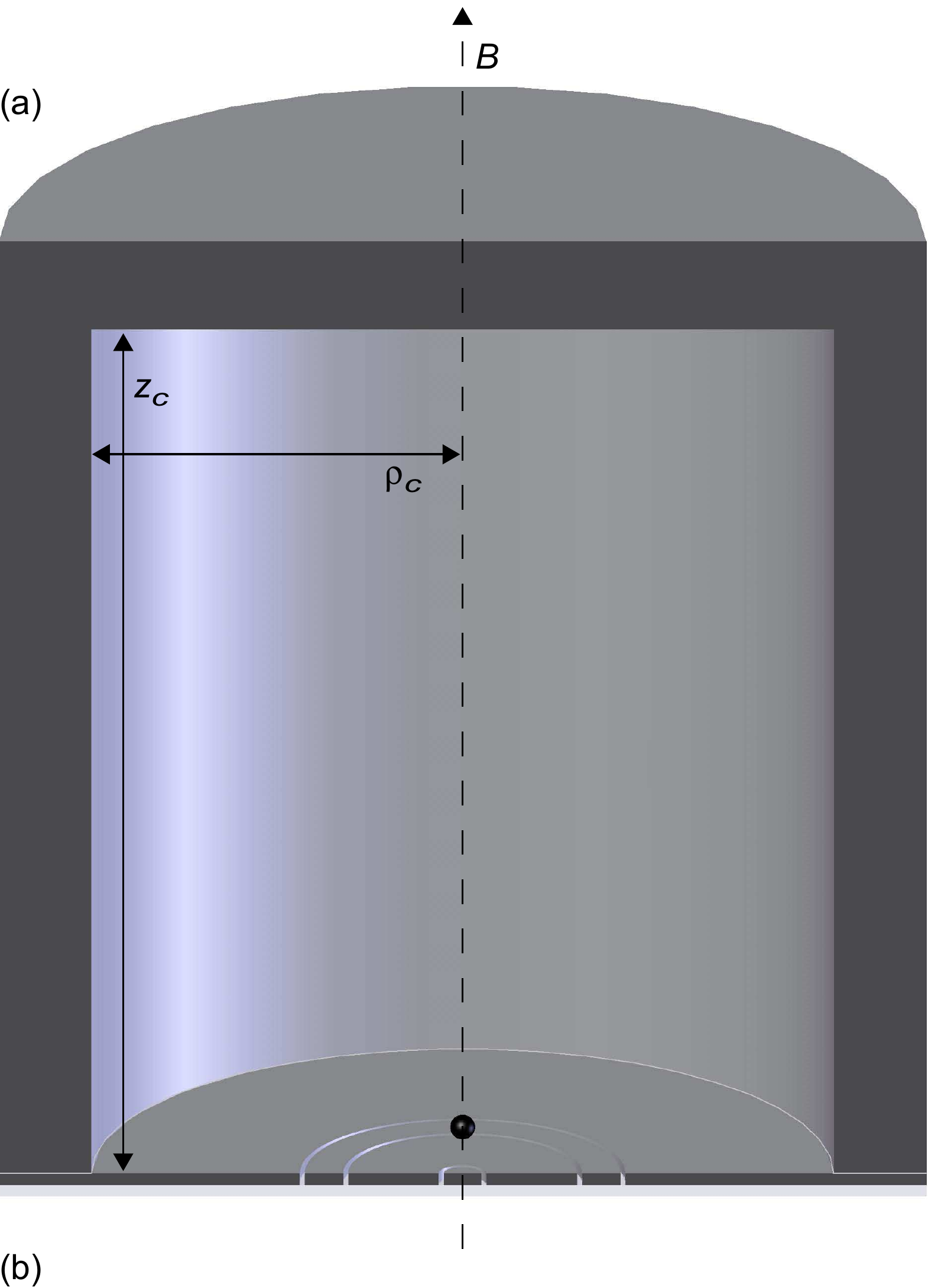}
\includegraphics[width=\w]{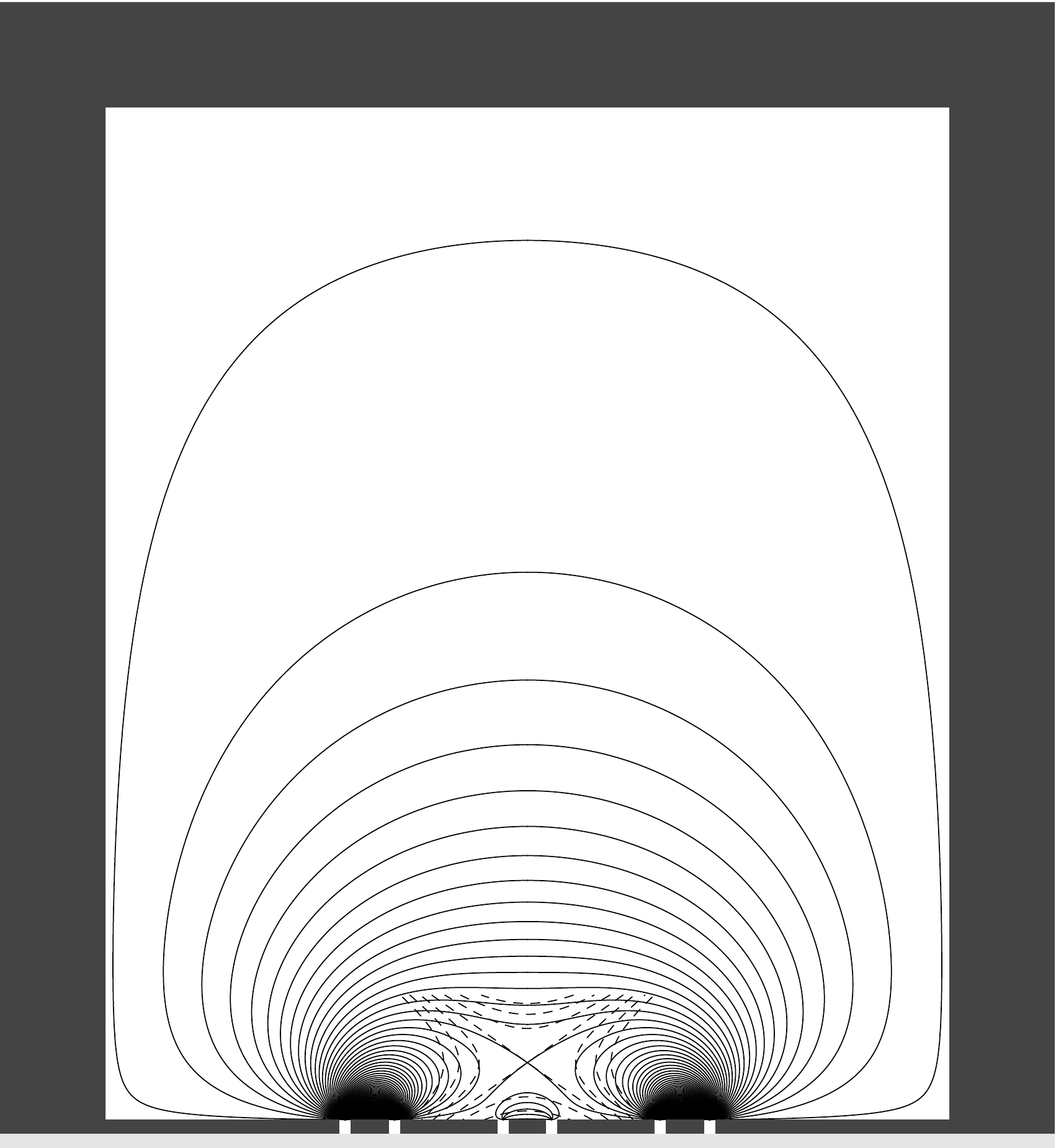}
\caption{(a) Planar trap enclosed within a conducting, capped cylinder. Particles can be loaded through a tiny axial hole in the cover (not visible). (b) Side view of trap electrodes and equipotentials spaced by $V_0$, with the infinitesimal gaps between the electrodes widened to make them visible. The equipotentials extend into the gaps between electrodes. The dashed equipotentials of an ideal quadrupole are superimposed near the trap center.}
\label{fig:FinitePlanarTrapThreeDimensions}
\end{figure}
}

\newcommand{\TuningWithImperfectionsFigure}{
\begin{figure}[htbp!]
\centering
\includegraphics*[width=\w]{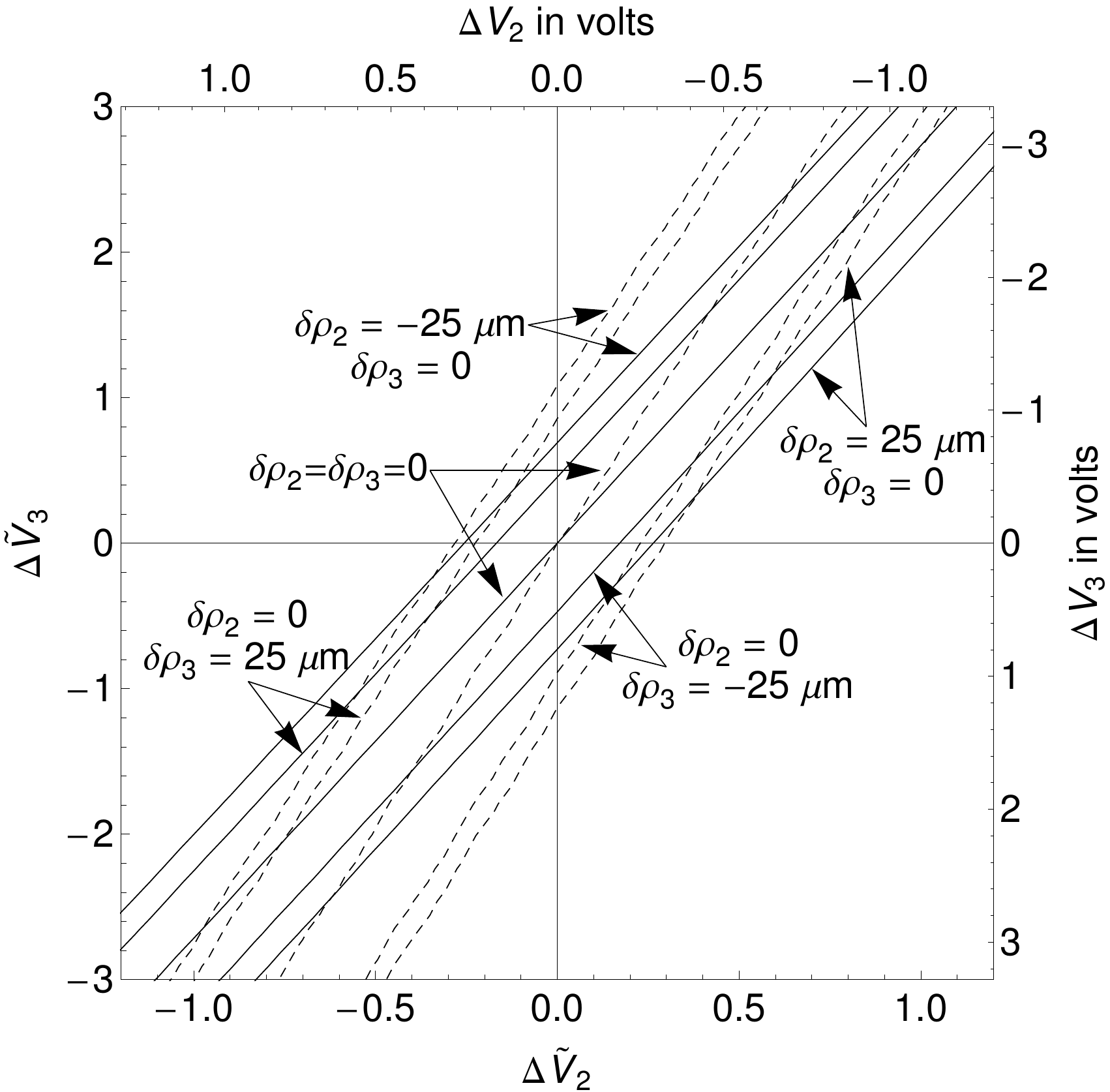}
\caption{The $a_2=0$ (solid) and $a_4=0$ (dashed) contours for the sample trap with one of its radii displaced by the indicated distance as a function of the potentials applied to the electrodes.  $V_2$ and $V_3$ are changed as plotted, and $V_1$ is adjusted to keep the axial frequency at 64 MHz.} \label{fig:TuningWithImperfections}
\end{figure}
}

\newcommand{\CoveredPlanarTrapFigure}{
\begin{figure}[htbp!]
\centering
\includegraphics*[width=\columnwidth]{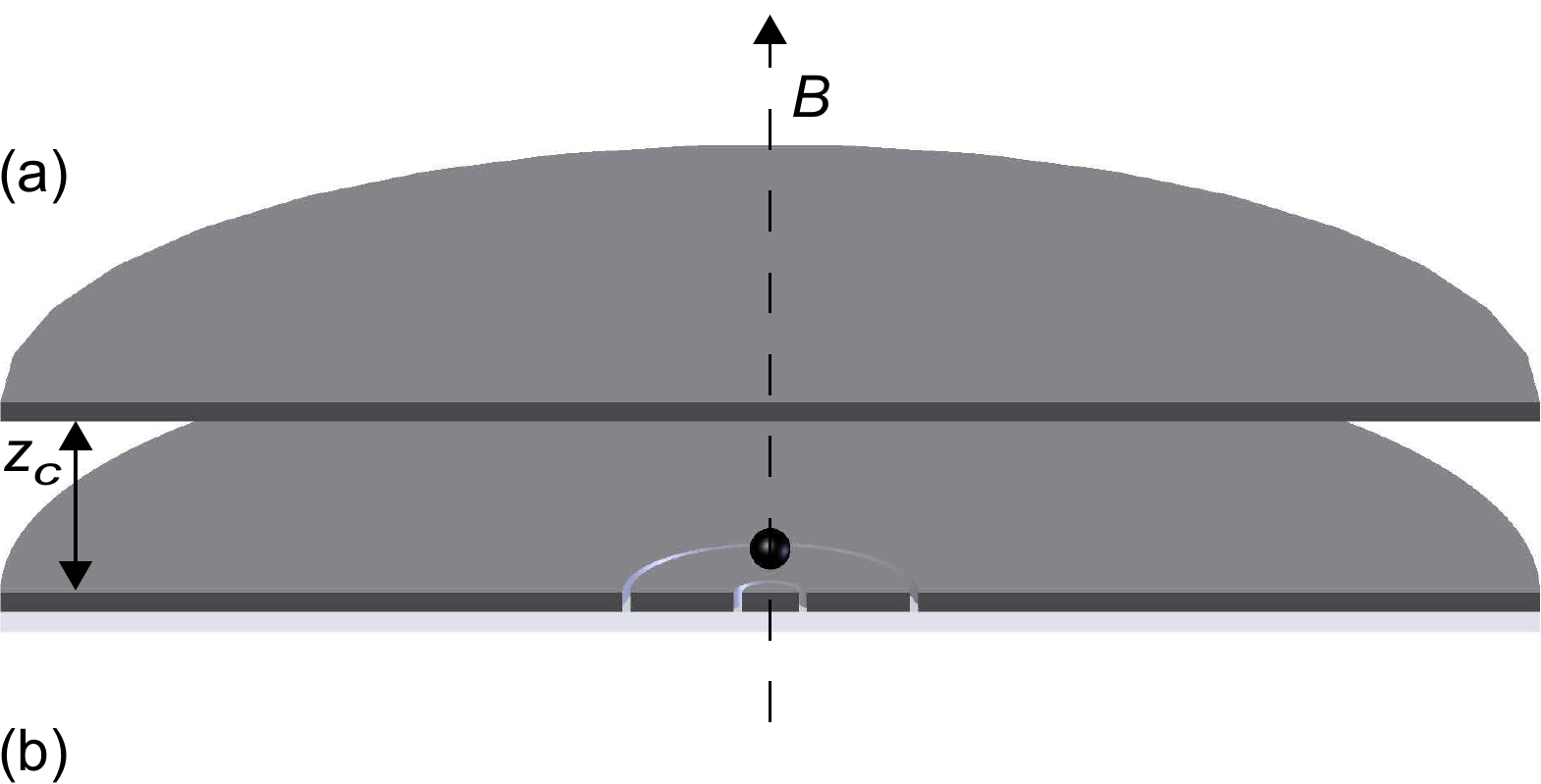}
\includegraphics*[width=\columnwidth]{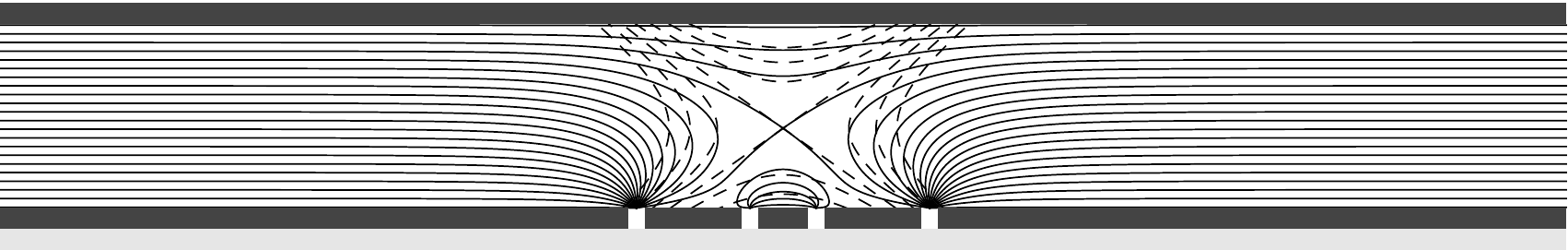}
\caption{(a) A covered planar Penning trap could be loaded through a tiny axial hole in the cover (not visible). (b) Side view of trap electrodes and equipotentials spaced by $V_0$, with the infinitesimal gaps between the electrodes widened to make them visible. Some  equipotentials extend into the gaps between electrodes and some terminate at infinity. The dashed equipotentials of an ideal quadrupole are superimposed near the trap center.}
\label{fig:CoveredPlanarTrap}
\end{figure}
}

\newcommand{\PossibleCoveredTrapGeometriesFigure}{
\begin{figure}[htbp!]
\centering
\includegraphics[width=\w]{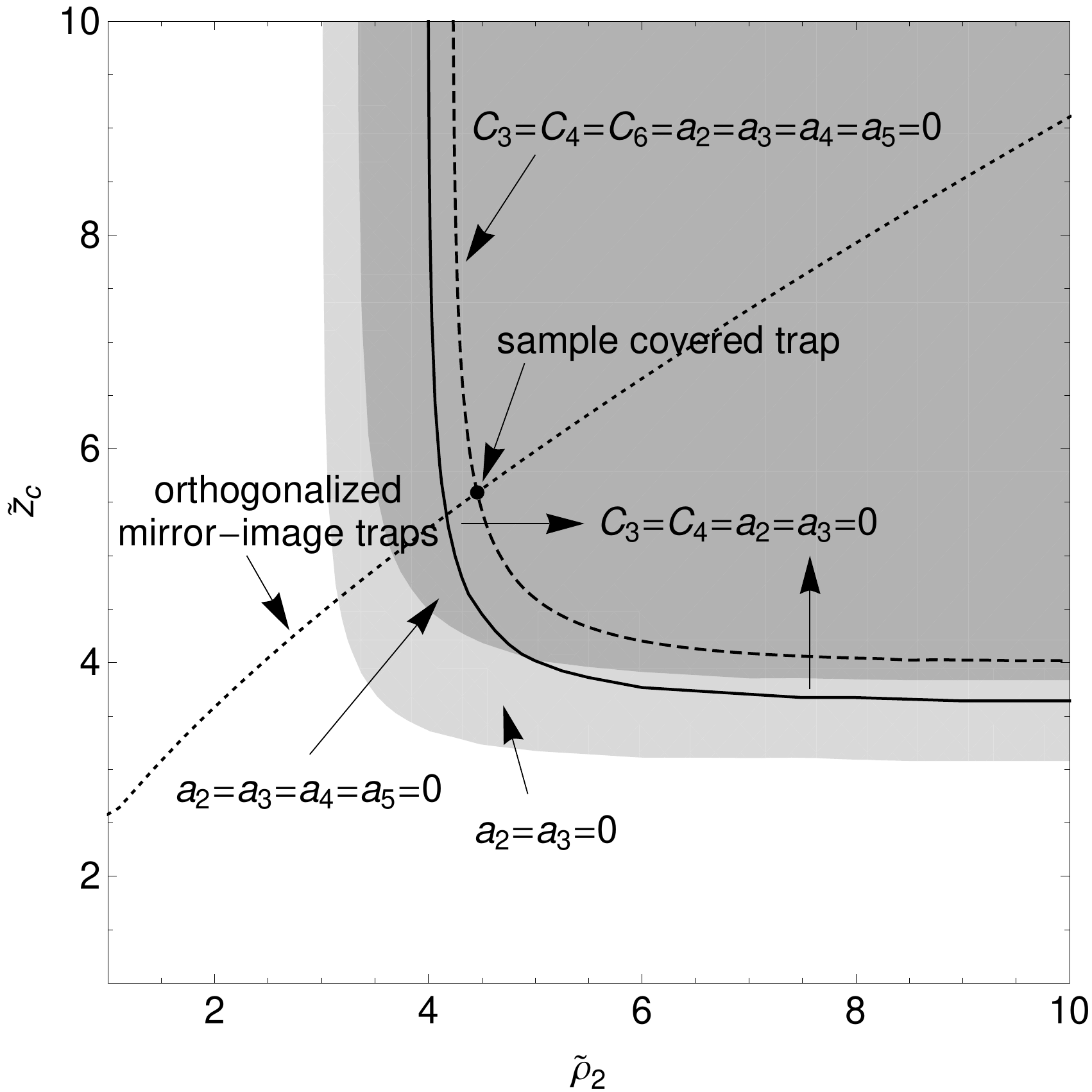}
\caption{Parameter space regions for which the indicated $a_k$ can be made to vanish for a two-gap planar trap with a cover, along with the region and the curve for which the indicated $C_k$ can alternatively be made to vanish.  No optimized traps are possible in the unshaded region.  The dotted line indicates orthogonalized mirror-image traps formed from two sets of two-gap planar trap electrodes, as described in Sec.~\ref{sec:MirrorImageTraps}}
\label{fig:PossibleCoveredTrapGeometries}
\end{figure}
}

\newcommand{\SampleCoveredTrapTable}{
\begin{table}
\newcommand\T{\rule{0pt}{2.6ex}}
\newcommand\B{\rule[-1.2ex]{0pt}{0pt}}
\begin{tabular}{|c|rr|rr|}
\cline{2-5}
\multicolumn{1}{c}{} & \multicolumn{4}{|c|}{\T $\{\wt{\rho}_i \} = \{1, 4.4572\}$, $\zt_c = 5.5914$ \B}\\
\cline{2-5}
\multicolumn{1}{c}{~} \T & \multicolumn{2}{|c}{$a_2=a_4=0$} & \multicolumn{2}{|c|}{$C_3=C_4=0$} \\
\multicolumn{1}{c}{} & \multicolumn{1}{|c}{I} & \multicolumn{1}{c}{II} &\multicolumn{1}{|c}{III} & \multicolumn{1}{c|}{IV} \\
\multicolumn{1}{c}{} & \multicolumn{1}{|c}{Eq.~\ref{eq:Constraintsa2a4}} & \multicolumn{1}{c}{Eqs.~\ref{eq:Constraintsa2a4}, \ref{eq:OptimizationHarmonic}~} &\multicolumn{1}{|c}{~Eqs.~\ref{eq:OptimizationC3C4}, \ref{eq:OptimizationHarmonic}} & \multicolumn{1}{c|}{Eq.~\ref{eq:OptimizationC3C4}} \\
\hline
$~\wt{V}_1~$ & \T 23.6322 & 23.9786 & 23.9786 & 24.2851\\
$\wt{V}_2$ & 19.0275 & 23.3251 & 23.3251 & 23.6609\\
$\wt{V}_c$ & \B 21.2413 & 29.8478 & 29.8478 & 32.4943\\
\hline
$\wt{z}_0$ & \T 2.4214 & 1.4338 & 1.4338 & 1.0306\\
\hline
$C_3$ & \T $-0.1532$ & 0.0000 & 0.0000 & 0.0000\\
$C_4$ & 0.0294 & 0.0000 & 0.0000 & 0.0000\\
$C_5$ & $-0.0143$ & $-0.0099$ & $-0.0099$ & 0.0208\\
$C_6$ &\B 0.0057 & 0.0000 & 0.0000 & $-0.0351$\\
\hline
$a_2$ &\T 0.0000 & 0.0000 & 0.0000 & 0.0000\\
$a_3$ & 0.0000 & 0.0000 & 0.0000 & 0.0000\\
$a_4$ & 0.0000 & 0.0000 & 0.0000 & $-0.0329$\\
$a_5$ & 0.0000 & 0.0000 & 0.0000 & 0.0000\\
$a_6$ &\B $-0.0003$ & $-0.0038$ & $-0.0038$ & $-0.0096$\\
\hline
$C_{11}$ & \T $-0.1155$ & $-0.3781$ & $-0.3781$ & $-0.6789$\\
$C_{12}$ & $-0.2558$ & $-0.0690$ & $-0.0690$ & 0.2056\\
$C_{1c}$ & 0.3577 & 0.3577 & 0.3577 & 0.3577\\
$C_{1d}^{(opt)}$ & $-$0.3715 & $-$0.5521 & $-$0.5521 & $-$0.7545 \B\\
\hline
$\gamma_1$ & \T $-142.78$ & 3.37 & 3.37 & 3.47\\
$\gamma_2$ & 6.72 & 2.34 & 2.34 & 4.28\\
$\gamma_c$ & \B 7.64 & 0.00 & 0.00 & 0.00\\
\hline
\end{tabular}
\caption{Scaled parameters for the sample two-gap covered planar trap geometry.  }
\label{table:SampleCoveredTrap}
\end{table}
}

\newcommand{\SampleCoveredTrapTableAbsolute}{
\begin{table}
\newcommand\T{\rule{0pt}{2.6ex}}
\newcommand\B{\rule[-1.2ex]{0pt}{0pt}}
\begin{tabular}{|c|rr|rr|c|}
\cline{2-5}
\multicolumn{1}{c}{~} & \multicolumn{4}{|c|}{\T $\{\rho_i \} = \{1, 4.4572\}$ mm, $z_c = 5.5914$ mm} & \multicolumn{1}{c}{ \B}\\
\cline{2-5}
\multicolumn{1}{c}{~} \T & \multicolumn{2}{|c}{$a_2=a_4=0$} & \multicolumn{2}{|c|}{$C_3=C_4=0$} & \multicolumn{1}{c}{} \\
\multicolumn{1}{c}{~} & \multicolumn{1}{|c}{I} & \multicolumn{1}{c}{II} &\multicolumn{1}{|c}{III} & \multicolumn{1}{c|}{IV} & \multicolumn{1}{c}{~}\\
\multicolumn{1}{c}{} & \multicolumn{1}{|c}{Eq.~\ref{eq:Constraintsa2a4}} & \multicolumn{1}{c}{Eqs.~\ref{eq:Constraintsa2a4}, \ref{eq:OptimizationHarmonic}~} &\multicolumn{1}{|c}{~Eqs.~\ref{eq:OptimizationC3C4}, \ref{eq:OptimizationHarmonic}} & \multicolumn{1}{c|}{Eq.~\ref{eq:OptimizationC3C4}} \\
\hline
$\rho_1$ \T & 1 & 1 & 1 & 1 & mm\\
$z_0$ & 2.4214 & 1.4338 & 1.4338 & 1.0306 & mm\\
$\rho_d^{(opt)}$ \B & 4.6396 & 2.1166 & 2.1166 & 1.4797 & mm\\
\hline
$f_z$ \T & 64 & 64 & 64 & 64 & MHz\\
$V_0$ \B & $-0.9194$ & $-0.9194$ & $-0.9194$ & $-0.9194$ & V\\
\hline
$V_1$ & \T $-21.7271$ & $-22.0456$ & $-22.0456$ & $-22.3274$ & V\\
$V_2$ & $-17.4936$ & $-21.4448$ & $-21.4448$ & $-21.7535$ & V\\
$V_c$ & \B $-19.5290$ & $-27.4416$ & $-27.4416$ & $-29.8749$ & V\\
\hline
$\Delta f_z$ \T & 0.0 & 0.0 & 0.0 & 0.0 & Hz \B\\
\hline
$1:~\gamma_z$ & \T $2\pi$\,1.50 & $2\pi$\,16.03 & $2\pi$\,16.03 & $2\pi$\,51.68 & s$^{-1}$\\
$2:~\gamma_z$ & $2\pi$\,7.34 & $2\pi$\,0.53 & $2\pi$\,0.53 & $2\pi$\,4.74 & s$^{-1}$\\
$c:~\gamma_z$ & $2\pi$\,14.35 & $2\pi$\,14.35 & $2\pi$\,14.35 & $2\pi$\,14.35 & s$^{-1}$\\
d:~$\gamma_z^{(opt)}$ & \B 2$\pi\,$15.47 & 2$\pi\,$34.18 & 2$\pi\,$34.18 & 2$\pi\,$63.83 & s$^{-1}$\\
\hline
\end{tabular}
\caption{A set of absolute values for the sample two-gap covered planar trap geometry.  }
\label{table:SampleCoveredTrapAbsolute}
\end{table}
}

\newcommand{\SampleMirrorImageTrapTable}{
\begin{table}
\newcommand\T{\rule{0pt}{2.6ex}}
\newcommand\B{\rule[-1.2ex]{0pt}{0pt}}
\begin{tabular}{|c|r|}
\hline
\multicolumn{2}{|c|}{\T $\{\wt{\rho}_i \} = \{1, 4.4572\}$, $\zt_c = 5.5914$ \B}\\
\hline
\multicolumn{2}{|c|}{\T $C_3=C_4=0$, Eq.~\ref{eq:OptimizationC3C4}}\\
\hline
$~\wt{V}_1 = \wt{V}_1^{top}~$ & \T 13.9582\\
$\wt{V}_2=\wt{V}_2^{top}$ & 12.3743\\
$\wt{V}_3=\wt{V}_3^{top}$ & \B 0.0000\\
\hline
$\T \wt{z}_0$ & 2.7957\\
\hline
$C_3$ & \T 0.0000\\
$C_4$ & 0.0000\\
$C_5$ & 0.0000\\
$C_6$ &\B 0.0015\\
\hline
$a_2$ &\T 0.0000\\
$a_3$ & 0.0000\\
$a_4$ & 0.0014\\
$a_5$ & 0.0000\\
$a_6$ &\B 0.0002\\
\hline
$\T C_{11}=-C_{11}^{top}$ & \T $-0.0811$\\
$C_{12}=-C_{12}^{top}$ & $-0.2624$\\
$C_{13}=-C_{13}^{top}$ & $-0.0142$\\
$C_{1d}^{(opt)}$ & $-$0.3577\\
$\rt_d^{(opt)}$ & \B $\infty$\\
\hline
$\gamma_1=\gamma_1^{top}$ & \T 4.35\\
$\gamma_2=\gamma_2^{top}$ & 0.00\\
$\gamma_3=\gamma_3^{top}$ & \B $-33.96$\\
\hline
\end{tabular}
\caption{Scaled parameters for the sample two-gap mirror-image planar trap geometry.  }
\label{table:SampleMirrorImageTrap}
\end{table}
}

\newcommand{\SampleMirrorImageTrapTableAbsolute}{
\begin{table}
\newcommand\T{\rule{0pt}{2.6ex}}
\newcommand\B{\rule[-1.2ex]{0pt}{0pt}}
\begin{tabular}{|c|r|c|}
\hline
\multicolumn{3}{|c|}{\T $\{\wt{\rho}_i \} = \{1, 4.4572\}$, $\zt_c = 5.5914$ \B}\\
\hline
\multicolumn{3}{|c|}{\T $C_3=C_4=0$, Eq.~\ref{eq:OptimizationC3C4}}\\
\hline
$\rho_1$ \T & 1 & mm\\
$z_0$ & 2.7957 & mm\\
$\rho_d^{|opt|}$ \B & $\infty$ & mm\\
\hline
$f_z$ \T & 64 & MHz\\
$V_0$ \B & $-0.9194$ & V\\
\hline
$\T V_1=V_1^{top}$ & $-12.8330$ & V\\
$V_2=V_2^{top}$ & $-11.3768$ & V\\
$V_3=V_3^{top}$ & 0.0000 & V\\
\hline
$\Delta f_z$ \T & 0.0 & Hz \B\\
\hline
$\T 1:~\gamma_z$ & $2\pi$\,0.74 & s$^{-1}$\\
$2:~\gamma_z$ & $2\pi$\,7.72 & s$^{-1}$\\
$3:~\gamma_z$ & $2\pi$\,0.02 & s$^{-1}$\\
d:~$\gamma_z^{(opt)}$ & $2\pi$\,14.35 & s$^{-1}$\\
\hline
\end{tabular}
\caption{A set of absolute values for the sample two-gap mirror-image planar trap geometry.}
\label{table:SampleMirrorImageTrapAbsolute}
\end{table}
}

\newcommand{\MirrorImagePlanarTrapThreeDimensionsFigure}{
\begin{figure}[htbp!]
\centering
\includegraphics*[width=\w]{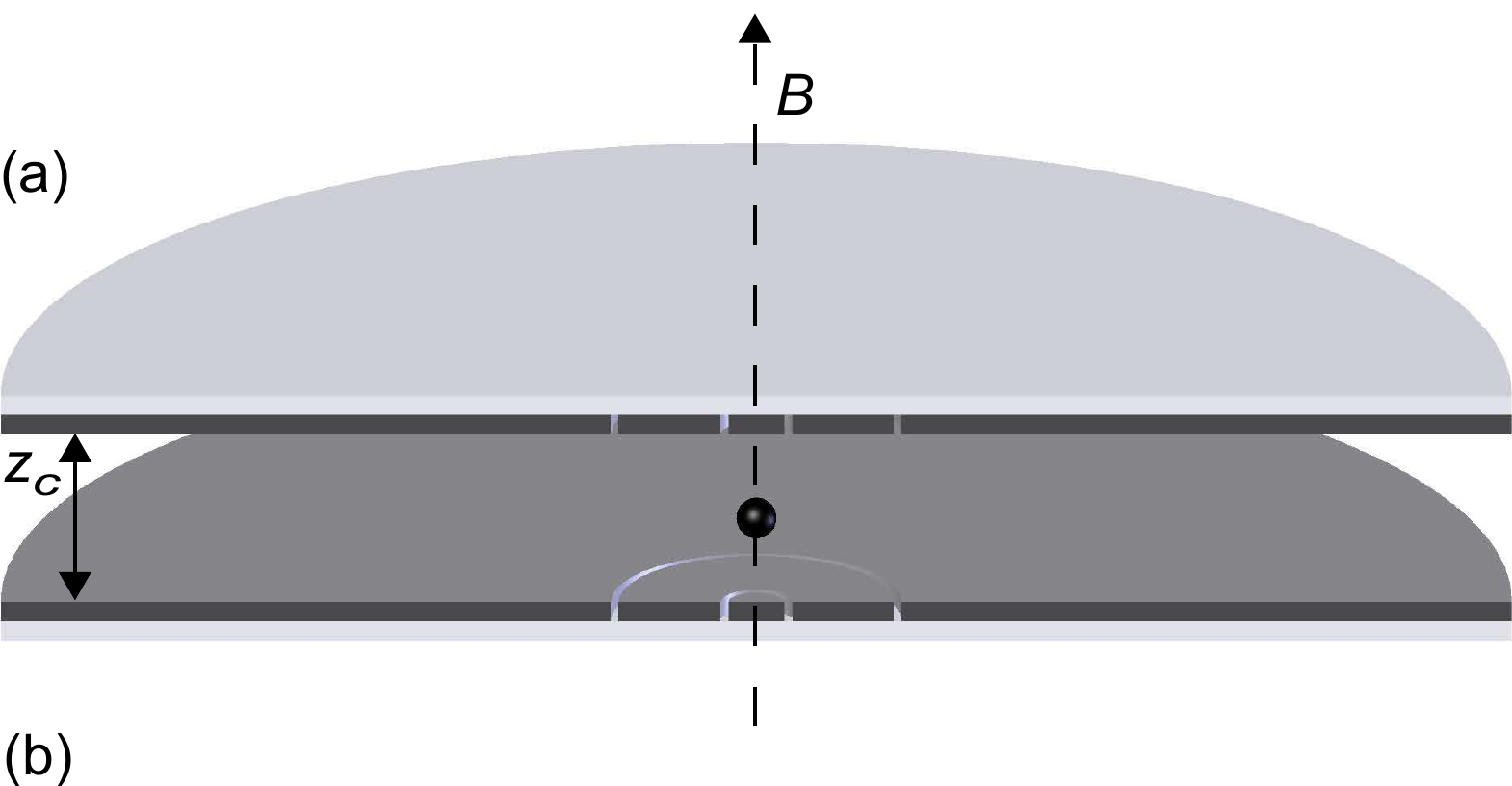}
\includegraphics*[width=\w]{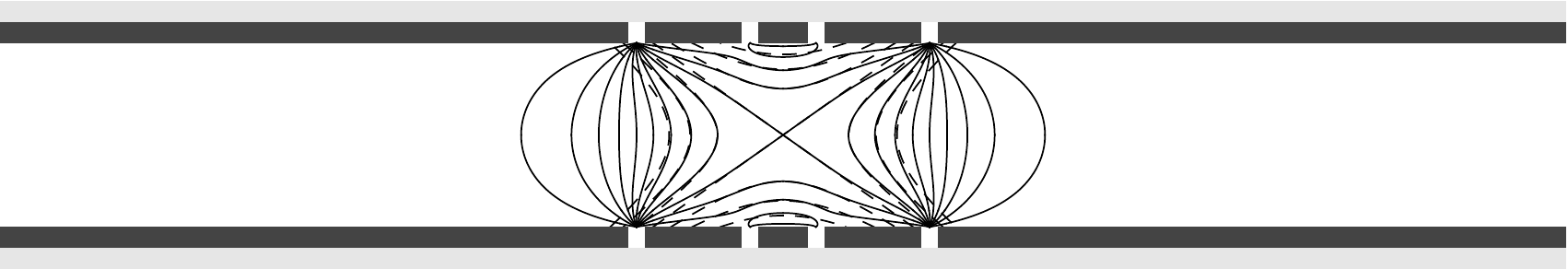}
\caption{(a) A mirror-image Penning trap is formed with two planar trap electrodes facing each other. Particles can be loaded through a tiny axial hole in one of the electrodes (not visible). (b) Side view of trap electrodes and equipotentials spaced by $V_0$, with the infinitesimal gaps between the electrodes widened to make them visible. The equipotentials extend into the gaps between electrodes. The dashed equipotentials of an ideal quadrupole are superimposed near the trap center. }
\label{fig:MirrorImagePlanarTrapThreeDimensions}
\end{figure}
}

\newcommand{\MirrorImagePotentialsFigure}{
\begin{figure}[htbp!]
\centering
\includegraphics*[width=\w]{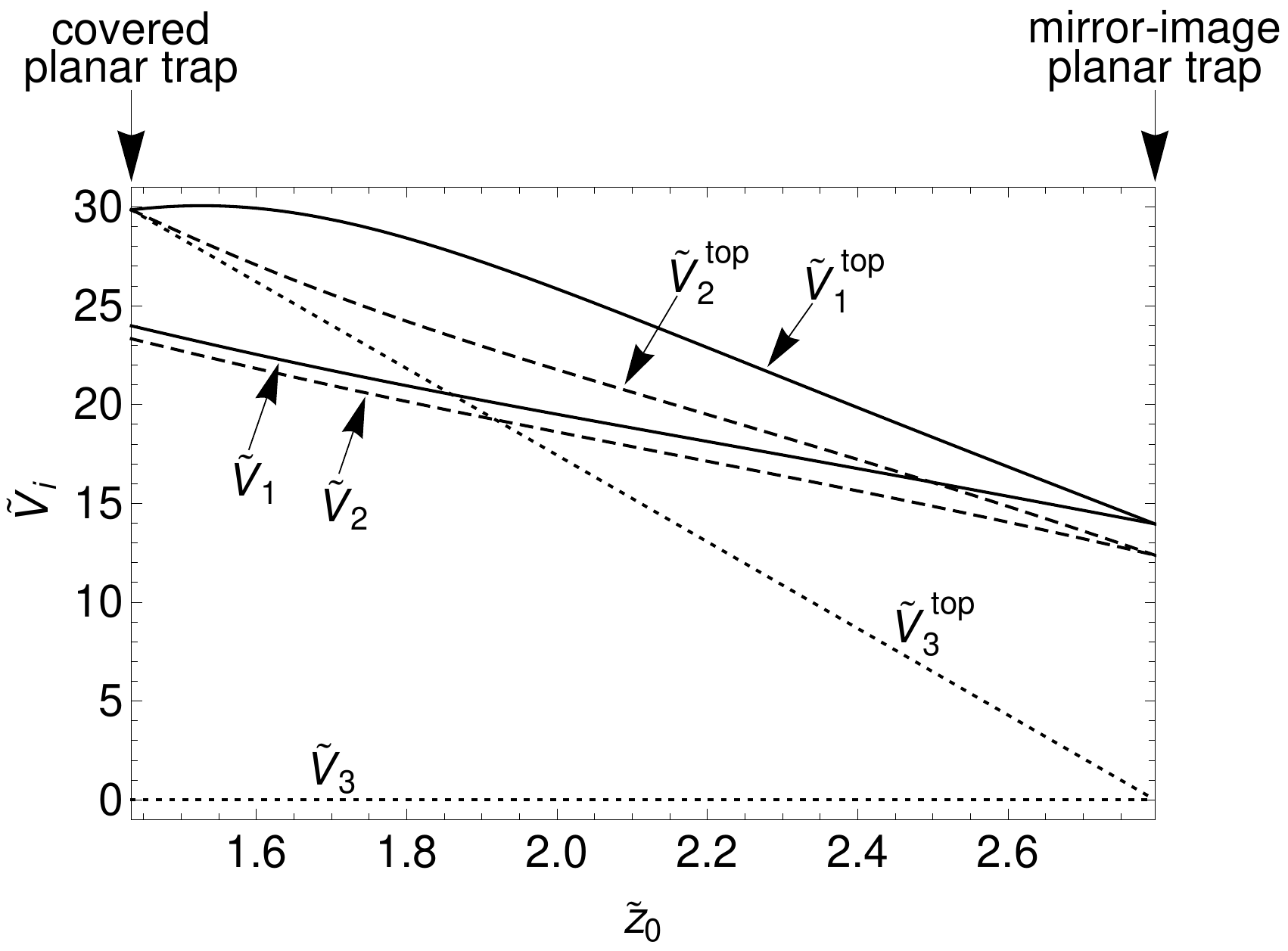}
\caption{One set of applied potentials that relocates an electron centered between the electrode planes of a mirror-image plane (far right) to a covered planar trap (far left) while keeping the axial frequency constant and keeping $a_2=a_3=C_3=C_4=0$.}
\label{fig:MirrorImagePotentials}
\end{figure}
}

\newcommand{\DetectingAndDampingCircuitFigure}{
\begin{figure}[htbp!]
\centering
\includegraphics*[width=\w]{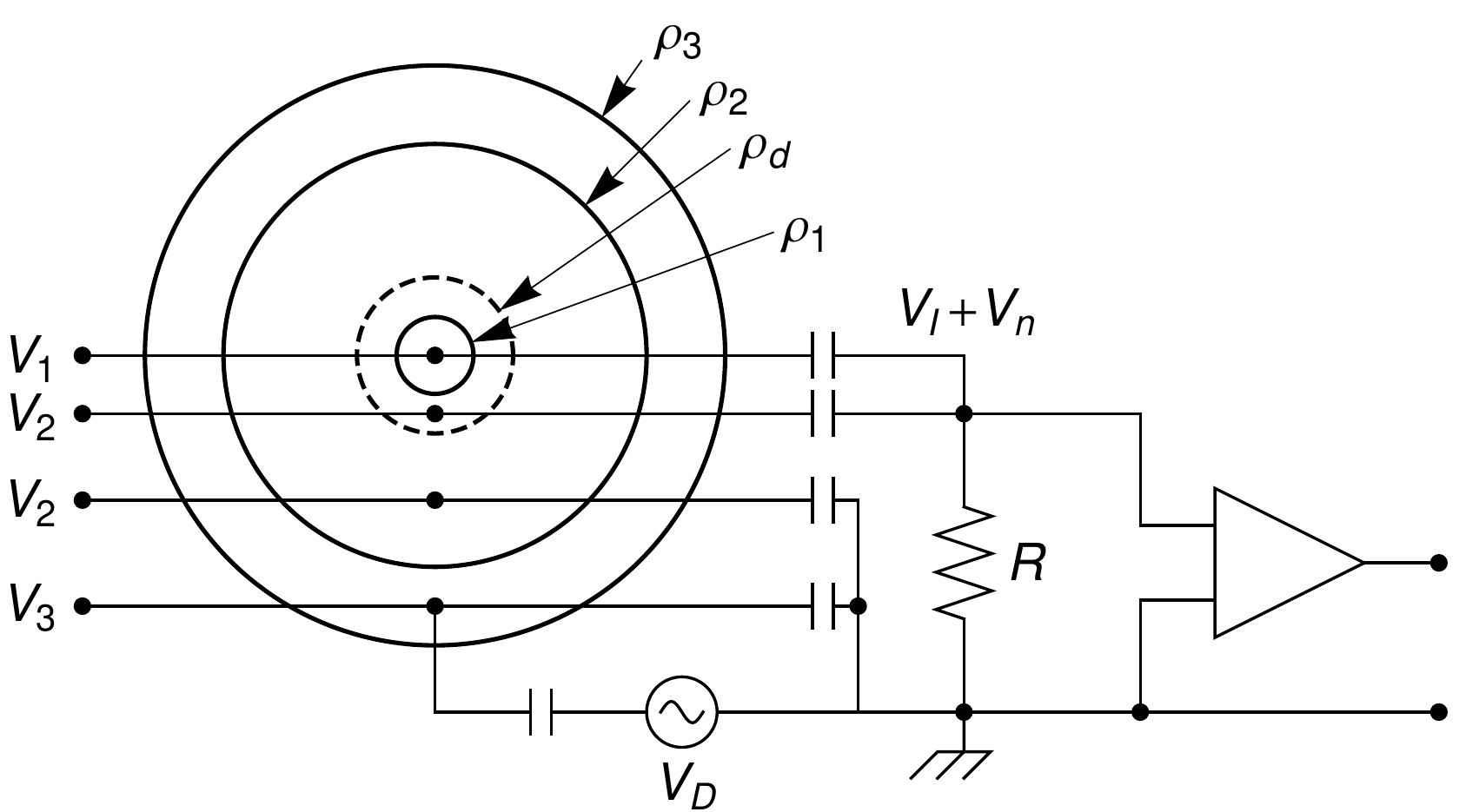}
\caption{Electrical circuit used to bias, detect, damp and drive a trapped particle's axial motion. An extra gap (dashed circle) in the electrodes of the planar trap can be added to optimize the damping and detection without changing the electrostatic properties of an optimized planar trap.  The relative trap geometry is that of  the sample trap.}
\label{fig:DetectingAndDampingCircuit}
\end{figure}
}

\newcommand{\MaximizedDampingFigure}{
\begin{figure}[htbp!]
\centering
\includegraphics*[width=\w]{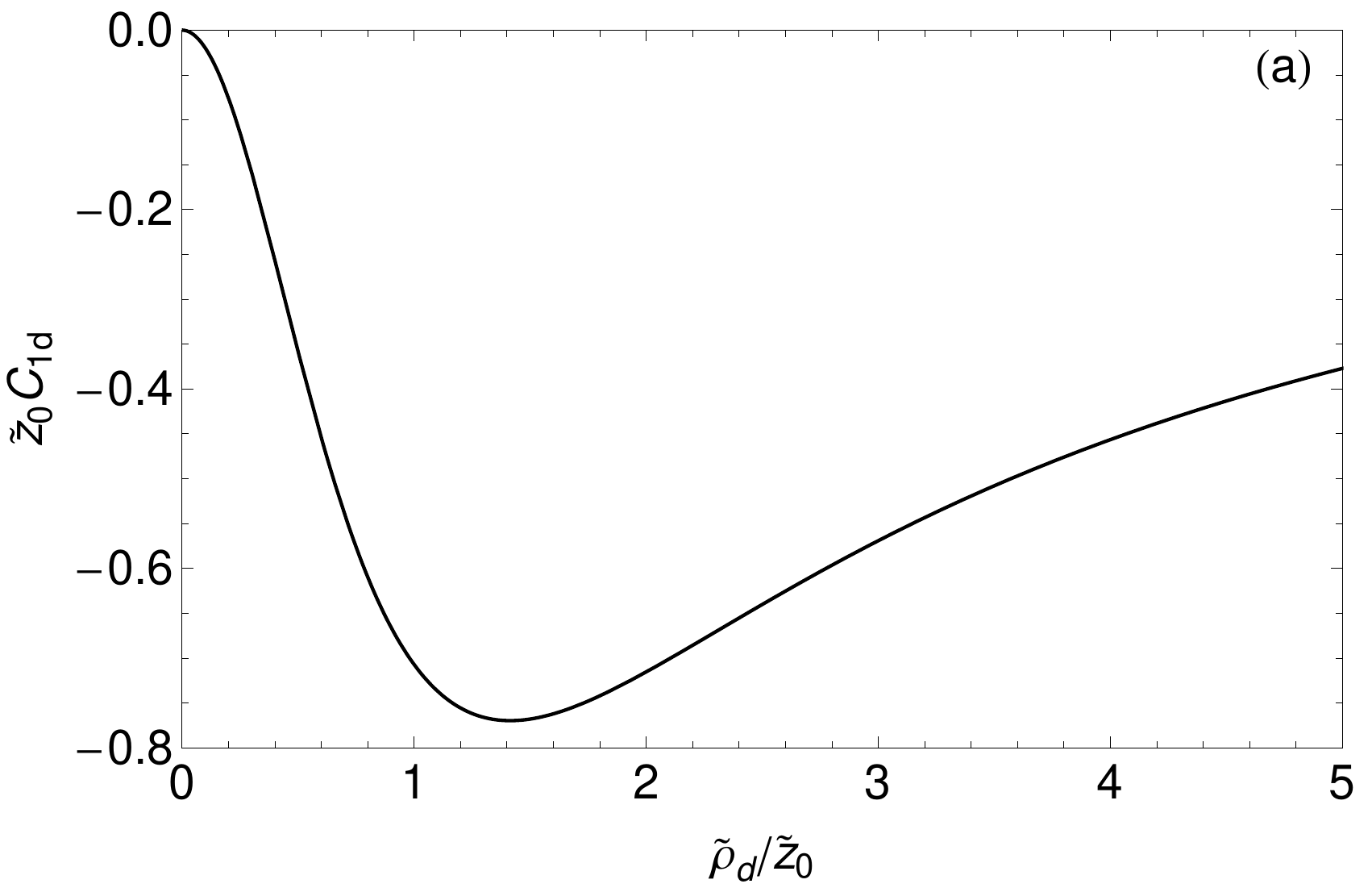}
\includegraphics*[width=\w]{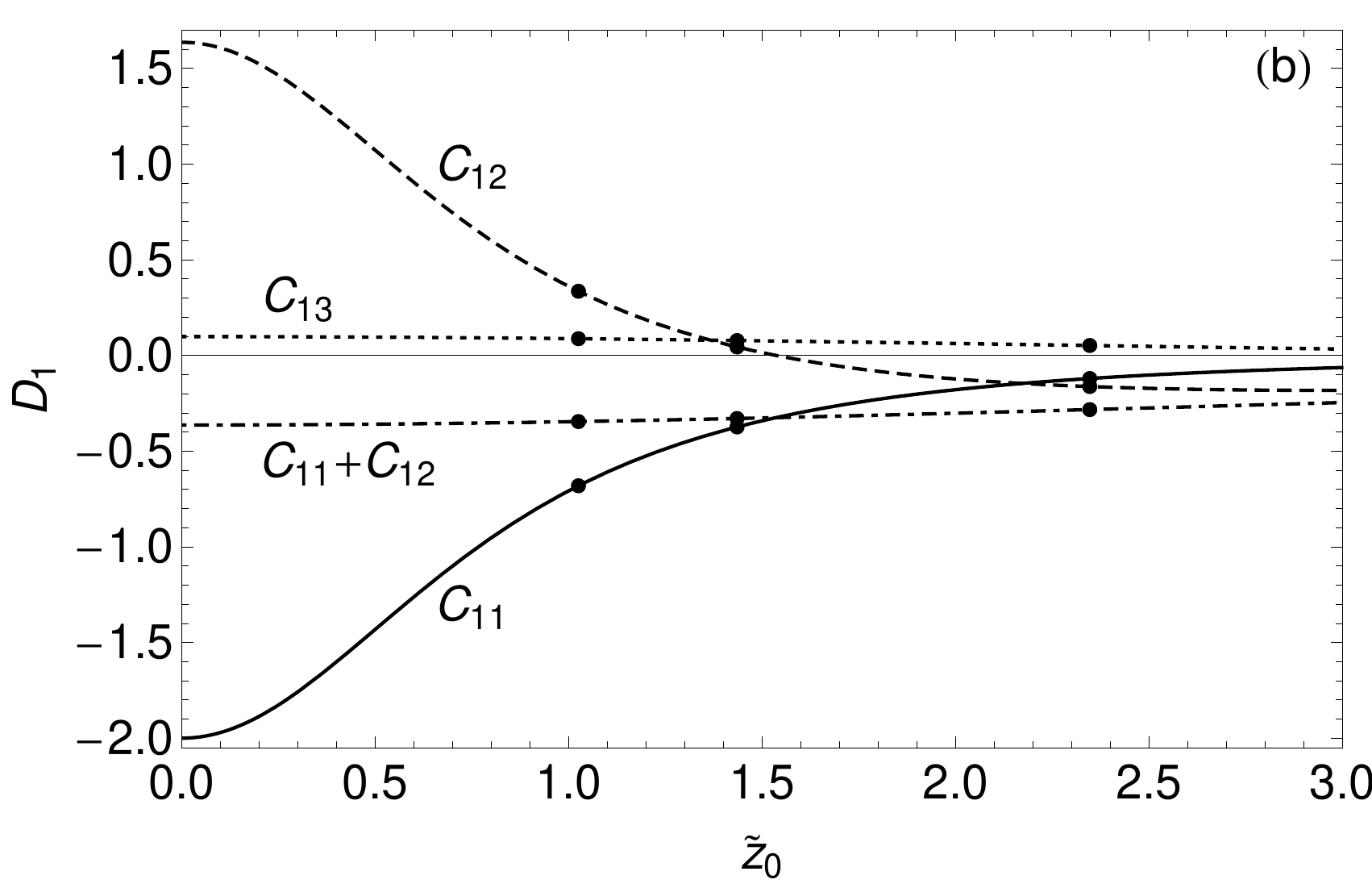}
\caption{(a) The coefficient $C_{1d}$ that describes the coupling, and the plotted product $\zt_0 C_{1d}$, both have a maximum magnitude at $\rt_d = \sqrt{2} \zt_0$. (b) Electric field coefficients that describe the damping rates and detection efficiency for the sample trap.}
\label{fig:MaximizedDamping}
\end{figure}
}

\newcommand{\UlmTrapScaledTable}{
\begin{table}
\newcommand\T{\rule{0pt}{2.6ex}}
\newcommand\B{\rule[-1.2ex]{0pt}{0pt}}
\begin{tabular}{|c|rrr|r|}
\cline{2-5}
\multicolumn{1}{c}{} & \multicolumn{4}{|c|}{\T $\{\wt{\rho}_i \} = \{1, 3, 5\}$  (Eq.~\ref{eq:UlmRho})\B}\\
\cline{2-5}
\multicolumn{1}{c}{} & \multicolumn{3}{|c|}{\T Used} & \multicolumn{1}{c|}{Mentioned} \\
\multicolumn{1}{c}{} & \multicolumn{1}{|c}{\B Eq.~\ref{eq:UlmV1}} & \multicolumn{1}{c}{\B Eq.~\ref{eq:UlmV2}} & \multicolumn{1}{c|}{\B Eq.~\ref{eq:UlmV3}} & \multicolumn{1}{c|}{\B Eq.~\ref{eq:UlmV4}}\\
\hline
$~\wt{V}_1~$ &\T 0 & 0 & 0 & 0 \\
$\wt{V}_2$ & $-$4.6075 & $-$5.2815 & $-$5.2040 & 22.3016\\
$\wt{V}_3$ &\B $-$4.6075 & $-$14.2194 & $-$15.1782 & $-$58.2296\\
\hline
$\wt{z}_0$ & \T 1.4761 & 1.9070 & 1.9493 & 3.9328\\
\hline
$C_3$ & \T $-$0.6386 & $-$0.4251 & $-$0.4082 & $-$0.3267\\
$C_4$ & 0.2718 & 0.1340 & 0.1211 & 0.0540\\
$C_5$ & $-$0.0766 & $-$0.0443 & $-$0.0389 & $-$0.0015\\
$C_6$ &\B $-$0.0021 & 0.0141 & 0.0129 & $-$0.0021\\
\hline
$a_2$ &\T $-$0.1785 & $-$0.0689 & $-$0.0654 & $-$0.0596\\
$a_3$ & 0.1140 & 0.0293 & 0.0267 & 0.0195\\
$a_4$ & $-$0.0447 & $-$0.0143 & $-$0.0127 & $-$0.0041\\
$a_5$ & 0.0088 & 0.0074 & 0.0065 & 0.0007\\
$a_6$ &\B 0.0026 & $-$0.0030 & $-$0.0026 & $-$0.0001\\
\hline
$C_{11}$ &\T $-$0.3529 & $-$0.2003 & $-$0.1902 & $-$0.0299\\
$C_{12}$ &   $-$0.1287 & $-$0.2004 & $-$0.2029 & $-$0.1188 \\
$C_{13}$ & 0.1287 & 0.0744 & 0.0696 & $-$0.0455\\
$C_{1d}^{(opt)}$ & $-$0.5215 & $-$0.4037 & $-$0.3949 & $-$0.1957\\
$\rt_d^{(opt)}$ & \B 2.0875 & 2.6969 & 2.7568 & 5.5618\\
\hline
$\gamma_1$ &\T $-$4.89 & 0.47 & 0.07 & 30.26 \\
$\gamma_2$ & $-$5.27 & $-$8.32 & $-$10.05 & $-$82.50 \\
$\gamma_3$ & $-$1.09 & \B 1.49 & 2.80 & $-$194.34 \\
\hline
\end{tabular}
\caption{Scaled parameters for the trap used in Ulm \cite{ElectronsInPlanarPenningTrapUlm}.} \label{table:UlmTrapScaled}
\end{table}
}

\newcommand{\UlmTrapAbsoluteTable}{
\begin{table}
\newcommand\T{\rule{0pt}{2.6ex}}
\newcommand\B{\rule[-1.2ex]{0pt}{0pt}}
\begin{tabular}{|c|rrr|r|c|}
\cline{2-5}
\multicolumn{1}{c}{} & \multicolumn{4}{|c|}{\T $\{\wt{\rho}_i \} = \{1, 3, 5\}$  (Eq.~\ref{eq:UlmRho})} & \multicolumn{1}{c}{ \B}\\
\cline{2-5}
\multicolumn{1}{c}{} & \multicolumn{3}{|c|}{\T Used} & \multicolumn{1}{c|}{Mentioned} \\
\multicolumn{1}{c}{} & \multicolumn{1}{|c}{\B Eq.~\ref{eq:UlmV1}} & \multicolumn{1}{c}{\B Eq.~\ref{eq:UlmV2}} & \multicolumn{1}{c|}{\B Eq.~\ref{eq:UlmV3}} & \multicolumn{1}{c|}{\B Eq.~\ref{eq:UlmV4}}\\
\hline
$\rho_1$ \T & 1 & 1 & 1 & 1 & mm\\
$z_0$ & 1.4761 & 1.9070 & 1.9493 & 3.9328 & mm\\
$\rho_d^{(opt)}$ \B & 2.0875 & 2.6969 & 2.7568 & 5.5618 & mm\\
\hline
$f_z$ \T & 82.2714 & 66.2300 & 64.1039 & 14.1339 & MHz\\
$V_0$ \B & $-$1.5193 & $-$0.9846 & $-$0.9224 & $-$0.0448 & V\\
\hline
$V_1$ & \T 0 & 0 & 0 & 0 & V\\
$V_2$ & 7 & 5.2 & 4.8 & $-$1 & V\\
$V_3$ & \B 7 & 14 & 14 & 2.611 & V\\
\hline
\T $\Delta f_z$ @ 5 K& 8.3 & 4.0 & 3.9 & 16.2 & kHz \\
 \B$\Delta f_z$ @ 300 K& 500 & 240 & 235 & 971 & ~kHz \B\\
\hline
$1:~\gamma_z$ & \T $2\pi$\,13.96 & $2\pi$\,4.50 &  $2\pi$\,4.06 & $2\pi$\,0.10 & s$^{-1}$\\
$2:~\gamma_z$ & $2\pi$\,1.86 & $2\pi$\,4.50 & $2\pi$\,4.61 & $2\pi$\,1.58 & s$^{-1}$\\
$3:~\gamma_z$ & $2\pi$\,1.86 & $2\pi$\,0.62 &  $2\pi$\,0.54 & $2\pi$\,0.23 & s$^{-1}$\\
d:~$\gamma_z^{(opt)}$ & \B $2\pi$\,30.49 & $2\pi$\,18.27 & $2\pi$\,17.49 & $2\pi$\,4.30 & s$^{-1}$\\
\hline
\end{tabular}
\caption{Absolute values for the trap used in Ulm \cite{ElectronsInPlanarPenningTrapUlm}.  Axial frequencies calculated here differ from those reported in Ref.~\cite{ElectronsInPlanarPenningTrapUlm}, which claims $f_z = 62.2$ MHz and 67.97 MHz for the potentials given in for columns 1 and 2, respectively, of this table.} \label{table:UlmTrapAbsolute}
\end{table}
}

\newcommand{\UlmTrapWithVOneEqualZeroFigure}{
\begin{figure}[htbp!]
\centering
\includegraphics*[width=\w]{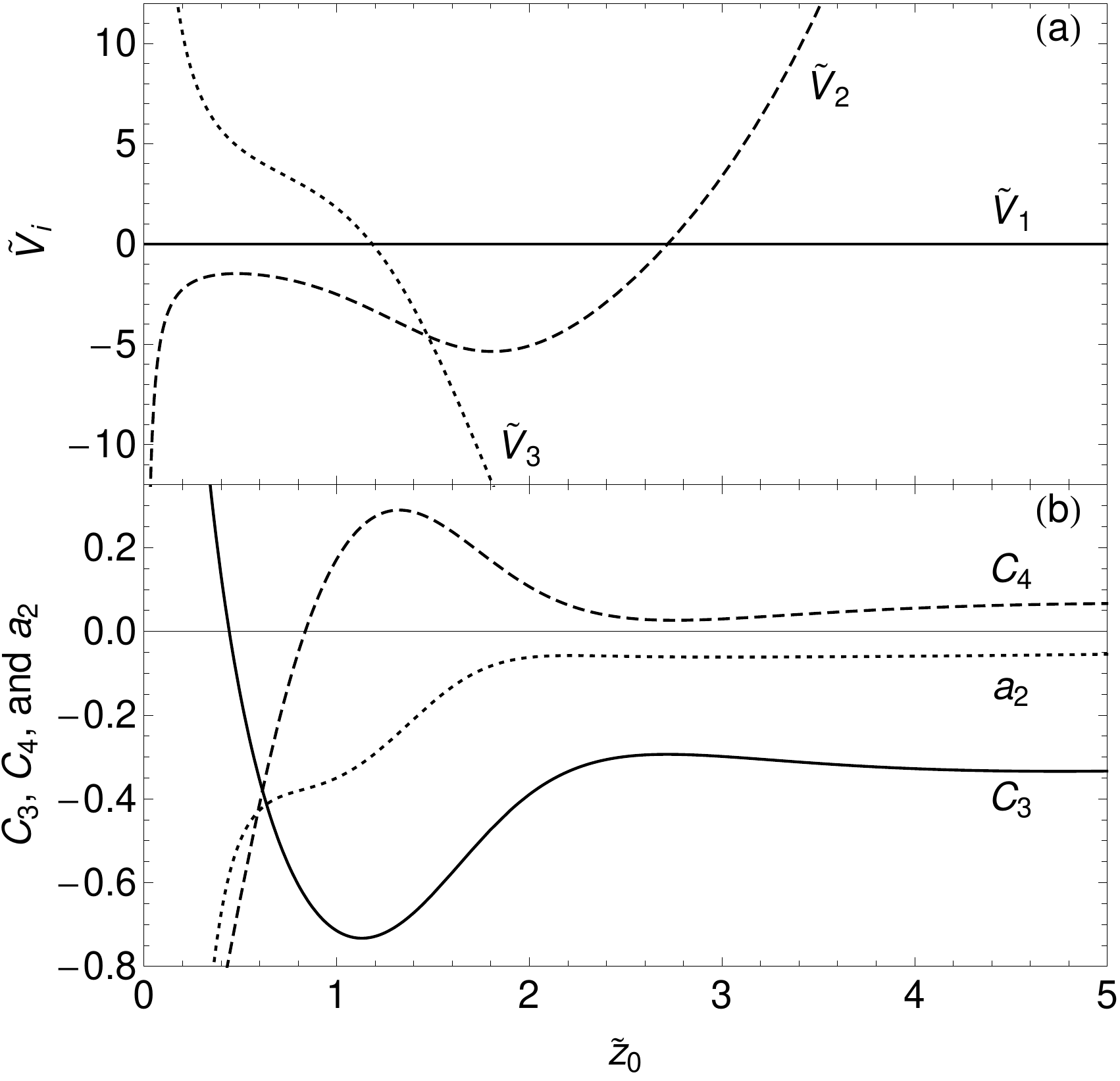}
 \caption{(a) Scaled potentials applied to the Ulm trap to get a particular $\zt_0$.  (b) Resulting trap properties for each $\zt_0$.    The trap is biased subject to the optional constraint $V_1=\Vt_1=0$.  It is not possible to make $a_2$ very small.} \label{fig:UlmTrapWithVOneEqualZero}
\end{figure}
}

\title{Optimized Planar Penning Traps for Quantum Information Studies}

\author{J.\ Goldman}
\affiliation{Department of Physics, Harvard University, Cambridge, MA 02138}

\author{G.\ Gabrielse}
\email[Email: ]{gabrielse@physics.harvard.edu} \affiliation{Department of Physics, Harvard University, Cambridge, MA 02138}

\date{Version 2: \today}

\begin{abstract}     
A one-electron qubit would offer a new option for quantum information science, including the possibility of extremely long coherence times.  One-quantum cyclotron transitions and spin flips have been observed for a single electron in a cylindrical Penning trap.  However, an electron suspended in a planar Penning trap is a more promising building block for the array of coupled qubits needed for quantum information studies.  The optimized design configurations identified here promise to make it possible to realize the elusive goal of one trapped electron in a planar Penning trap for the first time -- a substantial step toward a one-electron qubit.
\end{abstract}


\maketitle


\section{Introduction}
\label{sec:Introduction}

Quantum jumps \cite{QuantumCyclotron} have been observed between between the lowest cyclotron and spin states of an electron suspended in the magnetic field of a cylindrical Penning trap (Fig.~\ref{fig:QuantumJumps}).  These observations made possible the most precise measurements of the electron magnetic moment and the fine structure constant \cite{HarvardMagneticMoment2008}. The one-electron observations also triggered intriguing studies on  using one-electron qubits as building blocks for quantum information processing \cite{Tombesi1999,Tombesi2001,TombesiThreeQubit,Tombesi2003,TombesiLinearPRA,Tombesi2005,TombesiPlanarDecoherence,PlanarPenningTrapMainz,ElectronsInPlanarPenningTrapMainz,ElectronFrequenciesInPlanarPenningTrapMainz,ElectronsInPlanarPenningTrapUlm,UlmFinalElectronReview}.  The possibility of a very long coherence time is very attractive.

\QuantumJumpsFigure

A new trap design is needed to realize one-electron qubits for quantum information studies.  Although a single electron in a single trap is the focus of this work, a scalable array of coupled one-electron qubits is the long term goal.  Impressive progress has been made in the microfabrication of three-dimensional trap arrays \cite{SandaMicrofabricatedCylindricalTraps}, but it is still beyond these or more traditional methods to fabricate a large array of small cylindrical traps with the properties needed to observe one-quantum transitions with one electron. A scalable array of small traps seems more feasible with traps whose electrodes are entirely in a plane since these could be fabricated on a chip using variations on more standard microfabrication methods \cite{UlmFinalElectronReview}.  The chip could include electrical couplings between the traps, and could even include some detection electronics.  Secondary advantages of a planar trap would be an open structure that makes it easier to introduce microwaves (to modify or entangle electron spin states) and possibly to load electrons.

\PlanarTrapThreeDimensionsFigure

One possible planar Penning trap geometry (Fig.~\ref{fig:PlanarTrapThreeDimensions}) is a round center electrode with concentric rings \cite{PlanarPenningTrapMainz}.
Electrons were stored and observed in such a trap, first in Mainz \cite{ElectronsInPlanarPenningTrapMainz,ElectronFrequenciesInPlanarPenningTrapMainz} and then in Ulm \cite{ElectronsInPlanarPenningTrapUlm}.  The objective of the latter experiment was to duplicate in a planar trap the observations of the one-quantum transitions of a single electron in a cylindrical Penning trap (Fig.~\ref{fig:QuantumJumps}).  The final experimental report \cite{ElectronsInPlanarPenningTrapUlm} is not encouraging.  It concludes that the ``lack of mirror symmetry"  makes it ``impossible to create a genuinely harmonic potential'' and that it is thus ``impossible'' to detect a single electron within a planar Penning trap.  Whether the situation changes with much smaller planar traps is being considered \cite{UlmFinalElectronReview}.

We reach a more optimistic conclusion in this work, though the pessimism may be appropriate for the trap designs used so far.
The key to a successful planar trap for one electron is a design that minimizes  amplitude-dependent frequency shifts of the observed oscillation frequency.  This report focuses upon calculating the relationship of such shifts to the electrode geometry and applied potentials. (Appendix \ref{sec:EarlierCalculation} corrects an earlier calculation \cite{PlanarPenningTrapMainz} of the crucial amplitude-dependent frequency shifts needed to characterize and optimize planar traps.)   We identify optimized  planar Penning trap geometries and potentials that produce amplitude-dependent frequency shifts that are orders-of-magnitude smaller than for previous planar trap designs.

A high measurement precision with a single trapped particle is part of what will be required to observe a spin flip and realize a one-electron qubit. Several  of the most accurate measurements in physics illustrate the feasibility of attaining the needed precision when a trap design
that is optimized for the particular high-precision application is used.  The cylindrical Penning trap of Fig.~\ref{fig:QuantumJumps}c \cite{CylindricalPenningTrap} was designed so that its electrodes form a microwave cavity that inhibits spontaneous emission. This trap design enabled the observation of one-quantum transitions, which made possible the most accurate measurements of the electron magnetic moment and the fine structure constant \cite{HarvardMagneticMoment2008}.  An orthogonalized hyperbolic trap \cite{OrthogonalCompensate,Gabrielse84h} was designed to allow a trapping potential to be optimized without changing the trap depth.  The most precise mass spectroscopy (e.g.,\ \cite{PritchardIonBalance}) was carried out in such a trap with a single ion (or two).  An open-access Penning trap \cite{OpenTrap} was designed to allow antiprotons from an accelerator facility to enter the trap.  The most accurate comparison of $q/m$ for an antiproton and proton \cite{FinalPbarMass} was carried out with a single antiproton and a single H$^-$ ion in such a trap, as were the most accurate one-ion measurements of bound electron $g$ values \cite{Quint2000b,Quint2004} and the most precise proton-to-electron mass ratio \cite{MainzSummary2006}.  Inspired by these examples, this study of planar Penning traps for one-electron applications is carried out in the hope that a similar rigorous design approach will indicate the best route to observing one electron in a planar trap.

 Optimized geometries and biasing schemes identified here for planar Penning traps promise to reduce the amplitude dependence of the observed frequency by many orders of magnitude.  This reduction makes it much more likely that one electron can be observed in a planar Penning trap -- an important first step towards realizing a one-electron qubit.  Two related trap configurations, a covered planar trap and a mirror-image trap, offer improved shielding, new detection options, and easier trap loading.  It remains, of course, to demonstrate experimentally that the optimized planar trap designs proposed will approach the performance of the cylindrical Penning trap in which one-quantum transitions and spin flips of a single electron were observed.

\section{Outline}

Sec.~\ref{sec:PlanarPenningTraps} describes the potential and potential expansions for a planar Penning trap.
Sec.~\ref{sec:AxialOscillations} relates the amplitude dependence of the particle's axial oscillation frequency to the
potential expanded around the equilibrium location of the trapped particle. The axial oscillation of a trapped particle must
be detected to tell that a single particle is in the trap.  Small shifts in this frequency will reveal spin flips and
one-quantum cyclotron transitions.

Two-gap traps (with two biased electrodes surrounded by a ground plane) are shown in Sec.~\ref{sec:TwoGapTraps} to be
inadequate for the observation and the manipulation of a single electron.  The considerable promise of three-gap traps (with
three biased electrodes surrounded by a ground plane) is the subject of Sec.~\ref{sec:OptimizedThreeGapTraps}.  Optimized
planar trap configurations that make the particle's oscillation frequency essentially independent of oscillation amplitude
are identified and discussed, along with the detection and damping of the particle's motion.

Sec.~\ref{sec:LaboratoryPlanarTraps} estimates the size of the unavoidable deviations between ideal planar Penning traps
and the actual laboratory traps.  Real traps have gaps between electrodes, finite boundary conditions, and imperfections in the trap dimensions, all of which must be compensated by modifying the voltages applied to the trap electrodes.

A covered planar trap (a two-gap planar trap covered by a parallel conducting plane) is proposed in Sec.~\ref{sec:CoveredPlanarTraps} as a scalable way to make planar chip traps less sensitive to nearby apparatus.    An electron suspended midway between a mirror-image pair of planar electrodes  is shown in Sec.~\ref{sec:MirrorImageTraps} to be in a potential with much the same properties as is experienced by an electron centered in a cylindrical Penning trap.  For an electron initially loaded and observed in an ``orthogonalized'' mirror-image trap, we illustrate in Sec.~\ref{sec:Mirror-ImageToCoveredTrap} the possibility to adiabatically change the applied trapping potentials to move the electron into a covered planar Penning trap that is optimized.

The damping and detection of a particle in planar traps, covered planar traps and mirror image traps are considered in Sec.~\ref{sec:Damping}.  The  optimization of damping and detection is discussed, as are unique detection opportunities available with a covered planar trap.

A conclusion in Sec.~\ref{sec:Conclusion} is followed by three appendices.  Appendix \ref{sec:EarlierCalculation} corrects an earlier calculation of amplitude-dependent
frequency  shifts in a planar trap.  Appendices \ref{sec:MainzTrap} and \ref{sec:UlmTrap} use the calculations of this work
to analyze the properties of planar traps built at Mainz and Ulm.

\section{Planar Penning Traps}
\label{sec:PlanarPenningTraps}

\subsection{The Ideal to be Approximated}

An ideal Penning trap, which we seek to approximate, starts with a spatially uniform magnetic field,
\begin{equation}
\mathbf{B} = B \zhat. \label{eq:MagneticField}
\end{equation}
Superimposed is an electrostatic quadrupole potential, $V_2(\rho,z)$ in cylindrical coordinates, that is a harmonic
oscillator potential on the $\rho=0$ axis,
\begin{equation}
V_2(0,z) = \frac{1}{2} V_0 \left(\frac{z - z_0}{\rho_1}\right)^2,\label{eq:HarmonicPotential}
\end{equation}
where $\rho_1$ sets the size scale for the trap and $V_0$ sets the potential scale.

A particle of charge $q$ and mass $m$ on axis then oscillates at an axial angular
frequency
\begin{equation}
\omega_z =\sqrt{ \frac{q}{m} \frac{V_0}{ {\rho_1}^2}}\label{eq:AxialFrequency}
\end{equation}
about the potential minimum at $z_0$.  The potential will trap a particle only if $q V_0 > 0$.

The axial oscillation frequency $\omega_z$ is the key observable for possible quantum information studies. The one-quantum
cyclotron and spin flip transitions that have been observed (e.g., Fig.~\ref{fig:QuantumJumps}a-b) were detected using the
small shifts in $\omega_z$ caused by a quantum non-demolition (QND) coupling of the cyclotron and spin energies to
$\omega_z$.

For potentials that can be expressed on an axis of symmetry as a power series in $z-z_0$ (e.g.,
Eq.~\ref{eq:HarmonicPotential}), the general solution to Laplace's equation near the point $(0,z_0)$ is related to the
axial solution by the substitution,
\begin{equation}
(z-z_0)^k \rightarrow \left[{\rho^2 + (z-z_0)^2}\right]^{k/2} P_k\left[\cos (\theta)\right],\label{eq:Substitution}
\end{equation}
where $\cos (\theta ) = (z-z_0)/\sqrt{\rho^2+(z-z_0)^2}$ and $P_k$ is a Legendre polynomial. We will focus upon axial
potentials throughout this work, since this procedure can be used to obtain the general potential in the neighborhood of any
axial position when this is needed.

Applied to the harmonic axial potential of an ideal Penning trap,
\begin{equation}
V_2(\rho,z) = \frac{V_0}{2} \frac{\rho^2 + (z-z_0)^2}{{\rho_1}^2} P_2\left[\cos (\theta)\right].\label{eq:IdealQuadrupole}
\end{equation}
This quadrupole potential for an ideal Penning trap extends through all space.

\subsection{Electrodes in a Plane}

A planar Penning trap (Figs.~\ref{fig:PlanarTrapThreeDimensions}-\ref{fig:PlanarTrap}) starts with a  spatially uniform
magnetic field as in Eq.~\ref{eq:MagneticField}.  An electrostatic potential is produced by biasing $N$ ring electrodes in a
plane perpendicular to the symmetry axis of the electrodes, $\zhat$.  An electrode with an outer radius $\rho_i$ is biased
to a potential $V_i$, as illustrated in Fig.~\ref{fig:PlanarTrap}.  Without loss of generality, the potential
beyond the rings, $\rho > \rho_N$, is taken to be the zero of potential, $V_{N+1} = 0$.  The $N$ gaps between biased
electrodes are taken initially to be infinitesimal, but this condition is relaxed in Sec.~\ref{sec:Gaps}.

\PlanarTrapFigure

Two remaining boundary conditions,
\begin{subequations}\label{eq:BoundaryConditionsInfinity}
\begin{eqnarray}
V(\rho, z\rightarrow\infty) = 0\\
V(\rho\rightarrow\infty,z) = 0
\end{eqnarray}
\end{subequations}
will be assumed to derive the potential for $z\ge0$.  It is not always possible in real apparatus to keep all metal far
enough away from the trap electrodes so that these boundary conditions are accurately satisfied.  We consider the case of
finite boundary conditions in Sec.~\ref{sec:FiniteBoundaries}.

Throughout this work we will illustrate the basic features and challenges of a planar Penning trap using a three-gap ($N =
3$) sample trap with dimensions
\begin{equation}
\{ \rho_i \} = \{1, 5.5, 7.5426\}\,\rho_1, \label{eq:SampleTrapDimensions}
\end{equation}
for reasons discussed in Sec.~\ref{sec:OptimizedThreeGapTraps}.  Fig.~\ref{fig:PlanarTrap} shows this relative geometry (to scale).

\subsection{Scaling Distances and Potentials}

It is natural and often useful to scale distances of the radius of the inner electrode, $\rho_1$.  We will do so, using the
notation $\zt = z /\rho_1$ and $\rt = \rho /\rho_1$.  The relative geometry of a planar Penning trap is then given by the
set of dimensions $\{\rt_i \} = \{1,\rt_2,  \rt_3, \ldots \}$, for example.

It is natural and convenient to scale the trap potential $V$, along with the voltages $V_i$ applied to trap electrodes, in
terms of a voltage scale, $V_0$, to be determined.  We will then use scaled applied potentials, $\wt{V}_i = V_i/V_0$, and a
scaled trap potential, $\wt{V} = V/V_0$.

\subsection{Exact Superposition}

The potential produced by a planar Penning trap is a superposition
\begin{equation}
V(\rt,\zt)= \sum_{i=1}^N V_i \, \phi_i(\wt{\rho},\wt{z})\label{eq:Superposition}
\end{equation}
that is linear in the relative voltages applied to trap electrodes. The functions $\phi_i$ are solutions to Laplace's
equation with boundary conditions such that $\phi_i=1$ on the electrode that extends to $\rho_i$ and is otherwise zero on
the boundary.  More precisely,
\begin{subequations}\label{eq:PhiBoundaryConditions}
\begin{eqnarray}
\phi_i(\rt, 0) &=& \left\{
    \begin{array}{cc}
    0, & \rt < \rt_{i-1}\\
    1, & \rt_{i-1}<\rt<\rt_i\\
    0, & \rt>\rt_i
    \end{array}
     \right. ,  \\
\phi_i(\rt, \zt \rightarrow\infty) &=& 0\\
\phi_i(\rt \rightarrow\infty,\zt) &=& 0.
\end{eqnarray}
\end{subequations}
These potentials are independent of the voltages applied to the trap and depend only upon the relative geometry of the trap
electrodes.

Standard electrostatics methods \cite{Jackson3rdEd,Kusse} give the $\phi_i$ that satisfy Laplace's equation for $\zt \ge 0$ and the cylindrically
symmetric boundary  conditions above,
\begin{eqnarray}
\phi_i (\rt,\zt) &=&  \rt_i \int_0^\infty dk e^{- k \zt} J_1(k\rt_i) J_0(k \rt )\nonumber\\
&-&\rt_{i-1} \int_0^\infty dk e^{- k \zt} J_1(k\rt_{i-1}) J_0(k \rt ),\label{eq:PotentialOffAxis}
\end{eqnarray}
with the convention that $\rt_0=0$.  The integrals are over products of Bessel functions. On axis,
\begin{equation}
\phi_i(0,\zt) = \frac{\zt}{\sqrt{(\rt_{i-1})^2 + \zt^2}} - \frac{\zt}{\sqrt{(\rt_i)^2 + \zt^2}}.\label{eq:Exactphii}
\end{equation}
Most of the properties of a planar Penning trap can be deduced from just the potential on axis. Expressions equivalent to Eqs.~\ref{eq:PotentialOffAxis}-\ref{eq:Exactphii} are in Ref. \cite{PlanarPenningTrapMainz}.

To emphasize the role of the $N$ gaps of a planar trap we define the gap potential across gap $i$ as the difference, $\Delta V_i \equiv
V_{i+1}-V_i$. The axial potential is then given by
\begin{align}
V(0,\zt) &= \sum_{i=1}^N \Delta V_i \, \Phi_i(\zt) \label{eq:GapExpansion}\\
\Phi_i(\zt) &= \frac{\zt}{\sqrt{(\rt_i)^2 + \zt^2}} - 1,\label{eq:Phi}
\end{align}
a sum of contributions from the $N$ gap potentials.

The axial potential can be computed exactly using Eq.~\ref{eq:Superposition} and Eq.~\ref{eq:Exactphii}, or  alternately
from Eq.~\ref{eq:GapExpansion}.  Fig.~\ref{fig:AxialPotential} compares an ideal harmonic axial potential to examples of axial potentials for optimized planar Penning
trap configurations to be discussed.  Fig.~\ref{fig:PlanarTrapThreeDimensions}b shows equipotentials spaced by $V_0$ for a planar Penning trap (configuration I in Table~\ref{table:SampleTrap}).  The equipotentials are calculated for infinitesimal gaps, but the electrodes are represented with finite gaps to make them visible.  The equipotentials terminate in the gaps between electrodes.  The dashed equipotentials of an ideal quadrupole are superimposed near the trap center.

\AxialPotentialFigure

\subsection{Expansion of the Trap Potential}

To characterize the trap potential $V(\rt,\zt)$ for $\zt\ge 0$ it suffices to focus upon expansions of the potential on the
$\rt=0$ axis.  The potential near any expansion point $\zt_0$ on this axis can be obtained using the substitution of
Eq.~\ref{eq:Substitution}.  The axial potential due to one electrode (Eq.~\ref{eq:Exactphii}) can be expanded in a
Taylor series,
\begin{equation}
\phi_i (0,\zt ) = \frac{1}{2} \sum_{k=0}^\infty C_{ki} (\zt - \zt_0)^k.\label{eq:Phii}
\end{equation}
The expansion coefficients,
\begin{equation}
C_{ki} = \frac{2}{k!} \left[\frac{\partial^k \phi_i(0,\zt)}{\partial \zt^k}  \right]_{\zt = \zt_0},\label{eq:Cki}
\end{equation}
are analytic functions of the relative trap geometry, $\{ \rt_i \}$, and the relative location of the expansion point,
$\zt_0$.

The full trap potential can be similarly expanded as
\begin{equation}
V(0,\zt) = \frac{1}{2} V_0 \sum_{k=0}^\infty \,C_k \,(\zt - \zt_0)^k. \label{eq:ExpandV}
\end{equation}
The one expansion coefficient needed for $k=2$ is so far written as $V_0 C_2$.  With no loss of generality we are thus free
to choose $C_2=1$. This determines $V_0$ and the $C_k$,
\begin{eqnarray}
V_0 &=& \sum_{i=1}^N C_{2i} V_i\label{eq:V0}\\
C_k &=& \sum_{i=1}^N C_{ki} \Vt_i \label{eq:Ck}.
\end{eqnarray}
The latter equation, and the rest of this work, make frequent use of the scaled potentials $\Vt_i=V_i/V_0$.  For the scaled
potentials, Eq.~\ref{eq:V0} can be regarded as a constraint,
\begin{equation}
\sum_{i=1}^N C_{2i} \Vt_i = 1,\label{eq:C2Constraint}
\end{equation}
that an acceptable set of relative potentials must satisfy.

A trap is formed at $\zt=\zt_0$ only if there is a minimum in the potential energy $qV(0,\zt )$ for a particle with charge
$q$ and mass $m$.  The linear gradient in the potential must thus vanish at this point, whereupon
\begin{equation}
C_1 = \sum_{i=1}^N C_{1i} \Vt_i = 0.   \label{eq:C1Constraint}
\end{equation}
Near the minimum the potential energy will then have the form $m {\omega_z}^2 (z-z_0)^2/2$, where $\omega_z$ is the angular
oscillation frequency of the trapped particle in the limit of a vanishing oscillation amplitude.  Comparing to the quadratic
term in Eq.~\ref{eq:ExpandV} gives
\begin{equation}
{\omega_z}^2 = \frac{q V_0}{m {\rho_1}^2}, \label{eq:AxialFrequency2}
\end{equation}
the same as for the ideal case considered earlier because of our choice of $V_0$. Forming a trap thus requires that $q$ and
$V_0$ have the same sign at $\zt_0$.  The sign of $V_0$ can be flipped if it is wrong by simply flipping the sign of all of
the applied potentials.

\subsection{Two Viewpoints}
\label{sec:TwoViewpoints}

Two different viewpoints of the potential expansions and equations are useful.  The first is needed to analyze the
performance of an $N$-gap trap.  The second facilitates the calculation of optimized trap configurations.

The point of view that we take to analyze an $N$-gap trap starts with the $N$ radii $\{\rho_i\}$ and the $N$ applied
potentials $\{V_i\}$.  These are the $2N$ parameters that fully characterize such a trap. No interrelations constrain the
values of these parameters, so the difference of the number of parameters and constraints is $2N$.

The axial potential is then a superposition (from Eq.~\ref{eq:Superposition}) of the $\phi_i(0,\zt)$ from
Eq.~\ref{eq:Exactphii} with scaled radii $\{ \rt_i \}=\{ \rho_i \}/\rho_1$.  The extremum of $V(0,\zt)$ is the $\zt_0$
needed to evaluate the expansion coefficients $C_{ki}(\rt_i,\zt_0)$ using Eq.~\ref{eq:Cki}.  All of the properties of a trap
at $\zt=\zt_0$ can then be determined.  The potential scale $V_0(\rho_i,\zt_0,V_i)$ is determined using Eq.~\ref{eq:V0}, the
axial frequency from Eq.~\ref{eq:AxialFrequency}, and the expansion coefficients $C_k$ from Eq.~\ref{eq:Ck}. An example
analysis for two existing planar Penning traps is provided in the Appendices.

The point of view we take to identify optimized planar trap configurations instead uses $2N+2$ parameters to characterize a
planar trap.  The effect of the two additional parameters is compensated by the addition of the two constraints $C_1=0$ and
$C_2=1$ (from Eq.~\ref{eq:C2Constraint} and Eq.~\ref{eq:C1Constraint}).  The difference in the number of the parameters and
constraints is thus $2N$, just as for our analysis above.

We will first seek solutions for scaled trap configurations, for which there are $2N$ parameters and $2$ constraints.  The
parameters are the $N$ scaled potentials $\{ \Vt_i \}$, the $N-1$ scaled radii $\{ \rt_i \}$, and the scaled distance $\zt_0
> 0$.  The two constraints, $C_1=0$ and $C_2=1$, are from Eq.~\ref{eq:C2Constraint} and Eq.~\ref{eq:C1Constraint}. The
difference of the number of parameters and constraints for any scaled trap configuration is thus $2N-2$. We are thus free to
specify up to $2N-2$ additional constraints on the scaled radii and scaled potentials, though not all constraints will have
a set of parameters that satisfies them.

Once the $2N-2$ scaled potentials and radii are chosen, we are then free to choose two additional parameters to bring the
difference of the parameters and constraints back up to $2N$.  A convenient distance scale $\rho_1$ and a convenient
potential scale $V_0$ can be chosen to get a desired axial frequency (using Eq.~\ref{eq:AxialFrequency2}).  The radii are
then $\{ \rho_i \} = \rho_1 \{ \rt_i \}$, and the applied potentials are $\{ V_i \} = V_0 \,\{ \Vt_i \}$.

Before applying these general considerations to two- and three-gap traps, we discuss amplitude-dependent frequency shifts
since these will determine the additional constraint equations that we need to design optimized planar traps.

\section{Axial Oscillations}
\label{sec:AxialOscillations}

In this section we investigate the axial oscillation of a trapped particle near the potential energy minimum of a planar
Penning trap.  In the following sections, Sec.~\ref{sec:TwoGapTraps} and Sec.~\ref{sec:OptimizedThreeGapTraps}, we investigate optimized planar Penning traps,
realized by imposing additional requirements on the design of planar traps (in addition to the two above) to make the axial
oscillation of a trapped particle more harmonic.

The crucial observable for realizing a one-electron qubit is the frequency of the axial oscillation of a trapped electron.
One trapped particle will be observed in the planar Penning trap only if the oscillation frequency is well enough defined to
allow narrow-band radiofrequency detection methods to be used. Small changes in the particle's oscillation frequency will
signal one-quantum transitions of the qubit, as has been mentioned.

For a perfect quadrupole potential, the motion of a trapped particle on the symmetry axis of the trap is perfect harmonic
motion at a single oscillation frequency, $\omega_z$, independent of the amplitude of the oscillation.  For a charged
particle trapped near a minimum of the non-harmonic potential expanded in Eq.~\ref{eq:ExpandV}, near $\zt = \zt_0$, the
oscillation frequency depends upon the oscillation amplitude.

\subsection{Amplitude-Dependent Frequency}

The oscillation frequency for a particle trapped near a potential minimum within a planar Penning trap depends upon the
oscillation amplitude.  A derivation of this amplitude-dependence starts with applying Newton's second law to get the
equation of motion.  For a particle of charge $q$ and mass $m$ on the symmetry axis $\zhat$ of the trap,
\begin{equation}
\frac{d^2}{dt^2} u + (\omega_z)^2 u + \lambda \, (\omega_z )^2 \, \sum_{k=3}^\infty \frac{k C_k}{2} u^{k-1} = 0,\label{eq:EquationOfMotion}
\end{equation}
where $u=\zt-\zt_0$.  The harmonic restoring force is presumed to be larger than the additional (unwanted) terms.  The latter are labeled with a
dimensionless smallness parameter, $\lambda$, that is taken to be unity at the end of the calculation.

Solutions are sought in the form of series expansions of the amplitude and the oscillation frequency in powers of the
smallness parameter,
\begin{subequations}
\begin{align}
u &= u_0 + \lambda u_1 + \lambda^2 u_2 + \cdots \\
\omega &= \omega_z + \lambda \omega_1 + \lambda^2 \omega_2 + \cdots .
\end{align}
\end{subequations}
The lowest-order solution is a harmonic oscillation, with oscillation amplitude $\At \, \rho_1$, for which we chose the phase
$u_0 = \At \cos (\omega t)$ with $\At>0$.

By assumption, the lowest-frequency fourier component of the particle's axial motion is predominant.  Fourier components at harmonics of $\omega_z(\At)$, not shown explicitly in the formula, have smaller amplitudes.  The frequency contributions are determined by the requirement that no artificial driving terms resonant at angular
frequency $\omega$ are introduced.  This well-known method \cite{LandauMechanics,Mickens} is sometimes called  the
Linstedt-Poincar\'e method. The result is that the oscillation frequency $\omega = \omega_z(\At)$ is a function of
oscillation amplitude $\At$ for the harmonic fourier component, given by
\begin{eqnarray}
\omega_z(\At) &=& \omega_z \left[1 + \sum_{k=2}^\infty a_k \At^k \right].\label{eq:AmplitudeDependentFrequency}
\end{eqnarray}
At zero amplitude the oscillation frequency $\omega_z(\At) \rightarrow \omega_z$, of course.  There is no term linear in the
oscillation amplitude (contrary to Ref.~\cite{PlanarPenningTrapMainz}, see Appendix \ref{sec:EarlierCalculation}.)

The amplitude coefficients $a_k$ are functions of the potential expansion coefficients $C_k$, each of which in turn is a
function of the trap dimensions $\rho_i$ and the potentials $V_i$ applied to the trap electrodes.
\begin{subequations}\label{eq:ak}
\begin{eqnarray}
a_2 &=& -\frac{15 (C_3)^2}{16}+\frac{3 C_4}{4}.\label{eq:a2}\\
a_3 &=& -\frac{15 (C_3)^3}{16}+\frac{3 C_3 C_4}{4}\label{eq:a3one}\\
&=& C_3 a_2.\label{eq:a3}\\
a_4 &=& -\frac{2565 (C_3)^4}{1024}+\frac{645 (C_3)^2 C_4}{128}-\frac{21 (C_4)^2}{64} \nonumber \\
&&-\frac{105 C_3 C_5}{32}+\frac{15 C_6}{16}.\label{eq:a4}\\
a_5 &=& -\frac{2565 (C_3)^5}{512}+\frac{765 (C_3)^3 C_4}{64}-\frac{69 C_3 (C_4)^2}{32} \nonumber \\
&&-\frac{15 (C_3)^2 C_5}{2} +\frac{3 C_4 C_5}{4}+\frac{15 C_3 C_6}{8}\\
&=& (C_5-2C_3C_4) a_2 + 2C_3 a_4.\label{eq:a5}\\
a_6 &=& -\frac{205845 (C_3)^6}{16384}+\frac{159795 (C_3)^4 C_4}{4096}\nonumber\\
&& -\frac{21039 (C_3)^2 (C_4)^2}{1024} +\frac{81 (C_4)^3}{256}-\frac{13545 (C_3)^3 C_5}{512}\nonumber \\
&& +\frac{1995 C_3 C_4 C_5}{128} -\frac{315 (C_5)^2}{128}+\frac{3015 (C_3)^2 C_6}{256}\nonumber \\
&&-\frac{57 C_4 C_6}{64} -\frac{315 C_3 C_7}{64}+\frac{35 C_8}{32}.\\
a_7 &=& 3 {C_3} a_6 + \left[-3 (C_3)^3-4 {C_3} {C_4}+2 {C_5}\right]a_4\nonumber\\
 &+& \left[3 {C_3}^3 {C_4} + 4{C_3} (C_4)^2 - 2 {C_4} {C_5} - 3 {C_3} {C_6}+{C_7}\right] a_2.\nonumber\\ \label{eq:a7}
\end{eqnarray}
\end{subequations}
The exact expressions derived for $a_8$ and $a_9$ take too much space to display, and are not normally needed. A convention
other than $C_2=1$ would require that each $C_k$ in the previous equations be replaced by $C_k/C_2$.

Several properties of the relationships between the $C_k$ and $a_k$ will be exploited for designing planar traps. Two
combinations of potential expansion coefficients make $a_2=0$:
\begin{eqnarray}
C_3=C_4=0 ~~~&\Rightarrow&~~~ a_2=0\label{eq:a2VanishHarmonic}\\
C_4=\frac{5}{4}(C_3)^2 ~~~&\Rightarrow&~~~ a_2=0.\label{eq:a2VanishNonharmonic}
\end{eqnarray}
Relationships between the $a_k$ in Eqs.~\ref{eq:ak}a-f imply
\begin{eqnarray}
a_2=0 ~~~&\Rightarrow&~~~ a_3=0\label{eq:a3Froma2}\\
a_2=a_4=0 ~~~&\Rightarrow&~~~ a_2=a_3=a_4=a_5=0.\label{eq:a5Froma4}
\end{eqnarray}
One set of potential coefficients that produce this remarkable suppression of the low-order $a_k$ is
\begin{align}
C_3&=C_4=C_6=0\nonumber\\
&\Rightarrow~~~ a_2=a_3=a_4=a_5=0.\label{eq:OptimizedTrap1}
\end{align}
Another is
\begin{align}
C_4&=\frac{5}{4}(C_3)^2 ~~\rm{and}~~
C_6 = -\frac{7}{2}C_3((C_3)^3-C_5)\nonumber\\
&\Rightarrow~~~ a_2=a_3=a_4=a_5=0.\label{eq:OptimizedTrap2}
\end{align}
It remains to investigate whether and how any or all of these attractive combinations of $C_k$ values can be produced by
biasing a planar Penning trap.

\subsection{Tunabilities}

A change in the potential $V_i$ applied to each electrode will change the axial frequency $\omega_z$ and will also change
the amplitude dependence of the axial frequency by changing $a_2$.  The orthogonalized hyperbolic, cylindrical, and
open-access traps were designed so that the potential applied to one pair of electrodes changed the axial frequency very
little while changing $a_2$.  The potential on such compensation electrodes could then be changed to tune $a_2$ to zero
without shifting the axial frequency out of resonance with the detectors that were needed to monitor the improvement.

We define a tunability for each electrode,
 \begin{equation}
\gamma_i =
 \frac{1}{\displaystyle \omega_z} \frac{\displaystyle \partial \omega_z}{\displaystyle \partial V_i}
 /
 \frac{\displaystyle \partial a_2}{\displaystyle \partial V_i}
 . \label{eq:gammai}
\end{equation}
to quantify how useful the electrodes will be for tuning $a_2$.  The tunabilities are defined as generalizations of the
single tunability $\gamma$ used to optimize the design of the orthogonalized traps.

Ideally, and this ideal was closely approximated in the orthogonalized traps, there are compensation electrodes for which
$\gamma_i \approx 0$, and other electrodes for which $\gamma_i$ is very large in magnitude.   In
Sec.~\ref{sec:CylindricalTrap} we will review the tunabilities that were calculated and realized for the cylindrical trap.
In sections that follow we will compare these to what can be realized with a planar Penning trap.

\subsection{Harmonics of the Axial Oscillation}

The largest fourier components for the small-amplitude motion of the trapped particle are given by
\begin{equation}
\tilde{z} = \tilde{z}_0 + \At_0 + \At_1 \cos(\omega t) + \At_2 \cos(2 \omega t) + \At_3 \cos(3 \omega t) + \ldots\\
\end{equation}
By assumption, the harmonic fourier component at frequency $\omega$ has the larger amplitude $\At_1\approx\At$, with the harmonics then
given by
\begin{align}
\At_1 &= \At + \frac{C_3}{2}\At^2 + \left[\frac{29 (C_3)^2}{64} - \frac{C_4}{16}\right] \At^3 + \ldots\label{eq:A1}\\
\At_2 &= \frac{C_3}{4}\At^2 + \frac{(C_3)^2}{4}\At^3 +  \ldots \label{eq:A2}\\
\At_3 &= \left[\frac{3 (C_3)^2}{64}+\frac{C_4}{16}\right] \At^3 + \ldots . \label{eq:A3}
\end{align}
Insofar as $\At \ll 1$, these higher-order oscillation amplitudes are smaller, but they depend critically upon the low-order
potential expansion coefficients as well.

\subsection{Thermal Spread in Axial Frequencies}
\label{sec:ThermalBroadening}

The image current induced in nearby trap electrodes by a particle's axial motion is sent through the input resistance of a detection amplifier circuit.  The oscillating voltage across the resistor is detected with a very sensitive cryogenic amplifier.  Energy dissipated in the resistor damps the axial motion, with some damping time $(\gamma_z)^{-1}$.  Sec.~\ref{sec:Damping} shows how the damping rate $\gamma_z$ is related to the resistance for a three-gap trap.

The damping brings the axial motion of a trapped particle into thermal equilibrium at the effective temperature of the amplifier.   It is quite challenging to achieve a low axial temperature with an amplifier turned on.    For example, the electrodes of the cylindrical Penning trap were cooled to 0.1 K with a dilution refrigerator.  Even with very careful heat sinking of a MESFET amplifier that was run at an extremely low bias current, however, the axial temperature with the amplifier operating was still $T_z = 5.2$ K \cite{FeedbackCoolingPRL}. We then used feedback cooling to bring the axial temperature as low as 0.85 K \cite{FeedbackCoolingPRL}.  A lower axial temperature was obtained, but only by switching the amplifier off during critical stages of the measurement of the electron magnetic moment. For the estimates that follow we will assume an axial temperature of 5 K, but stress that much higher axial temperatures are very hard to avoid.

What has prevented the observation of one electron in a planar trap so far is the large amplitude dependence of the axial frequency in such traps.  Thermal fluctuations of the particle's axial energy make the particle oscillate at a range of fourier components, $\Delta \omega_z$.  In the cylindrical trap of Fig.~\ref{fig:QuantumJumps}c this spread in frequencies is less than the damping width, $\Delta \omega_z < \gamma_z$.  For planar traps so far the thermal spread of axial oscillation frequencies is much broader than the damping width, $\Delta \omega_z \gg \gamma_z$.

As a measure of the thermal damping width we will consider only the lowest-order contribution
\begin{equation}
\frac{\Delta \omega_z}{\omega_z} \approx |a_2| \frac{k_B T_z}{\frac12 m {\omega_z}^2{\rho_1}^2}.\label{eq:ThermalBroadeninga2}
\end{equation}
It should be possible to calculate neglected higher-order contributions if correlations are considered carefully, but this lowest expression suffices for our purposes.

The tables that follow report the lowest order thermal widths $\Delta f_z = \Delta \omega_z/(2\pi)$  and the damping widths $\gamma_z/(2\pi )$ in Hz.  Vanishing values of $\Delta f_z$ thus mean that $a_2=0$, whereupon there is typically not much thermal broadening of the damping width.  However, higher-order contributions ensure that there is always a nonvanishing thermal width.

\section{Two-Gap Traps}
\label{sec:TwoGapTraps}

A minimal requirement for a useful trap is that it be possible to bias its electrodes to make the leading contribution to
the amplitude dependence of the axial frequency vanish, $a_2=0$. We show here that this is not possible with a two-gap
($N=2$) planar trap.

A scaled two-gap planar trap is characterized by $2N=4$ parameters: $\rt_2$, $\Vt_1$, $\Vt_2$, and $\zt_0$.  These
parameters must satisfy the two constraints $C_1=0$ and $C_2=1$ (of Eqs.~\ref{eq:C2Constraint} and \ref{eq:C1Constraint}).
The difference of the number of parameters and constraints is thus $2N-2=2$.  Consistent with this we can solve for any two
of the parameters in terms of the other two.

Unfortunately, if the additional constraint $a_2=0$ is added then there are no sets of parameters that are solutions.  An
explicit demonstration that $a_2$ cannot be made to vanish comes from solving for $\Vt_1$ and $\Vt_2$ in terms of $\zt_0$
and $\rt_2$ using the two constraint equations.  These solutions determine
\begin{widetext}
\begin{eqnarray}
C_3 &=& \frac{-9 (\zt_0)^4+(\rt_2)^2-4 (\zt_0)^2 \left[1+(\rt_2)^2\right]}{3 \left[\zt_0+(\zt_0)^3\right] \left[(\zt_0)^2+(\rt_2)^2\right]}\label{eq:TwoGapC3}\\
C_4 &=& \frac{5 \left[15 (\zt_0)^6+12 (\zt_0)^4 \left[1+(\rt_2)^2\right]-3 \left[(\rt_2)^2+(\rt_2)^4\right]+(\zt_0)^2 \left[4-5(\rt_2)^2+4 (\rt_2)^4\right]\right]}
{12 \left[1+(\zt_0)^2\right]^2 \left[(\zt_0)^2+(\rt_2)^2\right]^2}\\
a_2 &=& -\frac{5 \left[36 (\zt_0)^8+(\rt_2)^4+36 (\zt_0)^6 \left[1+(\rt_2)^2\right]+(\zt_0)^2
\left((\rt_2)^2+(\rt_2)^4\right)+{\zt_0}^4 \left[4+29 (\rt_2)^2+4 (\rt_2)^4\right]\right]}
{48 (\zt_0)^2
\left[1+(\zt_0)^2\right]^2 \left[(\zt_0)^2+(\rt_2)^2\right]^2}
\end{eqnarray}
\end{widetext}
The amplitude coefficient $a_2$ is explicitly negative for all values of $\zt_0$ and $\rt_2$, and it only approaches zero in
the not-so-useful limit that $\zt_0 \rightarrow 0$.

The best that can be done with a two-gap trap is to use $C_3=0$ as a third constraint on the four parameters, $\Vt_1$,
$\Vt_2$, $\zt_0$ and $\rt_2$. Only the analytic solution for $\zt_0$ is simple enough to display here,
\begin{equation}
\zt_0 = \frac{1}{3}\sqrt{\sqrt{4 (\rt_2)^4+17 (\rt_2)^2+4} -2 (\rt_2)^2 -2 }.\label{eq:TwoGapzt0}
\end{equation}
Fig.~\ref{fig:TwoGap} shows how the parameters of two-gap traps depend upon $\rt_2$.

\TwoGapFigure

Two-gap traps are not so useful given that it is not possible to do better than make $C_3=0$.  It is not possible to make
$a_2=0$.    For the remainder of our discussion of planar traps we concentrate on three-gap traps since these have much better properties.

\section{Optimized Three-Gap Traps}
\label{sec:OptimizedThreeGapTraps}

\subsection{Overview}

The goal of our optimization of a planar trap is to reduce the amplitude dependence of the axial oscillation frequency of a
trapped particle to a manageable level so that the trapped particle's oscillation energy is in a narrow range of fourier
components.  Otherwise, the oscillation energy will be broadened by noise-driven amplitude fluctuations to a broader range
of fourier components. The signal induced by the more harmonic axial oscillation can then be filtered with a narrow-band
detector that rejects nearby noise components, making possible the good signal-to-noise ratio needed to detect the small
frequency shifts that signal one-quantum transitions.

The dependence of the axial frequency $\omega_z(\At)$ on the oscillation amplitude $\At=A/\rho_1$ is given by
\begin{align}
\frac{\omega_z(\At) - \omega_z}{\omega_z} =&   \nonumber \\
 1 + a_2 \At^2 + a_3 &\At^3 + a_4 \At^4 + a_5 \At^5 + a_6 \At^6 + \ldots ,
\end{align}
the low-order terms from Eq.~\ref{eq:AmplitudeDependentFrequency}.   Since $\At \ll1$, the lowest-order amplitude
coefficient $a_2$ is the most important, followed by $a_3$, etc.  Each of the coefficients $a_k$ is a function (given in
Eq.~\ref{eq:ak}) of the potential expansion coefficients.  Each of these is determined by the geometry and the applied
trapping potentials, which must then be determined.

As discussed more generally in Sec.~\ref{sec:TwoViewpoints}, a scaled three-gap planar Penning trap configuration is
specified by $2N=6$ parameters: $\rt_2$, $\rt_3$, $\Vt_1$, $\Vt_2$, $\Vt_3$, and $\zt_0$. These can be chosen to realize
desired properties of a trap.  These parameters must satisfy the two constraint equations $C_1=0$ and $C_2=1$
(Eqs.~\ref{eq:C2Constraint} and \ref{eq:C1Constraint}). The difference of the number of parameters and the number of
constraints is thus $2N-2=4$.  The challenge is to identify up to 4 useful sets of constraint equations for which solutions
exist.

\subsection{What is Needed?}
\label{sec:CylindricalTrap}

To estimate what is needed to observe a single trapped electron it is natural to look to the demonstrated properties of the cylindrical trap used to observe the one-quantum transitions we seek to emulate.  The electrodes of a cylindrical trap are invariant under reflections $z \rightarrow -z$ about the position of the trapped particle. This symmetry is never true for a planar Penning trap.

The first consequence of the reflection symmetry is that the odd-$k$ expansion coefficients $C_k$ vanish.  The second is that the low-order, odd-$k$ $a_k$ vanish as well, since these are proportional to the $C_k$ with odd $k$. For a cylindrical trap the frequency expansion coefficients $a_k$ of Eq.~\ref{eq:ak} thus simplify to

\begin{subequations}\label{eq:CylindricalTrap}
\begin{align}
a_2 &= \frac{3}{4}C_4\\
a_3 &= 0\\
a_4 &= \frac{15}{16}C_6 - \frac{21}{64} (C_4)^2\\
a_5 &= 0.
\end{align}
\end{subequations}
The odd-order $a_k$ thus vanish naturally for an ideal cylindrical Penning trap.

Care must be taken in making quantitative comparisons between the planar traps and the cylindrical trap.  Amplitudes and distances in the cylindrical trap were naturally scaled by the larger value of $d = 3.54$ mm \cite{CylindricalPenningTrap,CylindricalPenningTrapDemonstrated} rather than by $\rho_1 = 1.09$ mm as in the sample trap considered here, for trap configurations that produce the same axial frequency.  The conversion between the $C_k^{(cyl)}$ for the cylindrical trap \cite{Review,CylindricalPenningTrapDemonstrated} and the $C_k$ for the planar trap is given by
\begin{equation}
C_k = \left(\frac{\rho_1}{d}\right)^{k-2} \frac{C_k^{(cyl)}}{C_2^{(cyl)}}.
\end{equation}
We apply this conversion to reported values for the cylindrical trap for the rest of this section.

The amplitude dependence of the axial frequency is reduced for the cylindrical Penning trap by adjusting a single compensation potential applied to a pair of compensation electrodes.  The adjustment changes primarily $C_4$, but also $C_6$ to a lesser amount.  The adjustment continues until $C_4 \approx 0$, whereupon $C_6 \approx -0.0008$.  The frequency coefficients are then found using the appropriately converted $C_k$ in
Eq.~\ref{eq:ak}.  This gives $a_2=a_3=a_5=0$ and $a_4 = -0.0007$.

The resulting frequency-versus-amplitude curve is shown in Fig.~\ref{fig:OptimizedTrapFrequencyShift} for $\omega_z/(2\pi) = 64$ MHz, and the corresponding thermal spread of axial frequencies is $\Delta f_z = 0$ Hz since $a_2=0$.  The compensation potential is typically then adjusted slightly away from $C_4 = 0$ to make the axial frequency insensitive to small fluctuations about a particular oscillation amplitude \cite{SelfExcitedOscillator}.

In practice, $C_4 = 0$ is not realized exactly, but $|C_4| < 10^{-5}$ is typically achieved.  If $C_4 = -10^{-5}$, and $C_6 = -0.0008$ as before, then the amplitude coefficients are $a_2 = -8\!\times\!10^{-6},\ a_4 = -0.0007,\ a_3=a_5 = 0$, the thermal spread of axial frequencies is 0.5 Hz, and the frequency-versus-amplitude curve is as shown in Fig.~\ref{fig:OptimizedTrapFrequencyShift}.

\OptimizedTrapFrequencyShiftFigure

The cylindrical trap is designed so that the axial frequency is much more insensitive to the tuning compensation potential than to the potential applied to make the main trapping potential.  If we define the endcap electrode potential to be our zero of potential, the $\gamma_i$ factors are $\gamma_{ring} = -141$ and $\gamma_{comps} = 0.032$.  The latter would have the value $\gamma_{comps}=0$ for the ``orthogonalized'' design for the cylindrical trap except for the unavoidable imperfections of a real laboratory trap.

\subsection{Previous Three-Gap Traps}

In marked contrast to the cylindrical trap within which the one-quantum transitions of a single electron were observed,  the
planar traps attempted so far were not designed to make $a_2=0$ and were not biased to make even $C_3=0$.  It is thus not
so surprising that attempts to observe one electron in a planar Penning trap have not succeeded.

\PossibleThreeRingGeometriesFigure

In fact, Fig.~\ref{fig:PossibleThreeRingGeometries} shows that the three-gap trap geometries tried so far (crosses) are
outside of all of the shaded regions that we use to identify optimized trap geometries. The best that could have been done
for the earlier  planar traps would have been to make $C_3=0$. In fact, any trap geometry represented in the upper
triangular region of Fig.~\ref{fig:PossibleThreeRingGeometries} can be tuned to make $C_3$ vanish. Appendices \ref{sec:MainzTrap} and \ref{sec:UlmTrap} look more
closely at the design of earlier planar Penning traps, and suggests that these were not biased to make $C_3=0$.

\subsection{Optimize to $a_2=a_3=a_4=a_5=0$}

For a scaled three-gap trap we must choose six parameters ($\rt_2$, $\rt_3$, $\Vt_1$, $\Vt_2$, $\Vt_3$, and $\zt_0$) that
solve the constraints of Eqs.~\ref{eq:C2Constraint}-\ref{eq:C1Constraint}.  Our preferred path to optimization starts from
adding the constraints
\begin{subequations}\label{eq:Constraintsa2}
\begin{align}
C_1=0&\\
C_2=1&\\
a_2=a_3=0&.
\end{align}
\end{subequations}
What appear here to be four constraints are actually three constraints because $a_3=0$ follows from $a_2=0$ via
Eq.~\ref{eq:a3}.

The difference of the number of parameters and constraints is three.  Where solutions exist, we might thus expect them
to be functions of the two parameters that specify the relative geometry, and that a range of $\zt_0$ might be possible. The
shaded area in Fig.~\ref{fig:PossibleThreeRingGeometries} represents the relative geometries for which there are
solutions.  Solutions also do exist for a range of $\zt_0$ values, as illustrated in Fig.~\ref{fig:ThreeGapaTwoAndaFourVanishing}
for our sample trap geometry. The solutions are double-valued because the third constraint equation is quadratic in the scaled
potentials $\Vt_1$, $\Vt_2$, and $\Vt_3$. The solid and dashed curves distinguish the two branches.

\ThreeGapaTwoAndaFourVanishingFigure

The left and right points in Fig.~\ref{fig:ThreeGapaTwoAndaFourVanishing}, with detailed properties in columns I and II of
Tables~\ref{table:SampleTrap}-\ref{table:SampleTrapAbsolute}, are trap configurations that satisfy the more stringent set
of constraints
\begin{subequations}\label{eq:Constraintsa2a4}
\begin{align}
C_1=0&\\
C_2=1&\\
a_2=a_3=a_4=a_5=0&.
\end{align}
\end{subequations}
What appear to be six constraints on the six parameters are actually four constraints in light of Eq.~\ref{eq:a5Froma4}. The
difference of the number of parameters and constraints is thus two.  Where solutions exist we will thus regard them as
functions of the relative geometry, $\rt_2$ and $\rt_3$, which will then determine a particular value of $\zt_0$.  The
darkly shaded region in Fig.~\ref{fig:PossibleThreeRingGeometries} shows the relative geometries for which solutions can be
found.

\SamplePlanarTrapTable

\SamplePlanarTrapTableAbsolute

For the solution that is the right point in Fig.~\ref{fig:ThreeGapaTwoAndaFourVanishing} none of the potential coefficients
$C_3$, $C_4$, $C_5$ and $C_6$ vanish.  The axial potential in Fig.~\ref{fig:AxialPotential} is thus clearly different from a
harmonic oscillator potential.  Since $C_3 \ne 0$ the amplitude of the second harmonic of the axial oscillation is the
lowest-order term from Eq.~\ref{eq:A2},
    \begin{equation}
    \frac{\At_2}{\At_1} = \frac{C_3}{4} \At + \ldots\label{eq:A2I}
    \end{equation}
The amplitude of this second harmonic should still be relatively small insofar as $\At=A/\rho_1$ is small.

We discuss the solution that is the left point in Fig.~\ref{fig:ThreeGapaTwoAndaFourVanishing} and column II in
Tables~\ref{table:SampleTrap}-\ref{table:SampleTrapAbsolute} in Sec.~\ref{sec:OptimizedHarmonic}.

\subsection{Optimize to $C_3=C_4=a_2=a_3=0$}

A second path to optimizing the six parameters for a scaled trap configuration starts with adding the constraint $C_3=0$ to the two
requirements for a trap, $C_1=0$ and $C_2=1$ (Eqs.~\ref{eq:C2Constraint}-\ref{eq:C1Constraint}).  All three constraint
equations are then linear in the scaled potentials $\Vt_1$, $\Vt_2$, and $\Vt_3$, yielding single-valued solutions for a
given $\zt_0$, $\rt_2$, and $\rt_3$.  There are three more parameters than constraints.  Solutions that give $C_3=0$ are
possible for any relative geometry.  The traps can be biased to make a range of $\zt_0$ values.  For our sample trap
geometry, the scaled potentials are plotted as a function of $\zt_0$ in Fig.~\ref{fig:ThreeGapCThreeCFourVanishing}. The
resulting $C_k$ and $a_k$ are shown as well.

\ThreeGapCThreeCFourVanishingFigure

The axial potential is more harmonic at the two points in Fig.~\ref{fig:ThreeGapCThreeCFourVanishing}, both of which satisfy
the more stringent set of constraints,
\begin{subequations}\label{eq:OptimizationC3C4}
\begin{align}
C_1=C_3=C_4=0&\\
C_2=1&\\
a_2=a_3=0&.
\end{align}
\end{subequations}
What appear to be six constraints on the six parameters for the scaled trap are actually four (since
Eq.~\ref{eq:OptimizationC3C4}c follows from Eq.~\ref{eq:OptimizationC3C4}a-b via Eq.~\ref{eq:ak}a-c).  There are
thus only two more parameters than constraints.  Various relative trap geometries can thus be biased to satisfy this set of
constraints, as represented by the solid boundary and arrows labeled $C_3=C_4=a_2=a_3=0$ in
Fig.~\ref{fig:PossibleThreeRingGeometries}, and $\zt_0$ is thus determined for each relative geometry.  Note that although
this region lies within the shaded area for which $a_2=a_3=a_4=a_5=0$ can be realized, in general it is not possible to
satisfy both sets of constraints simultaneously.

For the sample trap geometry, two of the applied potentials are nearly the same, making this nearly a two-gap trap, but the
slight potential difference is needed.   More details about these solutions are in columns III
and IV of Tables~\ref{table:SampleTrap} and \ref{table:SampleTrapAbsolute}. Compared to the optimized configuration in
Eq.~\ref{eq:Constraintsa2a4}, the optimization of Eq.~\ref{eq:OptimizationC3C4} has a more harmonic potential
(Fig.~\ref{fig:AxialPotential}) but a less good suppression of the amplitude dependence of the axial frequency as long as
$a_4\ne0$ and $a_5\ne 0$.

Any solution with $C_3=C_4=0$ (including the two points in Fig.~\ref{fig:ThreeGapCThreeCFourVanishing} and columns III-IV in
Tables~\ref{table:SampleTrap} and \ref{table:SampleTrapAbsolute}, as well as the left solution point in Fig.~\ref{fig:ThreeGapaTwoAndaFourVanishing} column II in
Tables~\ref{table:SampleTrap} and \ref{table:SampleTrapAbsolute} described below) has a suppressed harmonic content compared to Eq.~\ref{eq:A2I}, with
    \begin{equation}
    \frac{\At_2}{\At_1} = \frac{5 C_5}{12} \At^3 + \ldots\label{eq:Harmonics2}
    \end{equation}
from Eqs.~\ref{eq:A1}-\ref{eq:A2}. The amplitude of higher harmonics is suppressed by additional powers of $\At$.

We discuss the solution that is the right point in Fig.~\ref{fig:ThreeGapCThreeCFourVanishing} and column III in
Tables~\ref{table:SampleTrap} and \ref{table:SampleTrapAbsolute} in the following section.

\subsection{Harmonic Optimization}
\label{sec:OptimizedHarmonic}

The highest level of optimization is for traps that are harmonic in that $C_3=C_4=0$, as well as having the remarkable
suppression of the amplitude dependence of the axial frequency that comes by adding $C_6=0$ (Eq.~\ref{eq:OptimizedTrap1}).
For this optimized harmonic configuration
\begin{subequations}\label{eq:OptimizationHarmonic}
\begin{align}
C_1=C_3=C_4=C_6=0&\\
C_2=1&\\
a_2=a_3=a_4=a_5=0&.
\end{align}
\end{subequations}
What appear to be nine constraints are actually five (because Eq.~\ref{eq:OptimizationHarmonic}c follows from
Eq.~\ref{eq:OptimizationHarmonic}a-b via Eq.~\ref{eq:ak}).  There is thus one more parameter to
chose ($\rt_2$, $\rt_3$, $\Vt_1$, $\Vt_2$, $\Vt_3$, and $\zt_0$) than there are constraints.  The free parameter leads to a range of possible
relative geometries (the dashed line in Fig.~\ref{fig:PossibleThreeRingGeometries}).  Missing from
Eq.~\ref{eq:OptimizationHarmonic} is $C_5=0$ since there are no solutions when this constraint is added.

Our sample trap (filled circle in the dashed line in Fig.~\ref{fig:PossibleThreeRingGeometries}) is one example.  For it, the
left solution point in Fig.~\ref{fig:ThreeGapaTwoAndaFourVanishing} and the right solution point in
Fig.~\ref{fig:ThreeGapCThreeCFourVanishing} are actually the same configuration, as is obvious from columns II and III
of Tables~\ref{table:SampleTrap} and \ref{table:SampleTrapAbsolute}.  This convergence of two solutions happens only for
traps with relative geometries on the dashed line in Fig.~\ref{fig:PossibleThreeRingGeometries}.  For other traps in the shaded region where $a_2=a_4=0$ can be satisfied, these two solutions remain distinct, and the highly optimized constraints of
Eq.~\ref{eq:OptimizationHarmonic} cannot be satisfied for any choice of the trap potentials.

The optimized harmonic trap configurations (Fig.~\ref{fig:OptimizedHarmonic}) involve only a very narrow range of scaled
distances $\zt_0$ from the electrode plane to the axial potential minimum.  The scaled potentials needed  are shown in
Fig.~\ref{fig:OptimizedHarmonic}b.  The leading departure from a harmonic potential is described by $C_5=-0.011$
(Fig.~\ref{fig:OptimizedHarmonic}b).  As mentioned, we find no solutions to the constraint equations if a vanishing $C_5$ is
required.

\OptimizedHarmonicFigure

The highly optimized properties of Eq.~\ref{eq:OptimizationHarmonic} are an optimized harmonic configuration in that the leading departures from a harmonic axial potential vanish because $C_3=C_4=0$ at the same time that the amplitude dependence of the axial frequency is strongly suppressed.  A particle's axial oscillation will thus have a very small amplitude at the overtones of the fundamental harmonic, as given by Eq.~\ref{eq:Harmonics2}, with the amplitude of higher harmonics suppressed by additional powers of $\At$.

\subsection{Comparing Amplitude-Dependent Frequency Shifts}

The optimized trap configurations greatly reduce the amplitude dependence of the axial oscillation frequency.   Avoiding
frequency fluctuations caused by noise-driven amplitude fluctuations is critical to resolving the small frequency shifts
that signal one-quantum cyclotron and spin transitions.

One way to compare the optimized configurations is in Fig.~\ref{fig:OptimizedTrapFrequencyShift}.  The axial frequency shift
is shown as a function of oscillation amplitude for the three optimized configurations of the sample trap. For one electron
in the cylindrical trap an oscillation amplitude of 0.1 mm was large and easily detectable.

Another figure of merit is the frequency broadening for the thermally driven axial motion of a trapped particle, which was
discussed in Sec.~\ref{sec:ThermalBroadening}.  We use the 5 K axial temperature realized and measured for a cylindrical
Penning trap cooled by a dilution refrigerator \cite{FeedbackCoolingPRL}, though it should be noted that realizing such a
low detector temperature is a challenging undertaking. Each trap configuration can thus be characterized by the thermal
broadening of the axial resonance frequency, as indicated in Fig.~\ref{fig:FrequencyWidthComparison}.  Imperfections in real
planar traps and instabilities in applied potentials will likely make it difficult to get thermal widths much less than 1
Hz for  an axial frequency of 64 MHz.

\FrequencyWidthComparisonFigure

\section{Laboratory Planar Traps}
\label{sec:LaboratoryPlanarTraps}

Planar Penning traps put into service in the laboratory will not have the ideal properties described in the previous sections of this work.  A real trap does not have gaps of negligible width, does not have an electrode plane that extends to infinity, does not have conducting boundaries at an infinite distance above the electrode plane and at an infinite radius, and will not have exactly the ideal dimensions and the perfect cylindrical symmetry that are being approximated.  None of these have large effects.  However, the result is that the potentials applied to the electrodes of a real laboratory trap will need to be adjusted a bit from the ideal planar trap values to compensate for the unavoidable deviations and imperfections.

The effect of non-negligible gaps is calculated in Sec.~\ref{sec:Gaps}.  The effect of a finite electrode plane and a finite conducting radial enclosure is discussed in Sec.~\ref{sec:FiniteBoundaries}. Imperfections in the trap dimensions and symmetry are dealt with in Sec.~\ref{sec:RadiusImprecision} using simple estimates that proved adequate for the design of earlier traps. These estimates are used to discuss the tunning of the trap potentials required to compensate for imperfections of this order (Sec.~\ref{sec:Tuning}).

\subsection{Gaps Between Electrodes}
\label{sec:Gaps}

Small gaps of some width $w$ between electrodes are unavoidable, of course.  As long as $w \ll z_0$, the potential variation caused by the gaps at the position of the trapped particle should be small since it should diminish exponentially with an argument that goes as $w/z_0$.  We assume that the  gaps between electrodes are  deeper than they are wide since this is needed to  screen the effect of any stray charges on the insulators that keep the electrodes apart.

Solving exactly for the trapping potential using boundary conditions that include deep gaps between the electrodes is a challenging undertaking.  Instead we use a simple and approximate boundary condition that was used to demonstrate the small effect of the gaps in a cylindrical trap \cite{CylindricalPenningTrap}.  We take the potential in the electrode plane across each gap to vary linearly between the potentials of the two electrodes.  The potential is thus determined everywhere in the electrode plane by the potentials on the electrodes.

The basis of this approximation is illustrated by the equipotentials shown for a planar trap in Fig.~\ref{fig:PlanarTrapThreeDimensions}b (and later in  Figs.~\ref{fig:FinitePlanarTrapThreeDimensions}b, \ref{fig:CoveredPlanarTrap}b, and  \ref{fig:MirrorImagePlanarTrapThreeDimensions}b)).
All the equipotentials from the trapping volume must connect to equipotentials within the gaps.  Deep within a small but deep gap the equipotentials will locally be similar to the equipotentials between parallel plates, the plates being the vertical electrode walls within the gap.  The equipotentials will remain roughly parallel until they rise above the electrode plane, whereupon they will spread.  We make the approximation that in the electrode plane the potential in the gap varies linearly with radius between the voltages applied to the two electrodes that are separated by the gap.  Since the effect of the gaps is already small a  better approximation should not be needed.

The completely specified electrode plane boundary is thus given by the electrode boundaries and the linear change of potential between them at the gaps of  width $w_i$ (with $\gt_i=w_i/\rho_1$) centered at radius $\rho_i$.  The solution to Laplace's equation on axis then becomes \begin{align}
V^{gap}(0, z) &= \sum_{i=1}^N \, \Delta V_i \, \Phi_i(\zt)\label{eq:PotentialWithGaps}\\
\Phi_i(\zt) &= \frac{\zt}{\gt_i}  \sinh^{-1}\left(\frac{\rt_i + \gt_i/2}{\zt}\right)\nonumber\\
 &- \frac{\zt}{\gt_i}  \sinh^{-1}\left(\frac{\rt_i - \gt_i/2}{\zt}\right)-1\label{eq:PhiWithGaps}.
\end{align}
In the limit of vanishing gap widths this potential becomes the potential of an ideal planar trap in Eqs.~\ref{eq:GapExpansion}-\ref{eq:Phi}.

Following the procedure outlined earlier (Sec.~\ref{sec:OptimizedThreeGapTraps}), this potential is expanded about $\zt=\zt_0$.  Sets of parameters that satisfy reasonable constraint equations identify the optimized trap configurations that greatly reduce the amplitude dependence of the axial frequency and make the trap potential more harmonic.  We get four optimized configurations, as before, but with applied potentials that are slightly shifted.

Biasing a trap with a finite gap width as if it was an ideal planar trap with no gap width is one approach.  Table~\ref{table:SampleTrapWithGapsBiasedForNoGaps} shows the $C_k$ and $a_k$ when ideal trap biases (from Tables~\ref{table:SampleTrap}-\ref{table:SampleTrapAbsolute}) are applied to the sample trap with $w=50~\mu \rm{m}$ gap widths.  The broadening of the axial frequency $\Delta f_z$ for a thermal distribution of axial frequencies is still small enough that it should not prevent observing one electron in a trap with such gaps.

\SamplePlanarTrapWithGapsBiasedForNoGapsTable

Shifting the potentials applied to the electrodes improves the $C_k$ and $a_k$, as indicated in Table~\ref{table:SampleTrapWithGaps}.  However, the predicted thermal widths then become smaller than what imperfections (discussed in Sec.~\ref{sec:RadiusImprecision}) will likely allow us to attain, so this adjustment is not really needed.

\SamplePlanarTrapWithGapsTable

The small size of these coefficients illustrates that realistic gaps
between the electrodes of a trap as large as our sample trap pose no
threat to realizing a planar Penning trap. Simply biasing the trap
as if it were a trap with vanishing gap widths suffices. However, as planar
traps get smaller the gaps will likely be relatively larger with
respect to the trap dimensions.  The use of
Eqs.~\ref{eq:PotentialWithGaps}-\ref{eq:PhiWithGaps} will then be required.

\subsection{Practical Limitations on Gaps}
\label{sec:PracticalLimitationsOnGaps}

Two practical considerations are associated with gaps between
electrodes.  Both can make the difference between a trap that works
and one that does not.  Both are difficult to calculate.

The first is that charges that accumulate on the insulators in the
gaps between the electrodes can substantially modify the trapping
potential.  When the trap is cooled to 4 K or below, these charges
can remain for days.  Such charges have made some traps in our lab
completely unusable, but no systematic study has been
undertaken.

There are only two solutions that we know of.  Careful loading and
operation procedures can minimize the number of charges that build
up on the insulators.  Also, thick metal electrodes with narrow gaps
make it more difficult for charges to reach the insulating substrate
at the bottom of the slits between electrodes.  Any charges that do
collect on the insulator will be screened by metal surfaces to
either side of the gaps.

As mentioned in the Appendices, both the Mainz and Ulm traps had
exposed insulators in gaps that were not screened because the gaps were wider than they were deep.
Charges on the insulating substrates that are exposed in the gaps of
these traps may well have contributed to the broad frequency spreads
that were observed.  Improved traps with much better screening of the insulator at the bottom of the gaps seem possible but have yet to be used with trapped particles.

The second practical consideration involving gaps between electrodes
becomes more serious with decreasing gap width. We have observed currents between polished gold-plated trap electrodes separated by small gaps.  These field emission currents \cite{FieldEmissionReview,FieldEmissionCleanCopperSurfaces,FieldEmissionVariousMaterials,SandiaFieldEmission} grow exponentially with the difference in
potential across the gap.

Trap designs that limit the size of the gap potentials are one solution.
For three-gap traps, $\Delta V_3=0-V_3$ is generally the largest
of the gap potentials.  It can generally be reduced by decreasing the
radial width of the second electrode, $\rho_2-\rho_1$, and increasing
the radial width of the third electrode, $\rho_3-\rho_2$. We will see
in Sec.~\ref{sec:CoveredPlanarTraps} that a planar trap with a
conducting plane above it will permit an optimized trap with lower
gap potentials.  Other solutions are to increase the gap width and
to make the metal surfaces within the gap as smooth as
possible.  As planar traps and planar trap arrays get smaller it
will be necessary to investigate these solutions further.

\subsection{Finite Boundaries}
\label{sec:FiniteBoundaries}

For laboratory traps it is difficult to approximate an infinite electrode plane and to keep all parts of the apparatus many trap diameters away from the trapping volume.  The effects of realistic finite boundary conditions are thus extremely important.  For smaller planar Penning traps the finite boundaries may be less important.

\FinitePlanarTrapThreeDimensionsFigure

One choice of finite boundary conditions come from locating a planar trap within a grounded conducting cylinder closed
with a flat plate (Fig.~\ref{fig:FinitePlanarTrapThreeDimensions}).  The boundary conditions in the electrode plane are still
given in Fig.~\ref{fig:PlanarTrap} for $\rho < \rho_c$.  The boundary conditions at infinity in Eq.~\ref{eq:BoundaryConditionsInfinity} are replaced by
\begin{eqnarray}
V(\rho_c,z) &= 0\\
V(\rho,z_c) &= 0.
\end{eqnarray}
Particles can be loaded into the trap through a hole through the conducting plate above that is small enough to negligibly affect the potential near the particle.

The solution to Laplace's equation for $z>0$ that satisfies these boundary conditions can be written as
\begin{align}
V(0,\zt) &= \sum_{i=1}^N \Delta V_i \, \Phi_i(\zt; \rt_c,\zt_c).\label{eq:Vgap}
\end{align}
Standard electrostatics methods \cite{Jackson3rdEd,Kusse} give dimensionless potentials,
\begin{align}
\Phi_i(\zt; \rt_c,\zt_c) &= \frac{\rt_i}{\rt_c}  \sum_{n=1}^\infty \frac{2J_1(\alpha_{0n}\frac{\rt_i}{\rt_c})}{\alpha_{0n}J_1\,\!^2(\alpha_{0n})} \frac{\sinh\left(\alpha_{0n} \frac{\zt-\zt_c}{\rt_c}\right)}{\sinh\left(\alpha_{0n} \frac{\zt_c}{\rt_c}\right)},
\label{eq:PhiiGrounded}
\end{align}
that are functions of zeros of the lowest-order Bessel function, with $J_{\,0}(\alpha_{0n})=0$.    The potential off the axis is given by substituting $V(\rt,\zt)$ for $V(0,\zt)$ in Eq.~\ref{eq:Vgap}, and inserting  $J_{\,0}(\alpha_{0n}\rt/\rt_c)$ to the far right in Eq.~\ref{eq:PhiiGrounded}.  The planar trap described in Eq.~\ref{eq:GapExpansion} is recovered in the limit of large $\rt_c$ and $\zt_c$ insofar as the $\Phi_i(\zt; \rt_c,\zt_c)$ reduce to the $\Phi_i(\zt)$ of Eq.~\ref{eq:Phi}.

The simplest approach is to bias the electrodes of the enclosed trap as if it was an ideal planar trap with no enclosure, using the potentials tabulated in Tables~\ref{table:SampleTrap}-\ref{table:SampleTrapAbsolute}.  The size of the resulting $C_k$ and $a_k$ coefficients are then displayed in Table~\ref{table:SamplePlanarTrapWithEnclosureBiasedForNoEnclosure} for the conducting enclosure shown to scale in Fig.~\ref{fig:FinitePlanarTrapThreeDimensions} with dimensions $\rho_c = 19.05$ mm and $z_c = 45.72$ mm, both substantially larger than $\rho_3 = 8$~mm. The resulting thermal frequency shifts for a 5 K axial motion are large enough that this broadening will make it hard to observe one electron and realize a one-electron qubit.

\SamplePlanarTrapWithEnclosureBiasedForNoEnclosureTable

It is possible to do much better by shifting the potentials applied to the trap electrodes, without changing the relative geometry of the electrodes.  Table~\ref{table:SamplePlanarTrapWithEnclosure} shows the required potential shifts and the calculated $C_k$ and $a_k$ that result for each of the four optimized planar trap configurations for an ideal planar trap (summarized in Tables~\ref{table:SampleTrap}-\ref{table:SampleTrapAbsolute}).  The frequency broadening is small enough that it should be possible to observe one electron within such a trap.

\SamplePlanarTrapWithEnclosureTable

The configurations in columns II and III of Table~\ref{table:SamplePlanarTrapWithEnclosure} no longer coincide exactly, however, even though both trap configurations still have very attractive properties.  The finite boundary conditions effectively shift the dashed line in Fig.~\ref{fig:PossibleThreeRingGeometries} that represents the possible geometries for which an optimized three-gap trap can be realized, so that the relative geometry of the sample trap no longer allows this highest level of optimization.  What could be done is to slightly change one of the trap radii to compensate for the calculated effect of the finite boundary conditions.   However, the shift of geometry is often less than the size of the typical imprecision with which the electrode radii of a real trap can be fabricated (discussed in the next section) so in practice this makes little sense.

\subsection{Imprecision in Trap Dimensions and Symmetry}
\label{sec:RadiusImprecision}

A fabricated laboratory trap will not have exactly the intended dimensions and symmetry because of unavoidable fabrication imprecision.  Such effects can only be estimated. The simple estimation method used here has proved itself to be adequate for the design of cylindrical traps \cite{CylindricalPenningTrap} and open access traps \cite{OpenTrap}.

We start with an achievable fabrication tolerance of $0.001~\rm{in}=25~\mu \rm{m}$ that is realistic for existing traps of the size of our sample trap.  (Whether smaller traps can be constructed with better fractional tolerances is being investigated \cite{UlmPixelTrap}.) Adding and subtracting the achievable tolerance to the radii $\rho_2$ and $\rho_3$ of a three-gap planar trap makes variations (Table~\ref{table:OptimizedHarmonicTrapWithChangeRadii}) from the design ideal (Tables~\ref{table:SampleTrap}-\ref{table:SampleTrapAbsolute}).

These variations do not have the exact ratios of the trap radii needed to make an optimized harmonic trap configuration (Eq.~\ref{eq:OptimizationHarmonic}) that is specified by the dashed line in Fig.~\ref{fig:PossibleThreeRingGeometries} and in Fig.~\ref{fig:OptimizedHarmonic}a). The variations have better properties than what has been observed to date with a laboratory planar trap.  However, the imprecision in the radii still makes the predicted broadening of an electron's axial resonance for a 5 K thermal distribution of axial energies to be too large to observe one trapped electron very well.  It would be virtually impossible to realize a one-electron qubit.

The solution must be to slightly adjust the potentials on the electrodes to recover properties closer to the ideal, if this is possible. In the cases of the gaps and the conducting enclosure we saw that this could be done, at least in principle, by calculating what the improved set of potentials should be.  For imprecision in the trap radii, however, the effective radii for the electrodes will be unknown and hence no such calculation is possible.  What is required is a procedure for tuning the potentials of the trap to narrow the thermal broadening.  For trap design we must make sure that the trap potentials can be tuned to compensate for imperfections of this order.  The tuning procedure and range is the subject of the next section.

\OptimizedHarmonicTrapWithChangeRadiiTable

Imperfections that are not cylindrically symmetric are no doubt present. While it is possible with some effort to make calculations of potential configurations that are not cylindrically symmetric \cite{UlmPixelTrap}, the input from imperfections that should be used in such a calculation is difficult to estimate. Fortunately, the experience with earlier traps suggests that this is not necessary for trap design.

\subsection{Tuning a Laboratory Trap}
\label{sec:Tuning}

The point of carefully designing the optimized traps for which the lowest order $a_k$ vanish, preferably along with $C_3$ and $C_4$, is not that we actually expect to realize this performance in a real laboratory trap.  The previous section illustrates that radius imprecision alone will keep this from happening.  The reason for the careful optimized designs is to make sure that imprecision alone will make these crucial coefficients differ from zero.  Notice in Table~\ref{table:OptimizedHarmonicTrapWithChangeRadii} that the imperfections considered do not make either $a_3$ or $a_5$ deviate much from zero, and $a_4$ stays at an acceptably low value.

To make a useful trap we need a way to tune the trap {\it in situ} to make $a_2=0$.  The other important coefficients will remain small enough because of the optimized design.  To tune out the effect of radius imperfections in our sample trap, for example, the trap must be tuned to change the size of $a_2$ by about $\pm 0.003$.  After each adjustment the width of the axial resonance line can be measured to see if the thermal broadening has been reduced or increased.

For the cylindrical Penning trap used to observe one-quantum transitions of one electron, tuning of the trap was essential to the observations that were made.  In that trap, like every trap within which precise frequency measurements are made, the effect of imperfections could never be calculated well enough to be useful.  {\it In situ} tuning of a compensation potential was always needed.

For a cylindrical trap, tuning is a straightforward (if a bit tedious) matter.  To a good approximation, the potential applied to the compensation electrodes (Fig.~\ref{fig:QuantumJumps}c) changes $a_2$, while the potential applied between the endcap and ring electrodes changes $V_0$ and $\omega_z$.   The axial resonance line is measured after every adjustment of the compensation potential to see if the thermal broadening increased or decreased.  The ``orthogonalized design'' of this trap kept the change in the compensation potential from changing the axial frequency very much at all.  The axial resonance line was thus easy to keep track of during trap tuning, and the axial oscillation never comes close to going out of resonance with the detection circuit.

The tunability defined in Eq.~\ref{eq:gammai} quantifies how much the axial frequency changes for a given change in $a_2$ when the potential on a particular electrode is changed. For the compensation electrodes of the cylindrical trap the tunability was 0 for a perfect trap, and $\gamma_{comps}=0.03$ was realized for a laboratory trap (Sec.~\ref{sec:CylindricalTrap}).  The much larger $\gamma_{ring}=-141$ indicates that this electrode is for changing the axial frequency of the trap rather than for tuning $a_2$.

For a planar Penning trap such an orthogonalization is unfortunately not possible.  Changing the potential on each electrode will change both $a_2$ and $\omega_z$, as indicated by tunabilities in Table~\ref{table:SampleTrap} that are not small and which do not vary much from electrode to electrode in most cases (e.g.,~$|\gamma_i| \!\approx\! 3$ in one example).  The result is that it is necessary to adjust two or three of the potentials applied to the electrodes of a 3-gap trap for each step involved in tuning the trap.  Adjustments of the applied potentials must be chosen to vary $a_2$ by a reasonable amount while keeping $V_0$ and $\omega_z$ fixed.

Fig.~\ref{fig:TuningWithImperfections} identifies the potentials for which $a_2=0$ and $a_4=0$ for our sample trap with and without the radius imperfections of Table~\ref{table:OptimizedHarmonicTrapWithChangeRadii}.  For each point on this plot $V_1$ has been adjusted so that $V_0$ and hence the axial frequency $\omega_z$ remain fixed.  In this example it would be necessary to change $V_2$ or $V_3$ (along with $V_1$ to keep $V_0$ fixed) to achieve $a_2=0$.  However, by changing both $V_2$ and $V_3$ (along with $V_1$) it would be possible to make $a_2=0$ while at the same time making $a_4$ much smaller in magnitude.

\TuningWithImperfectionsFigure

\section{Covers and Mirrors}

\subsection{Covered Planar Trap}
\label{sec:CoveredPlanarTraps}

\CoveredPlanarTrapFigure

A covered planar Penning trap (Fig.~\ref{fig:CoveredPlanarTrap}) is a planar trap that is electrically shielded by a nearby conducting plane.  The   covered planar trap has some very attractive features.
\begin{enumerate}
\item The electrodes are in a single plane that can be fabricated as part of a single chip.
\item The conducting plane provides an easily controlled boundary condition above the electrode plane that needs no special fabrication, nor any alignment beyond making the planes parallel.
\item A trap that is radially infinite is well-approximated if the radial extent of the two planes beyond the electrodes is large compared to their spacing.
\item A covered planar trap is naturally scalable to an array of traps.
\item The axial motion of electrons in more than one trap could be simultaneously detected with a common detection circuit attached to the cover.
\item The axial motions of electrons in more than one trap could be coupled and uncoupled as they induce currents across a common detection resistor by tuning the axial motions of particular electrons into and out of resonance with each other.
\end{enumerate}
Three possible additional advantages emerge when the properties of the trapping potential in a covered planar trap are considered.
\begin{enumerate}
\item A two-gap covered planar trap can be optimized in much the same way as a three-gap infinite planar trap.
\item Smaller gap potentials can sometimes be used to achieve optimized configurations, permitting smaller gap widths and better screening of the exposed insulator between electrodes.
\item In some cases a smaller $a_6$ can be realized for trap configurations with $a_2=a_3=a_4=a_5=0$
\end{enumerate}
These possibilities are illustrated using an example.

The secondary advantages for planar Penning traps (mentioned in the introduction) may be diminished when a cover is used.  Microwaves of small wavelength can be introduced between the electrode plane and the cover.  However, the added complication of small striplines \cite{QuantumDotESR} is likely required for longer wavelengths.  It should not be significantly more difficult to load electrons with typical methods via small holes in the electrodes, but if other loading mechanisms are used then the electron trajectories may be obstructed by the cover.

The potential between the electrode plane and the cover plane is a superposition of terms proportional to the potentials applied to the electrodes, $V_i$, and the potential applied to the cover plane, $V_c$,
\begin{align}
V(0,\zt) = \sum_{i=1}^N \Delta V_i \, &\Phi_i(\zt;\zt_c)\nonumber\\
+ V_c \, &\Phi_c(\zt;\zt_c).
\end{align}
The grounded cover plane makes  the $\Phi_i(\zt)$ of Eq.~\ref{eq:Phi} dependent  upon $\zt_c$,
\begin{equation}
\Phi_i(\zt;\zt_c)  =  \rt_i \int_0^\infty\!\!\! dk \, \frac{\sinh\!\left[k (\zt-\zt_c)\right]}{\sinh(k \zt_c)}\, J_1(k\rt_i),\label{eq:Phiip}
\end{equation}
which approaches $\Phi_i(\zt)$ for large $\zt_c$.  Biasing the cover plane at a nonvanishing $V_c$ superimposes a uniform electric field, described by
\begin{equation}
\Phi_c(\zt;\zt_c) = \zt / \zt_c,\label{eq:UniformFieldLimit}
\end{equation}
between the large electrode and cover planes.

The scaled geometry and potentials of a two-gap covered Planar traps are characterized by six parameters ($\rt_2$, $\zt_c$, $\Vt_1$, $\Vt_2$, $\Vt_c$ and $\zt_o$).  This is the same number of parameters that characterize a three-gap planar trap with no cover electrode, the optimization of which was discussed in detail in Sec.~\ref{sec:OptimizedThreeGapTraps}.

The trap geometries that can be optimized are represented in Fig.~\ref{fig:PossibleCoveredTrapGeometries}.
The six parameters can be chosen to satisfy the same sets of four constraints considered in Sec.~\ref{sec:OptimizedThreeGapTraps}, giving the various shaded regions in Fig.~\ref{fig:PossibleCoveredTrapGeometries}.
The six parameters can be chosen to satisfy the five constraints of Eq.~\ref{eq:OptimizationHarmonic} on the dashed curve in Fig.~\ref{fig:PossibleCoveredTrapGeometries} for which $C_3=C_4=C_6=a_2=a_3=a_4=a_5=0$.

The optimized harmonic configuration represented by the dot in Fig.~\ref{fig:PossibleCoveredTrapGeometries} has its scaled parameters listed in Table~\ref{table:SampleCoveredTrap}.  One set of possible absolute parameters is listed in Table \ref{table:SampleCoveredTrapAbsolute}. In the following section we discuss other attractive features of this particular configuration.

\PossibleCoveredTrapGeometriesFigure

Fig.~\ref{fig:CoveredPlanarTrap}b shows
 equipotentials spaced by $V_0$ for a covered planar Penning trap (configuration I in Table~\ref{table:SampleCoveredTrap}).  The equipotentials are calculated for infinitesimal gaps, but the electrodes are represented with finite gaps to make them visible.  The equipotentials terminate in the gaps between electrodes or at infinity.  The dashed equipotentials of an ideal quadrupole are superimposed near the trap center.

Covered planar traps are scalable in that an array of traps can share the same covering plane at potential $V_c$, with the axial frequency and the harmonic properties of each trap being tuned by the potentials applied to the other electrodes. This is analogous to Fig.~\ref{fig:TuningWithImperfections} in which $a_2$ can be tuned at constant frequency by changing only $V_1$ and $V_2$ while leaving $V_3$ fixed.

\SampleCoveredTrapTable

\SampleCoveredTrapTableAbsolute

The effect of a grounded radial boundary at $\rt_c$ (rather than at infinity) can also be calculated.
The superposition
\begin{align}
V(0,\zt) = \sum_{i=1}^N \Delta V_i \, &\Phi_i(\zt;\rt_c,\zt_c)\nonumber\\
+ V_c \, &\Phi_c(\zt;\rt_c,\zt_c)
\end{align}
has dimensionless potentials that depend the distance to the radial boundary, $\rt_c$, as well as upon $\zt_c$.  The first of these,
\begin{equation}
\Phi_i(\zt; \rt_c,\zt_c) = \frac{\rt_i}{\rt_c}  \sum_{n=1}^\infty \frac{2J_1(\alpha_{0n}\frac{\rt_i}{\rt_c})}{\alpha_{0n}J_1\,\!^2(\alpha_{0n})} \frac{\sinh\left(\alpha_{0n} \frac{\zt-\zt_c}{\rt_c}\right)}{\sinh\left(\alpha_{0n} \frac{\zt_c}{\rt_c}\right)},\tag{\ref{eq:PhiiGrounded}}
\end{equation}
was used earlier in Eq.~\ref{eq:PhiiGrounded} to describe a grounded enclosure around a planar trap.  The second,
\begin{equation}
\Phi_c(\zt;\rt_c,\zt_c) = \sum_{n=1}^\infty \frac{2}{\alpha_{0n}J_1(\alpha_{0n})} \frac{\sinh\left(\alpha_{0n} \frac{\zt}{\rt_c}\right)}{\sinh\left(\alpha_{0n} \frac{\zt_c}{\rt_c}\right)},
\end{equation}
goes to the uniform field limit of Eq.~\ref{eq:UniformFieldLimit} in the limit of large $\rt_c$.
These potentials can be used to investigate radial boundary effects as needed, though we will not give examples here.

\subsection{Mirror-Image Trap}
\label{sec:MirrorImageTraps}

\MirrorImagePlanarTrapThreeDimensionsFigure

A mirror-image planar trap (Fig.~\ref{fig:MirrorImagePlanarTrapThreeDimensions}) is a set of two planar electrodes that are biased identically and face each other.   The axial potential,
\begin{align}
V(0, \zt) = \sum_{i=1}^N &\Delta V_i \left[ \Phi_i(\zt; \zt_c) + \Phi_i(\zt_c-\zt; \zt_c)\right],\label{eq:MirrorImageTrapPotential}
\end{align}
is a function of the dimensionless potentials defined in Eq.~\ref{eq:PhiiGrounded}.

For a two-gap mirror-image trap there are only four scaled parameters to be chosen ($\rt_2$, $\zt_c$, $\Vt_1$, $\Vt_2$).  The mirror-image symmetry of the electrodes ensures that the potential minimum is midway between the electrode planes and that all odd-order $C_k$ vanish.  The constraints are $C_2=1$, $C_4=0$ and $C_{22}=0$, the latter giving the orthogonality property discussed below.  With one more parameter than constraints, the possible geometries for a two-gap mirror-image trap are given by the dotted curve in Fig.~\ref{fig:PossibleCoveredTrapGeometries}.  The filled circle on this curve represents the trap geometry that is used to illustrate the properties of a mirror-image trap in Fig.~\ref{fig:MirrorImagePlanarTrapThreeDimensions} and in Tables \ref{table:SampleMirrorImageTrap}-\ref{table:SampleMirrorImageTrapAbsolute}.

The properties of a mirror-image trap are similar to those of the cylindrical Penning trap (Fig.~\ref{fig:QuantumJumps}c) used to suspend one electron and to observe its one-quantum cyclotron transitions and spin flips. A charged particle suspended midway between the two electrode planes sees a potential that is symmetric under reflections across this midplane, in which case all odd-order potential coefficients ($C_3$, $C_5$, etc.) vanish, as for the cylindrical trap (Sec.~\ref{sec:CylindricalTrap}).  Also, as for a cylindrical trap, we can choose the potentials applied to the trap electrodes to make a trap with a very small $C_4$, whereupon $a_2$ and $a_3$ are very small.

\SampleMirrorImageTrapTable

\SampleMirrorImageTrapTableAbsolute

A useful property of mirror-image traps and cylindrical traps is that both of these can be ``orthogonalized'' in a way that a planar trap cannot.  A single potential (applied to two electrodes with mirror-image symmetry) is tuned to minimize the amplitude-dependence of the axial frequency.  The trap is orthogonalized in that this tuning does not change the axial frequency, which in general would take it out of resonance with the detection circuit.

Fig.~\ref{fig:MirrorImagePlanarTrapThreeDimensions}b shows equipotentials spaced by $V_0$ for the mirror image Penning trap of Table~\ref{table:SampleMirrorImageTrap}.  The equipotentials are calculated for infinitesimal gaps, but the electrodes are represented with finite gaps to make them visible.  The equipotentials terminate in the gaps between electrodes.  The dashed equipotentials of an ideal quadrupole are superimposed near the trap center.

\bigskip

\subsection{Mirror-Image Trap Transformed\\ to a Covered Trap}
\label{sec:Mirror-ImageToCoveredTrap}

At least for initial studies it may be useful first to load an electron into the center of an orthogonalized mirror-image trap. The presence of a single electron can be established with established methods used with cylindrical Penning traps.  The challenge is then to adiabatically change the potentials applied to the electrodes to turn the mirror-image trap into a covered planar trap.  It is crucial that the electron not be lost.  If a high quality trapping well can be maintained throughout the transfer, then it may even be possible to monitor the electron at intermediate points between the two configurations.

We investigate the feasibility of transferring from the mirror-image trap discussed above (Tables \ref{table:SampleMirrorImageTrap}-\ref{table:SampleMirrorImageTrapAbsolute}) to the optimized, covered planar trap discussed in the last section (Tables \ref{table:SampleCoveredTrap}-\ref{table:SampleCoveredTrapAbsolute}). The electrode geometry chosen for our example is the lone point in Fig.~\ref{fig:PossibleCoveredTrapGeometries} for which it is possible to make an orthogonalized mirror-image trap and also to make a the most highly optimized covered planar Penning trap.  The potentials applied to  achieve the mirror-image trap are those to the far right in Fig.~\ref{fig:MirrorImagePotentials}.  The potentials applied to realize the covered planar trap are those to the far left in Fig.~\ref{fig:MirrorImagePotentials}.

\MirrorImagePotentialsFigure

For these traps there are six parameters to choose: five relative trap potentials ($\Vt_1$, $\Vt_2$, $\Vt^{top}_1$, $\Vt^{top}_2$, $\Vt^{top}_3$) and $\zt_0$.  During the transfer we can choose a particular $\zt_0$ as a constraint, along with four others that we have discussed earlier, $C_1=C_3=C_4=0$ and $C_2=1$.  Since there are more parameters than constraints there is some freedom in the choice of potentials during the transfer, provided that solutions exist.  Our choice of intermediate potentials in Fig.~\ref{fig:MirrorImagePotentials}
was made to avoid large potential differences between electrodes (discussed in Sec.~\ref{sec:PracticalLimitationsOnGaps}).  The axial oscillation frequency does not change during the transfer.  Also, the trap remains optimized during every point in the transfer, with $a_2=a_3=C_3=C_4=0$.  It may thus be possible to detect the electron's axial oscillation at every step of the transfer.

\section{Damping and Detecting an Axial Oscillation}
\label{sec:Damping}

\subsection{Damping and Detection in a Planar Trap}

The damping rate $\gamma_z$ for the axial motion of a trapped particle is the observed resonance linewidth for the axial motion in the limit of a vanishing oscillation amplitude.  A thermal distribution of axial oscillation amplitudes broadens the observed resonance linewidth when the axial frequency is amplitude dependent.  When the oscillation energy has fourier components that extend well beyond the damping linewidth, it is difficult to detect the oscillation with the narrow-band detection methods needed to observe the small signal from a single particle. For the cylindrical Penning trap used to observe one-quantum transitions of a single trapped electron \cite{QuantumCyclotron} this was not a problem.  The thermal anharmonicity contribution to the linewidth was less than the damping linewidth.  For the Ulm planar trap the situation was very different.  The thermal width was $10^5$ times larger than the damping linewidth, making it impossible to observe a single electron at all \cite{ElectronsInPlanarPenningTrapUlm}.

In preceding sections we focused upon minimizing the amplitude-dependence of the axial frequency so that the thermal broadening could be reduced.  Just as important is increasing the axial damping linewidth.  Here we discuss what is needed to maximize the electron's damping rate.  Maximizing the damping maximizes the detected signal as well.

The usual method to probe the axial oscillation of a single trapped particle is to detect the current that its axial motion induces in a resistor $R$ connected to its electrodes \cite{FirstSingleElectron1973,Review}.   This resistance also damps the motion.  The energy dissipated in the resistor comes from the axial motion of the trapped particle, which is thereby damped to the bottom of the axial potential well.  In practice the resistor is a tuned circuit that is resonant at the axial oscillation frequency, at which frequency it acts as a pure resistance.

\DetectingAndDampingCircuitFigure

For a planar Penning trap, Fig.~\ref{fig:DetectingAndDampingCircuit} illustrates how the AC connections between the circuit and the electrodes can be made to the same electrodes that are DC biased to form the trapping potential.  Alternatively, an extra gap (e.g.,\ the dashed circle labeled $\rho_d$ in Fig.~\ref{fig:DetectingAndDampingCircuit}) can be added to one of the trap's electrodes to maximize the damping and detection, as will be discussed.  This damping-detection gap can coincide with one of the gaps already chosen to minimize the amplitude-dependence of the axial frequency.  When the extra gap does not coincide with one of the others, the extra gap will not change the electrostatic trapping potential insofar as the same DC bias voltage is applied to either  side of the  additional gap.

The circuit in Fig.~\ref{fig:DetectingAndDampingCircuit}  represents one way to connect the detection and damping resistance, $R$, to the electrodes of a three-gap planar trap.   The current induced by the particle's axial oscillation makes an instantaneous voltage $V_I$ across the resistance.  This induced voltage exerts a reaction force on the trapped particle. The thermal Johnson noise from random electron motions within the resistor induces an additional instantaneous noise voltage $V_n$ across the resistor and electrodes.   The oscillatory voltage $V_I + V_n$ on the effective damping electrode for which $\rho < \rho_d$ both drives the particle's axial motion and is detected.

The particle is in the near field of the potential
\begin{equation}
V_{osc}=(V_I + V_n)\phi_d
\end{equation}
produced by this oscillatory voltage, where the electrostatic potential
\begin{equation}
\phi_d(\zt) = 1-\frac{\zt}{\sqrt{(\rt_d)^2+\zt^2}}
\end{equation}
follows from Eq.~\ref{eq:Exactphii}.

For a potential $V_i$ applied to electrode $i$, the instantaneous electric field on a particle oscillating near its equilibrium position at $\zt = \zt_0$ is
\begin{equation}
E_i(\zt_0) \approx  -\frac{D_1}{2\rho_1} V_i.
\end{equation}
The factor $D_1$ depends upon electrodes to which a voltage is applied to make the field.  For a voltage applied just to the damping and detection electrode,
\begin{equation}
D_1 = C_{1d} =  \frac{-2(\rt_d)^2}{\left[(\zt_0)^2+(\rt_d)^2\right]^{3/2}},
\end{equation}
where the potential expansion coefficient $C_{1i}$ is defined in Eqs.~\ref{eq:Phii}-\ref{eq:Cki}.

There is a maximum coupling of the circuit and a trapped particle insofar as $C_{1d}$ has a maximum magnitude at
\begin{equation}
\rho_d = \sqrt{2}\,\zt_0.\label{eq:MaximumDamping}
\end{equation}
The coupling coefficient is then given by
\begin{equation}
C_{1d}^{(opt)} = \frac{-4 \times 3^{-3/2}}{\zt_0} \approx \frac{-0.77}{\zt_0}.
\end{equation}
Fig.~\ref{fig:MaximizedDamping}a illustrates the maximum for all values of $\zt_0$.

\MaximizedDampingFigure

If instead the electrodes of the optimized planar trap are attached to the resistance, without adding an extra gap at $\rho_d$, then $D_1$ is the sum of the $C_{1i}$ for the electrodes attached to the resistor.  If the central two electrodes are attached to the detection circuit, for example, then $D_1=C_{11}+C_{12}$.  The coefficients $C_{1i}$ for the optimized configurations of our sample trap are listed in Table \ref{table:SampleTrap}, as is $C_{1d}^{(opt)}$.
Fig.~\ref{fig:MaximizedDamping}b shows how the various possibilities for these coefficients and sums depend upon $\zt_0$.

The induced signal,
\begin{equation}
V_I =\frac{q D_1}{2\rho_1} R \dot{z}, \label{eq:InducedSignal}
\end{equation}
is proportional to the axial velocity of the oscillating particle, as well as to $D_1$ and $R$ \cite{Review}.
The damping force that arises from this induced potential produces the damping rate for a particle of charge $q$ and mass $m$,
\begin{equation}
\gamma_z = \left(\frac{q D_1}{2\rho_1}\right)^2 \frac{R}{m},\label{eq:DampingRate}
\end{equation}
that goes as the square of $D_1$ \cite{Review}.  One power of $D_1$ arises because the induced current is proportional to $D_1$.  The second power arises
because a potential on the electrodes induces a damping force that is also proportional to $D_1$.  The damping rates for a
resistor connected between a single electrode and ground are listed in Table~\ref{table:SampleTrapAbsolute} for $R=100
~\rm{k}\Omega$. The maximum damping rate $\gamma_z^{(opt)}$ that pertains for Eq.~\ref{eq:MaximumDamping} is also tabulated for comparison.  As noted above, if the resistor is connected to more than one electrode then the appropriate coefficients $C_{1i}$ for the connected
electrodes must be summed to make $D_1$ before squaring.

The thermal Johnson noise in the resistor drives a particle that is near to its equilibrium location with a driving force
\begin{equation}
F_n \approx \frac{q D_1}{2\rho_1} V_n. \label{eq:NoiseForce}
\end{equation}
For the circuit shown in Fig.~\ref{fig:DetectingAndDampingCircuit} we have  $D_1=C_{1d}$.

An external driving force may be added to drive the axial motion of a trapped particle.
Such a driving force has the advantage that a larger oscillation amplitude and hence a larger induced signal is produced at just the frequency of the drive in the steady state.  A larger oscillation amplitude, of course, makes it more important to minimize the amplitude-dependence of the axial frequency.
One choice is to apply an oscillatory driving voltage $V_D$ to the third electrode, between $\rho_2$ and $\rho_3$, as indicated in Fig.~\ref{fig:DetectingAndDampingCircuit}.  The applied driving potential, $V_D$, produces a driving force on a particle near its equilibrium location that is given by
\begin{equation}
F_D \approx \frac{q D_1}{2\rho_1} V_D. \label{eq:DrivingForce}
\end{equation}
For $V_D$ applied to the third electrode we have  $D_1=C_{13}$.

\subsection{Damping and Detection in a Covered Trap}

For a covered planar trap, the damping and the detected signal is  maximized by introducing an extra gap at radius $\rho_d$ and connecting the damping resistor to each of the electrodes with radius $\rho \le \rho_d$.  The choice $\rho_d^{(opt)}$ that gives the maximum damping and detection signal, along with the corresponding $C_{1d}^{(opt)}$ and $\gamma_z^{(opt)}$, are displayed in Tables~\ref{table:SampleCoveredTrap} and \ref{table:SampleCoveredTrapAbsolute}. This detection configuration offers an appreciable detection efficiency.  For some achievable values of $z_0$ the detection efficiency is nearly maximized without the need to make an extra gap in the electrode plane.

As mentioned earlier, covered planar traps are scalable in that an array of traps can share the same covering plane, with the axial frequency and the harmonic properties of each trap being tuned separately by the potentials applied to the other electrodes. Multiple traps can also share the same detection circuit if the detection resistor is attached to the covering plane -- a great simplification in practice. Many trapped electrons could be simultaneously detected with one circuit if their axial frequencies are tuned to be slightly different but within the detector's bandwidth.  A coupling between two electrons takes place during the time that their two traps are tuned to make their axial frequencies the same.

\subsection{Damping and Detection in a Mirror-Image Trap}

For a mirror-image trap, the damping and hence the signal is maximized by connecting all of the electrodes in one plane to the damping resistor (i.e., $\rho_d^{(opt)} \rightarrow \infty$).  Choosing $\rho_d = z_c/2$ gives $\gamma_z$ that is 64\% of the total possible damping.  Choosing $\rho_d = z_c$ gives $\gamma_z$ that is 98\% of the total possible damping.  For the sample mirror-image planar trap (Tables~\ref{table:SampleMirrorImageTrap} and \ref{table:SampleMirrorImageTrapAbsolute}), connecting the damping resistor to the first two electrodes of one of the planes (i.e., choosing $\rt_d = \rt_2$) results in $\gamma_z = 2\pi\,13.23$ s$^{-1}$, which is 92\% of the damping that would result from detecting the signal induced on the entire electrode plane.

\section{Conclusion}
\label{sec:Conclusion}

A cylindrical Penning trap has been used to observe one-quantum spin flip and cyclotron transitions of a single trapped electron.  Attempts to make similar observations in a planar Penning trap did not succeed, generating some pessimism about whether this is possible.  In this report we show how to optimize the properties of a planar Penning trap to reduce the deadly amplitude dependence of the monitored axial frequency by orders of magnitude, and how to optimize the damping and detection.  We introduce a covered planar trap that is well-isolated from its environment, readily scalable to an array of one-electron traps, with one detector promising to suffice for the efficient simultaneous detection of multiple particles.   We also introduce mirror-image planar traps that are an attractive option because of their reflection symmetry.  Mirror-image traps can be electrically transformed into a covered planar trap while a particle is stored within. The optimized planar trap designs that are proposed offer new  routes toward observing a single electron in a planar trap, realizing a one-electron qubit, and using a scalable array of such qubits for  quantum information studies.

\begin{acknowledgments}
We are grateful for the support of the NSF and AFOSR.  Part of this work was completed while G.G.\ was supported by a Alexander von Humboldt Research Award. Some computations were carried out using the Odyssey cluster of the Harvard FAS Research Computing Group.   We are grateful to P.\ Bushev, F.\ Galve, F.\ Schmidt-Kaler, and G.\ Werth for helpful comments on this report.
\end{acknowledgments}

\bigskip
\centerline{--------------------------------------}

\appendix
\addcontentsline{toc}{section}{Appendices}

\section{An Earlier Calculation}
\label{sec:EarlierCalculation}

Minimizing the amplitude dependence of the axial frequency is the key to designing a planar trap within which a single electron can be suspended and used to realize a one-electron qubit. An accurate description and prediction of the properties of a planar Penning trap configuration thus requires a calculation of the amplitude dependence of the axial frequency.  Here we correct an earlier calculation \cite{PlanarPenningTrapMainz} of the amplitude dependent frequency shifts.

Earlier in this work the amplitude dependence of the axial frequency was shown to have the form
\begin{equation}
\omega_z(\At) = \omega_z \left[1 + \sum_{k=2}^\infty a_k \At^k \right].\tag{\ref{eq:AmplitudeDependentFrequency}}
\end{equation}
Ref.~\cite{PlanarPenningTrapMainz} differs by starting this sum with $k=1$, suggesting that the dominant axial frequency shift is first-order in the oscillation amplitude.  We find no first order shift. The substantial disagreement between the $a_k$ in Eq.~\ref{eq:ak} with the much simpler
\begin{equation}
a_k=\left|\frac{C_{k+2}}{2}\right| \label{eq:MainzMistake2}
\end{equation}
from Ref.~\cite{PlanarPenningTrapMainz} (translated into our notation) is illustrated in Table~\ref{table:FormulaComparison}.   The differing expressions for $a_3$, for example, are not even functions of the same $C_k$.  Higher order $a_k$ differ more.   (Eq.~2 of Ref.~\cite{ElectronsInPlanarPenningTrapMainz} and also Ref.~\cite{ElectronFrequenciesInPlanarPenningTrapMainz} repeat the Ref.~\cite{PlanarPenningTrapMainz} results.)

\begin{table}[h!]
\newcommand\T{\rule{0pt}{2.6ex}}
\newcommand\B{\rule[-1.2ex]{0pt}{0pt}}
\begin{tabular}{|c|c|}
  \hline
  \T This work & Ref.~\cite{PlanarPenningTrapMainz} \B \\
  \hline
  ~ & ~ \\
  $a_1 = 0$                                               &     $a_1 = \frac{|C_3|}{2}$      \\
  \T $a_2 = -\frac{15 (C_3)^2}{16}+\frac{3 C_4}{4}$          &      $a_2 = \frac{|C_4|}{2}$      \\
  \T~~~~~$a_3 = -\frac{15 (C_3)^3}{16}+\frac{3 C_3 C_4}{4}$~~~~~      &      ~~~~~$a_3 = \frac{|C_5|}{2}$~~~~~ \\
  ~ & ~ \\
  \hline
\end{tabular}
\caption{Lowest-order coefficients that describe the amplitude dependence of the axial frequency.  This work and Ref.~\cite{PlanarPenningTrapMainz} differ considerably in amplitude and sign, most notably in the leading lowest order.}
\label{table:FormulaComparison}
\end{table}

The amplitude dependence of the axial frequency must be calculated by solving the equation of motion, Eq.~\ref{eq:EquationOfMotion}, to get the $a_k$ in Eq.~\ref{eq:ak}, as outlined between these two equations. Except for the absolute value whose origin is not clear, Eq.~\ref{eq:MainzMistake2} from Ref.~\cite{PlanarPenningTrapMainz} is instead consistent with equating $(1/2) m \omega_z^2 (z-z_0)^2$ to  $q[V(0,z)-V(0,z_0)]$, solving for $\omega_z$,  expanding the square root in $z-z_0$, and identifying the latter with $A$.

Finally, the relationship between energy and amplitude in Eq.~11 of Ref.~\cite{PlanarPenningTrapMainz}, repeated in Eq.~3 of Ref.~\cite{ElectronsInPlanarPenningTrapMainz}, is missing a factor of two.  It should read $A = \sqrt{2 E/(m\omega_z^2)}$.

\section{Mainz Trap}
\label{sec:MainzTrap}

The first planar Penning trap used to store electrons was demonstrated at Mainz \cite{ElectronsInPlanarPenningTrapMainz,ElectronFrequenciesInPlanarPenningTrapMainz}.  A large number of electrons (estimated to be between 100 and 1000 electrons) were stored and the three motions of the electrons' center-of-mass were observed.  The trap dimensions and potentials are given by
\begin{align}
\rho_i &= \{3.15, 6.3, 9.45\}~\rm{mm}\\
V_i &= \{0, 16, -38.5\}~\rm{V},
\end{align}
with radii taken to extend to the center of the 0.3 mm wide gaps between electrodes.  The potentials come from the caption of Fig.~8 of Ref.~\cite{ElectronsInPlanarPenningTrapMainz}, but with signs reversed compared to what is reported since this is necessary to approximately replicate the curves in Fig.~8 of that work.  Finite boundaries are not included in our analysis, even though Sec.~\ref{sec:FiniteBoundaries} illustrates that they can be important, because the needed information is not available in the experimental accounts.

We calculate that $V_3 = -36.2$ V would make $C_3=0$, which seems to have been the goal, whereupon $a_2 = -0.8$ and $C_4 = -1.0$.  This gives a calculated single-particle thermal frequency width of $\Delta f_z = 310$ kHz for $T_z = 300$ K.  This thermal width is smaller than the calculated and measured widths of 1--6 MHz reported in Fig.~7 of Ref.~\cite{ElectronsInPlanarPenningTrapMainz}.
Unless the calculation is corrected as described in Appendix~\ref{sec:EarlierCalculation}, the calculated width should not agree with what we calculate.  The measured width may be wider than expected  because there are more trapped electrons than was estimated. Experimental experience in our lab also suggests that it is likely that the observed width is broadened by charges accumulated on the insulator within
the gaps, since the gaps were not deep enough to screen the potential from such charges.

A minimal requirement for a trap that could be used to observe a single electron is that $a_2$ be close to zero.  Fig.~\ref{fig:PossibleThreeRingGeometries} shows that for the relative geometry used in the Mainz trap there is no set of applied potentials that could make $a_2=0$.   In fact, $|a_2|<0.1$ cannot be achieved for any reasonable values of $V_i$ and $z_0$.  This is true even if the artificial and unnecessary constraint $V_1=0$ is relaxed.

\section{Ulm Trap}
\label{sec:UlmTrap}

The serious effort made at Ulm to try to observed a single electron trapped in a planar Penning trap \cite{ElectronsInPlanarPenningTrapUlm} did not succeed.  Given that we conclude above that optimized planar traps could likely be used to observe one trapped  electron, we examine here the trap geometry and the applied potentials that were used. The object is to check whether the performance of the Ulm trap is consistent with our calculations. Given that the observed linewidth is broader than calculated, we also discuss some practical considerations that may have affected the performance.

The trap geometry is described in Sec.~2.1 of Ref.~\cite{ElectronsInPlanarPenningTrapUlm}: ``The diameter of the central electrode and the width of the trapping electrodes are equal to 2 mm.''  In our notation, this is
\begin{equation}
\{\rho_1,\rho_2,\rho_3\} = \{1,3,5\}~\mathrm{mm}.\label{eq:UlmRho}
\end{equation}
Several bias configurations are mentioned in Sec.~2.2 and Fig.~2 of Ref.~\cite{ElectronsInPlanarPenningTrapUlm}:
\begin{align}
\{V_1,V_2,V_3\} &= \{0,7,7\} ~\rm{V},\label{eq:UlmV1}\\
\{V_1,V_2,V_3\} &= \{0,5.2,14\} ~\rm{V},\label{eq:UlmV2}\\
\{V_1,V_2,V_3\} &= \{0,4.8,14\} ~\rm{V}.\label{eq:UlmV3}
\end{align}
The penultimate paragraph of Sec.~2.3 of Ref.~\cite{ElectronsInPlanarPenningTrapUlm} mentions a set of ``optimized control voltages'',
\begin{equation}
\{V_1,V_2,V_3\} = \{0,-1,2.611\} ~\mathrm{V},\label{eq:UlmV4}
\end{equation}
presumably for the same geometry.  However, this last set of potentials seems not to have been used successfully, perhaps because of the greatly reduced trap depth that is produced.

The calculated properties for each of these four configurations are summarized in Tables \ref{table:UlmTrapScaled} and \ref{table:UlmTrapAbsolute}.  None of these trap configurations make $a_2$ close to zero, the likely minimal requirement for observing and controlling one trapped electron.  The mentioned ``optimized'' biasing scheme is actually worse than the others.  Finite boundaries are not included in our analysis, even though Sec.~\ref{sec:FiniteBoundaries} illustrates that they can be important, because the needed information is not provided in the experimental account.

\UlmTrapScaledTable

\UlmTrapAbsoluteTable

A choice was made to keep the center electrode and the plane outside the electrodes at the same potential.  This is an added constraint, $\Vt_1=0$, upon the four scaled parameters that determine the behavior of a 3-gap trap:  $\Vt_1$, $\Vt_2$, $\Vt_3$, and $\zt_0$. Two additional constraints, $C_1=0$ and $C_2=1$, are required to form a trap.  With the optional constraint there is one more parameter than there are constraints.   If we choose $\zt_0$ as the corresponding free parameter, Fig.~\ref{fig:UlmTrapWithVOneEqualZero}a shows the bias potentials that must be applied to realize each possible value of $\zt_0$.  Fig.~\ref{fig:UlmTrapWithVOneEqualZero}b gives the corresponding $a_k$ and $C_k$.   For reasonable $\zt_0 \le \tilde{\rho}_N$ we find that $|a_2| \ge 0.05$.
Having explored what trap performance is possible with the optional constraint, we note that there is no compelling reason to make this choice.  In fact such a choice would make it impossible to identify optimized planar Penning trap configurations.

\UlmTrapWithVOneEqualZeroFigure

It is possible to bias the Ulm trap electrodes to make $C_3=0$ by choosing $V_3/V_2 = -3.533$ for any $V_2 > 0$.  However, this choice also results in $C_4 = -0.77$ and $a_2 = 0.58$, the latter being worse than for the configurations in Table \ref{table:UlmTrapScaled}.  For this relative trap geometry, relaxing the optional constraint $V_1=0$ does not improve the trap performance.

Observing a single electron will be difficult if the thermal broadening of the axial frequency, $\Delta f_z$, is very large compared to the damping linewidth, $\gamma_z/(2\pi)$.
Table \ref{table:UlmTrapAbsolute} shows that this is indeed the case if the axial temperature is as low as 5 K, the lowest effective axial temperature that has been achieved without feedback cooling \cite{FeedbackCoolingPRL}.  The heat generated in the detection amplifier makes it very difficult to achieve such a temperature, even for 0.1 K surroundings, so the effective axial temperature could easily have been much higher than 5~K.  The table shows the thermal broadening for an effective axial temperature of 300~K.

Ref.~\cite{ElectronsInPlanarPenningTrapUlm} says that a still broader width of 3 MHz ``is expected and is in agreement with the measured data.''  Why this particular width should be expected is not specified.
However, it is not surprising that the observed frequency width is larger than we calculate because the width grows with the large and unknown number of electrons in the trap.  Experimental experience in our lab also suggests that it is likely that the observed width is broadened by charges accumulated on the insulator within the gaps since nothing in the circuit board technology used screens the potential from such charges.

One further item may be worth mentioning even though it is not completely understood. Many years ago, the first trap cooled with a dilution refrigerator  was located on the still of the refrigerator, within the refrigerator's inner vacuum container (IVC).  The vacuum was expected to be extremely good once the helium gas used to precool the IVC was pumped out.  However, no good one-electron signals were ever observed, for reasons not clearly understood, but seemingly related to the cryopumped gas on the surface of the electrodes. Only when the trap vacuum was separated from the IVC vacuum did we get the clean signals used to resolve one-quantum transitions with one electron \cite{QuantumCyclotron}.  The Ulm trap was also located within the IVC vacuum.  The long trapping lifetime observed with many trapped particles confirmed the expectation of a very good vacuum.  Whether isolating the trap vacuum from the IVC would improve the observed signals has not been investigated at Ulm.

Is it possible to observe a single electron in a planar trap?  Ref.~\cite{ElectronsInPlanarPenningTrapUlm} concludes that it is not because the anharmonicity will always be too great, and it also reports that a thermal width narrower than 5 kHz could not be calculated for any $N$-gap trap where $N$ is anything between 1 and 6. The only hope offered was that a much smaller trap might make the damping rate large enough to observe one electron despite the anharmonicity inherent in planar traps
\cite{ElectronsInPlanarPenningTrapUlm,UlmFinalElectronReview}. Indeed, Eq.~\ref{eq:DampingRate} shows that when the trap dimension is decreased the damping rate increases as the square of dimension.  With microfabrication methods it should be possible to fabricate smaller traps that thus will have a much larger damping.  What remains to be demonstrated is that the anharmonicity will not become large enough for small traps to offset the damping advantage.

Our conclusion is different and  much more optimistic.  We agree that it should be very difficult to observe a single electron in the planar Penning trap used at Ulm. However, the fundamental problem is the relative geometry of the Ulm trap, not its size.  The thermal broadening is too great as a result of a choice of trap geometry that cannot be optimized to make $a_2=0$.  Nonetheless, the conclusion \cite{ElectronsInPlanarPenningTrapUlm} that it is ``impossible'' to observe a single electron in a planar Penning trap with a mm size scale now seems much too strong. The new optimized geometries and applied potentials that we present here for planar Penning traps of any size offer the possibility of a very large reduction in the critical amplitude dependence of the axial frequency.  Experimental trials are warranted.


%

\end{document}